\documentclass[11pt]{article}

\usepackage[T1]{fontenc}
\usepackage{lmodern}
\usepackage[width=17cm,height=23.5cm,centering]{geometry}
\usepackage{lipsum}
\usepackage{amsfonts}
\usepackage{graphicx}
\usepackage{epstopdf}
\usepackage{algorithmic}
\usepackage{amssymb}
\usepackage{mathtools}
\usepackage[bb=libus]{mathalpha}
\usepackage{bm,mathrsfs,xcolor,subfig}
\usepackage{verbatim}
\usepackage{hyperref}
\usepackage{stmaryrd}

\makeatletter
\newcommand*{\transpose}{%
	{\mathpalette\@transpose{}}%
}
\newcommand*{\@transpose}[2]{%
	\raisebox{\depth}{$\m@th#1\intercal$}%
}
\makeatother

\newtheorem{remark}{Remark}
\newcommand*{\email}[1]{\href{mailto:#1}{\nolinkurl{#1}} } 
\newcommand{\bs}{\boldsymbol}

 \def\rot{\mathop{\rm curl}\nolimits}  
 
\ifpdf
  \DeclareGraphicsExtensions{.eps,.pdf,.png,.jpg}
\else
  \DeclareGraphicsExtensions{.eps}
\fi
\graphicspath{{Figures/}}

\title{High-order homogenisation of the time-modulated wave equation: non-reciprocity for a single varying parameter}

\author{
  Marie Touboul\thanks{Department of Mathematics, Imperial College London, London SW7 2AZ, UK (\email{m.touboul@imperial.ac.uk}).}
\and 
Bruno Lombard \thanks{Aix-Marseille Univ, CNRS, Centrale Marseille, LMA UMR 7031, Marseille, France (\email{lombard@lma.cnrs-mrs.fr}).}
\and 
Rapha{\"e}l Assier\thanks{Department of Mathematics, University of Manchester,
Manchester, UK (\email{raphael.assier@manchester.ac.uk }).}
\and 
S{\'e}bastien Guenneau\thanks{The Blackett Laboratory, Department of Physics, 
UMI 2004 Abraham de Moivre-CNRS,Imperial College London, London SW7 2AZ, UK (\email{s.guenneau@imperial.ac.uk}).}
\and 
Richard V. Craster\thanks{Department of Mathematics, 
UMI 2004 Abraham de Moivre-CNRS, 
Department of Mechanical Engineering, Imperial College London, London SW7 2AZ, UK. 
(\email{r.craster@imperial.ac.uk}). }
}

\usepackage{amsopn}

\newcommand{\beq}{\begin{equation}}
\newcommand{\eeq}{\end{equation}}
\newcommand{\ba}{\begin{eqnarray}}
\newcommand{\ea}{\end{eqnarray}}
\definecolor{colorcb}{rgb}{0 0.4470 0.7410}

\definecolor{colorcbq}{rgb}{0.4660 0.6740 0.1880}

\newcommand{\ds}{\displaystyle}

\begin{document}

\maketitle

\begin{abstract}
Laminated media with material properties modulated in space and time in the form of travelling waves have long been known to exhibit non-reciprocity. 
However, when using the method of low frequency homogenisation, it was so far only possible to obtain non-reciprocal effective media when both material properties are modulated in time, in the form of a Willis-coupling (or bi-anisotropy in electromagnetism) model. If only one of the two properties is modulated in time, while the other is kept constant, it was thought impossible for the method of homogenisation to recover the expected non-reciprocity since this Willis-coupling coefficient then vanishes. Contrary to this belief, we show that effective media with a single time-modulated parameter are non-reciprocal, provided homogenization is pushed to the second order. This is illustrated by numerical experiments (dispersion diagrams and time-domain simulations) for a bilayered modulated medium.
  \end{abstract}


\section{Introduction}

Modulation of the electric permittivity in space and time was proposed 60 years ago in seminal papers by Simon, Oliner, Hessel and Cassedy  \cite{Oliner1961,Simon1960,Cassedy1963,Cassedy1967}. It is observed that when modulating the properties in space and time, band diagrams are rotated in {reciprocal} space, and this is a general feature observed in all kinds of physical contexts, from optics \cite{galiffi2022photonics}, to acoustics and elastic waves in solids and plates \cite{nassar2020nonreciprocity}, including surface Rayleigh waves \cite{palermo2020surface} and even flexural-gravity waves \cite{farhat2021spacetime}.

The physical interpretation of rotated, thus non-reciprocal, band diagrams is that directionality of space–time modulations such as travelling-wave modulations breaks time-reversal symmetry. The breaking of reciprocity induced by a time-harmonic modulation of the refractive index has been exploited in the realisation of photonic isolators and circulators \cite{fang2012photonic} without the need of an applied external static magnetic field \cite{smigaj2010magneto}.
Similar effects have been explored in the context of acoustic and elastic waves \cite{Nassar2017} or for acoustic modulated gratings \cite{Pham2022}.

In-phase modulations of two parameters with the same strength within the acoustic, elastic and electromagnetic wave equations (e.g. the shear modulus and the compressional modulus, or the permittivity and
the permeability) result in the closing of
the high-frequency band gaps, at the same time as keeping the
nonreciprocal character of these systems (induced by the rotated k-space). Interestingly, it has been shown in the electromagnetic context that giant linear non-reciprocity can be achieved at high-frequency within space-time modulated media \cite{taravati2018giant}. 

A successful method to approximate wave propagation in heterogeneous media, for periodic media, is the two-scale asymptotic expansion method and the notion of slow and fast variables  \cite{Bensoussan2011,Bakhvalov1989,cioranescu1999,Conca1997}. Low-frequency homogenisation of laminates with in-phase modulations of two parameters leads to Willis-coupling \cite{Nassar2017} in acoustics and magneto-electric coupling \cite{Huidobro2019,Huidobro2021} in electromagnetics.  A leading-order asymptotic analysis of the low-frequency regime for the time-modulated wave equation is also addressed in \cite{lurie2007introduction}. Homogenisation of a larger class of periodic multiscale space-time media has been studied with various techniques in the context of linear and nonlinear parabolic and hyperbolic equations \cite{Holmbom2005,floden2007homogenization,ariel2014iterated,dehamnia2022multiscale}.
This makes possible interesting features ranging from frequency-momentum transitions to
compression and amplification of electromagnetic signals, one-way hyperbolic metasurfaces, and even non-Hermitian and topological phenomena. However, while non-reciprocity is expected with only one parameter modulated in time, the leading-order low-frequency homogenisation does not encapsulate this behaviour \cite{lurie2007introduction}.  Therefore, leading-order homogenisation is able to describe non-reciprocity when two parameters are modulated in time, but not as soon as one of them is kept constant.

However, there is no clear physical reason for this necessity of in-phase modulations of two parameters to get non-reciprocity at low frequency. Moreover, while in elasticity the modulation of the elastic constants is possible \cite{Danas2012,Swinteck2015,Deymier2017,Matar2012,Vasseur2011}, it seems much more complicated to modulate the mass density. In a similar fashion, modulating the permittivity in electromagnetism has been achieved \cite{Tirole2022,Shaltout2019,Caloz2020} but modulating the permeability while natural materials are non magnetic is much more demanding. This is also the case for acoustic waves propagating in a fluid, where a modulation of both physical parameters cannot be done without generating a background flow or breaking mass conservation \cite{Martin2020}. This motivates the present work that establishes the second-order homogenised model (see \cite{Boutin1993,Chen2001,Fish2001,Fish2004,LAMACZ2011,Allaire2018} in the non modulated case) of the time-modulated scalar wave equation and proves that non-reciprocity is achieved even if only one of the parameters is modulated.

In Section \ref{Sec:Asymp}, the configuration of a time-modulated laminate is recalled in the framework of antiplane elasticity and homogenisation is performed up to the second order. For each order, we study the limit case of only one parameter being modulated and it is proved that non-reciprocity still happens in this case provided we consider the second-order model. In Section \ref{SecNum}, dispersion diagrams obtained by Bloch-Floquet theory and by homogenisation are compared, to validate the second-order model and to quantify the non-reciprocal effects. Time-domain simulations are also performed in the microstructured medium to validate the conclusions obtained through homogenisation. Finally, Appendix \ref{Sec:EM} explains why these results presented in the framework of antiplane elasticity, i.e. of non-reciprocity even if only one of the parameters is modulated in time, still hold when considering transverse electromagnetic waves. 
\section{Homogenisation of a time-modulated laminate in antiplane elasticity}\label{Sec:Asymp}
\begin{figure}[htbp]
\begin{center}
\includegraphics[height=2.5cm,trim={0cm 
 0cm 0cm 0cm },clip]{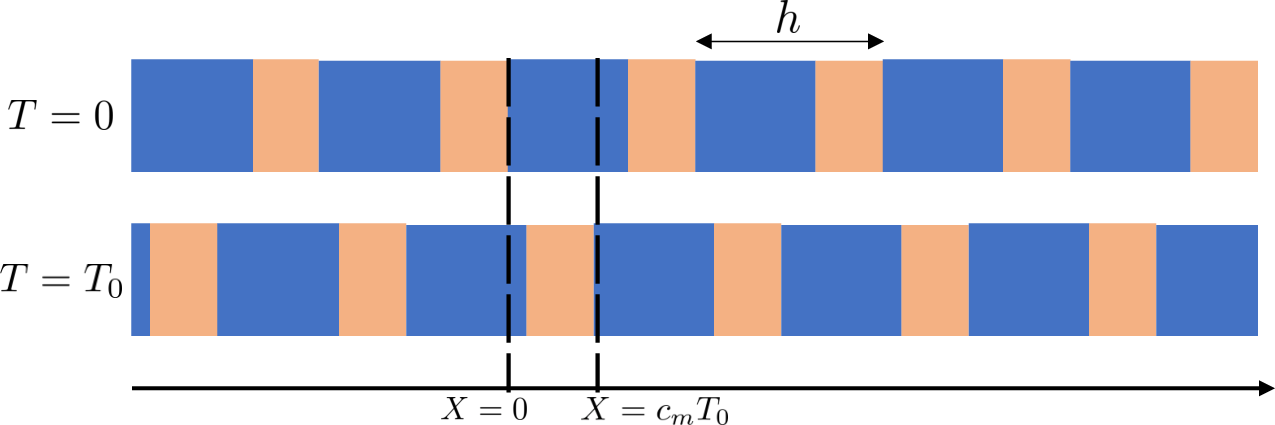}
\end{center}
\caption{Bilaminate modulated in a wave-like fashion at $T=0$ (top) and $T=T_0$ (bottom).} 
\label{fig:laminate}
\end{figure}

We first consider a 1D $h$-periodic elastic medium
of shear modulus $\mu_h$ and mass density $\rho_h$ in the antiplane elasticity framework. Therefore the equations for the elastic displacement $U_h(X,T)$ and the stress $\Sigma_h(X,T)$ read
\begin{equation}
    \label{dim_syst_1D}
    \partial_T(\rho_h \partial_T U_h) = \partial_X\Sigma_h+F \quad \text{ with } \quad \Sigma_h =\mu_h\partial_X U_h,
\end{equation}
where $X$ {(in m)} and $T$ {(in s)} denote the space coordinate and the time, respectively, and $F(X,T)$ is a source term. The partial derivatives with respect to space and time are denoted by $\partial_X$ and $\partial_T$, respectively.
Moreover, the physical parameters $\rho_h$ {(in kg m$^{-3}$)} and $\mu_h$ {(in Pa)} are modulated in space and time in a wave-like fashion of modulation speed $c_m$:
\begin{equation}
    \label{rho_and_kappa}
    \rho_h(X,T) = \rho(X-c_mT) \quad \text{ and } \quad \mu_h(X,T) = \mu(X-c_m T)
\end{equation}
with $\rho$ and $\mu$ $h$-periodic functions assumed to be piecewise continuous, see Figure \ref{fig:laminate}. \\
In order to satisfy some smoothness assumptions \cite{Cassedy1963,Cassedy1967}, we further assume that $c_m$ satisfies:
\begin{equation}
\label{eq:sub_or_sup}
    c_m < \min_{\xi\in(0,h)} c(\xi)\text{ or } c_m > \max_{\xi\in(0,h)} c(\xi) \quad \text{ with } \quad c=\sqrt{\frac{\mu}{\rho}}.
\end{equation}
\vspace*{-1cm}
\subsection{Moving frame}\label{SecMove}
As in \cite{Nassar2017}, we reformulate the problem in the moving frame, which will be useful to determine the continuity conditions for the fields, and for time-domain numerical simulations in Section \ref{SecNum}. \\
The momentum equation and the constitutive law \eqref{dim_syst_1D} can be written
\begin{equation}
\ds \partial_X V_h=\partial_T\left(\frac{\Sigma_h}{\mu_h}\right) \quad \text{ with } \quad \partial_X\Sigma_h=\partial_T\left(\rho_h V_h\right)-F, 
\label{Momentum-CL}
\end{equation}
where $V_h(X,T)=\partial_TU_h(X,T)$ is the velocity. Setting
\begin{equation}
\bs{W}_h=\left(
\begin{array}{c}
V_h\\
\Sigma_h
\end{array}
\right),\qquad
\bs{N}_h=\left(
\begin{array}{cc}
0 & -1/\mu_h\\
-\rho_h & 0
\end{array}
\right),\qquad
\bs{F}=\left(
\begin{array}{c}
0\\
-F
\end{array}
\right),
\label{NotaU}
\end{equation}
the equations \eqref{Momentum-CL} are written
\begin{equation}
\partial_X\bs{W}_h(X,T)+\partial_T(\bs{N}_h(X,T)\,\bs{W}_h(X,T))=\bs{F}(X,T).
\label{dXUdTNU}
\end{equation}
From \eqref{rho_and_kappa}, it follows that 
\begin{equation}
    \bs{N}_h(X,T)=\bs{N}(X-c_m T)=\left(
\begin{array}{cc}
0 & -1/\mu(X-c_m T)\\
-\rho(X-c_m T) & 0
\end{array}
\right).
\end{equation} One introduces a moving frame with coordinates $(\xi,T)=(X-c_mT,T)$; the fields in this frame are denoted by $\tilde{g}(\xi,T)=g(X,T)$ and satisfy the chain rules
\begin{equation}
    \label{chain_rule_g}
     \partial_\xi\tilde{g}=\partial_X {g} \quad \text{ and }\quad \partial_T\tilde{g}= \partial_T g +  c_m \partial_X g .
\end{equation}

One then defines
\begin{equation}
\begin{array}{l}
\bs{\Psi}(\xi,T)=\left(\bs{I}-c_m\,\bs{N}(\xi)\right)\,\tilde{\bs{W}}_h(\xi,T)=\left(
\begin{array}{c}
\ds \tilde{V}_h(\xi,T)+\frac{c_m}{\mu(\xi)}\,\tilde{\Sigma}_h(\xi,T)\\
\ds \tilde{\Sigma}_h(\xi,T)+c_m\,\rho(\xi) \tilde{V}_h(\xi,T)
\end{array}
\right)
\equiv \left(
\begin{array}{c}
\Psi_1(\xi,T)\\
\Psi_2(\xi,T)
\end{array}
\right),\\ [12pt]
\ds \bs{B}(\xi)=\bs{N}(\xi)\,\left(\bs{I}-c_m\,\bs{N}(\xi)\right)^{-1}=\frac{1}{c(\xi)^2-c_m^2}\left(
\begin{array}{cc}
c_m & -1/\rho(\xi) \\
-\mu(\xi) & c_m
\end{array}
\right),
\end{array}
\label{PsiB}
\end{equation}
where $\bs{I}$ is the $2\times2$ identity matrix. In the limit-case $c_m=0$, $\bs{\Psi}=\bs{W}_h$. Using the chain rules \eqref{chain_rule_g} on \eqref{dXUdTNU} yields
\begin{equation}
\partial_\xi \bs{\Psi}(\xi,T)+\bs{B}(\xi)\,\partial_T \bs{\Psi}(\xi,T)=\tilde{\bs{F}}(\xi,T).
\label{dXiPsi}
\end{equation}
Let us consider an interface at $\xi_n$. Integrating \eqref{dXiPsi} on $[\xi_n-\varepsilon,\xi_n+\varepsilon]$ with $\varepsilon\rightarrow 0$ gives $\llbracket \bs{\Psi}\rrbracket_{\xi_n}=\bs{0}$, where $\llbracket g \rrbracket_a=g(a^+)-g(a^-)$ refers to the jump across $a$. Using \eqref{PsiB}, these jump conditions in the moving frame read
\begin{equation}
\left\llbracket \tilde{V}_h+\frac{c_m}{\mu}\tilde{\Sigma}_h\right\rrbracket_{\xi_n}=0,\qquad \llbracket\tilde{\Sigma}_h+c_m\,\rho\,\tilde{V}_h\rrbracket_{\xi_n}=0,
\label{JCmove}
\end{equation}
which are the jump conditions one would obtain by Reynolds' transport theorem.
\begin{remark}
Since $\partial_T \tilde{U}_h(\xi,T) = V_h(X,T)+\frac{c_m}{\mu(\xi)}\Sigma_h(X,T)$ with \eqref{chain_rule_g} and \eqref{Momentum-CL}, the first jump condition in \eqref{JCmove} leads to 
$$
\left\llbracket{\partial_T}\tilde{U}_h\right\rrbracket_{\xi_n}={\partial_T}\llbracket \tilde{U}_h\rrbracket_{\xi_n}=0
$$
in the moving frame where the interfaces are at fixed positions $\xi_n$. Since the jump is zero at $T=0$, then we get $\llbracket \tilde{U}_h\rrbracket_{\xi_n}$=0. Equations \eqref{JCmove} are therefore equivalent to continuity for both $\tilde{U}_h$ and $\tilde{\Sigma}_h+c_m\,\rho\,\tilde{V}_h$ in the moving frame. Therefore, the same holds for both $U_h$ and $\Sigma_h+c_m\rho_h V_h$ in the fixed frame.
\label{rk:continuity}
\end{remark}
Setting
\begin{equation}
\bs{A}(\xi)=\bs{B}^{-1}(\xi)=\left(
\begin{array}{cc}
-c_m & -1/\rho(\xi)\\
-\mu(\xi) & -c_m
\end{array}
\right),
\label{A}
\end{equation}
and 
\begin{equation}
\bs{\Phi}(\xi+c_m T,T)=\bs{B}^{-1}(\xi)\tilde{\bs{F}}(\xi,T)=\left(
\begin{array}{c}
F(\xi+c_m T, T)\\
c_m\,F(\xi+c_m T, T)
\end{array}
\right),
\label{Phi}
\end{equation}
the system \eqref{dXiPsi} leads to 
\begin{equation}
\partial_T\bs{\Psi}(\xi,T)+\bs{A}(\xi)\,\partial_\xi \bs{\Psi}(\xi,T)=\bs{\Phi}(\xi+c_m T,T).
\label{dTPsi}
\end{equation}
The source term is therefore no longer fixed in the moving frame.  The eigenvalues of $\bs{A}$ are $\pm c_0-c_m$. Lastly, an energy balance of \eqref{dTPsi} without source term leads to 
\begin{equation}
\frac{d \cal E}{dT}{}=0,\quad \mbox{ with } {\cal E}(T)=\frac{1}{2}\int_\mathbb{R}\left(\rho(\xi)\Psi_1(\xi,T)^2+\frac{1}{\mu(\xi)}\,\Psi_2(\xi,T)^2\right)\,d\xi.
\label{NRJ}
\end{equation}
This can be interpreted as the conservation of the sum of a kinetic and a potential energy in the moving frame. However, in the fixed frame, there is no conservation of energy 
 due to the time modulation of the constitutive properties \cite{Nassar2017}.

\subsection{Non-dimensionalisation}
In order to simplify the mathematical notations, we start by
non-dimensionalising the physical problem \eqref{dim_syst_1D}. To do so, we introduce a characteristic dimensional density $\rho^{\star} $, Young's modulus $\mu^{\star}$, wavespeed $c^{\star} = \sqrt{\mu^{\star} / \rho^{\star}}$, and wavenumber $k^\star$. 
These can be used in order to define the following non-dimensional quantities
\begin{equation}
\begin{aligned}
\eta = k^\star h, \ x=k^\star X, \ t=(k^\star c^\star)T, \ u_\eta=k^\star U_h, \ \sigma_\eta=\frac{1}{\mu^{\star}}\Sigma_h, \ \alpha=\frac{\rho}{\rho^\star}, \ \beta=\frac{\mu}{\mu^{\star}}, \ c=\frac{c_m}{c^\star}, \ f=\frac{F}{k^\star \mu^\star}.
  \label{eq:nondimparam}
\end{aligned}
\end{equation}
Using these quantities, \eqref{dim_syst_1D} can be rewritten as the non-dimensional governing equation
\begin{equation}
    \label{non_dim_equation} 
    \partial_t(\alpha \partial_t u_\eta) = \partial_x(\beta\partial_x u_\eta)+f
\end{equation}
together with continuity for $u_\eta$ and $\sigma_\eta+c\,{\partial_t u_\eta}$ at interfaces (see Remark \ref{rk:continuity}).
\begin{remark}
    Due to \eqref{eq:sub_or_sup}, $ \beta-c^2\alpha$ is either strictly positive or strictly negative everywhere.
\color{black}
\end{remark}
\subsection{Two-scale analysis}
We now make the assumption that the characteristic wavelength $\lambda^\star=\frac{2\pi}{k^\star}$ is much larger than the periodicity $h$ so that $\eta\ll 1$ and is the small positive parameter of the low-frequency setting. The material parameters $\alpha$ and $\beta$ vary on a fine scale associated with the rescaled coordinate $y=k^\star\xi=\frac{x-ct}{\eta}$. \\
Following the two-scale asymptotic technique, we further assume that the field $u_\eta$ have small-scale variations that are described by $y$, and slow continuous features as well, which can be described by the variable $x$. Accordingly the field $u_\eta$ is expanded using the following ansatz
\begin{equation}
    \label{eq:ansatz_u}
    u_\eta(x,t) = \sum_{j\geq 0} \eta^ju_j(x,y,t)
\end{equation}
where $x$ and $y$ are assumed to be independent variables implying that  ${\partial_x} \leftrightarrow
{\partial_x} + \frac{1}{\eta}{\partial_y} \text{ and }{\partial_t} \leftrightarrow
{\partial_t} - \frac{c}{\eta} {\partial_y}.$ \\
 We introduce the stress and momentum
 \begin{equation}
 \label{eq:sigma_m}
     \sigma_j=\beta(\partial_x u_j+\partial_y u_{j+1}) \text{ and  } m_j=\alpha(\partial_t u_j-c\partial_y u_{j+1}). 
 \end{equation}
To ensure the correct continuity and periodicity conditions for $u_\eta$ and $\sigma_\eta$, the quantities $u_j$, $\sigma_j$ and $m_j$ are assumed to have the following properties: $u_j$ continuous with respect to the first variable, 
     $u_j$ 1-periodic with respect to the second variable, 
     $u_j$ and $\sigma_j+c m_j$ continuous with respect to the second variable (see Remark \ref{rk:continuity}).

The non-dimensional problem can hence be rewritten as the governing equation 
\begin{equation}
    \label{eq:non_dim_two_variables}
    \left(\partial_t-\frac{c}{\eta}\partial_y\right)\left(\alpha(y)\left(\partial_t-\frac{c}{\eta}\partial_y\right)\sum_{j\geq 0} \eta^ju_j \right)=\left(\partial_x+\frac{1}{\eta}\partial_y \right)\left(\beta(y) \left(\partial_x+\frac{1}{\eta} \partial_y\right)\sum_{j\geq 0} \eta^ju_j\right)+f(x,t),
\end{equation}
together with the properties given above.
\subsection{Homogenised equation at leading order}
The homogenised model at leading order has been obtained in various references, see \cite{Nassar2017} for elasticity. We summarize the procedure here for sake of completeness. \\
We start by introducing the mean value $\langle g \rangle =\int_0^1 g(y)\mathrm{d}y$ for a function $g(y)$.
Identifying the terms of order $\eta^{-2}$ in \eqref{eq:non_dim_two_variables} we get  
\begin{equation}
    \partial_y((\beta-c^2\alpha)\partial_y u_0)=0,
\end{equation}
which leads to 
$$ \partial_y u_0(x,y,t)=\frac{1}{\beta(y)-c^2\alpha(y)}q(x,t),$$ for some unknown function $q(x,t)$. Integrating this equation on a unit cell leads to $q(x,t)=0$, implying that
\begin{equation}
    \label{order_zero_macro}
    u_0(x,y,t) = \mathcal{U}_0(x,t).
\end{equation}
Collecting now the terms of order $\eta^{-1}$ in \eqref{eq:non_dim_two_variables}, we get
\begin{equation}
    \label{eq:collect_order1}
\partial_y \left[(\beta-c^2\alpha)\partial_y u_1 \right]=-c\partial_y(\alpha\partial_t u_0)-\partial_y(\beta\partial_x u_0).
\end{equation}
We can then write $u_1$ as 
\begin{equation}
    \label{decompo_u1}
    u_1(x,y,t)=\mathcal{U}_1(x,t)+P(y)\partial_x\mathcal{U}_0(x,t)+Q(y)\partial_t\mathcal{U}_0(x,t),
\end{equation}
with $P$ and $Q$ satisfying the following cell problems
\begin{equation}
    \label{cell_pb_P}
    \left\lbrace
    \begin{aligned}
    &\partial_y \left[(\beta-c^2\alpha)P'+\beta \right]=0 \\
    & \text{1-periodicity of } P \text{, } \langle P \rangle = 0, \text{ and } \text{ continuity of $P$ and $(\beta-c^2\alpha)P'+\beta$},
    \end{aligned}
    \right.
\end{equation}
and
\begin{equation}
    \label{cell_pb_Q}
    \left\lbrace
    \begin{aligned}
    &\partial_y \left[(\beta-c^2\alpha) Q'+c\alpha\right]=0 \\
    &  \text{1-periodicity of } Q \text{, } \langle Q \rangle = 0, \text{ and continuity of $Q$ and $(\beta-c^2\alpha) Q'+c\alpha$},
    \end{aligned}
    \right.
\end{equation}
where hereafter $g'$ will denote the derivative of any function $g$ of one variable. Then, collecting terms of order $\eta^0$ we get
\begin{equation}
    \label{eq:gov_eq_order0}
    \partial_t m_0 -c\partial_y m_1=\partial_x\sigma_0+\partial_y\sigma_1+f,
\end{equation}
which, upon integration on a unit cell and due to continuity of $\sigma_1+c m_1$, leads to 
\begin{equation}
    \partial_t \langle m_0 \rangle = \partial_x \langle\sigma_0\rangle+f.
    \label{eq:identif_0}
\end{equation}
Using \eqref{decompo_u1}, we get 
\begin{equation}
    \langle m_0 \rangle = \langle \alpha(1-c\,Q')\rangle \partial_t\mathcal{U}_0-c\,\langle \alpha\, P'\rangle \partial_x\mathcal{U}_0
\quad \text{ and } \quad
    \langle \sigma_0\rangle = \langle \beta\,Q'\rangle\partial_t\mathcal{U}_0+\langle\beta(1+P')\rangle\partial_x\mathcal{U}_0.
\end{equation}
We can prove by integrations by parts (see reciprocity identities in Appendix \ref{Sec:App}, and equation \eqref{eq:IPP6} in the present case) that the so-called Willis coupling coefficient satisfies
\begin{equation}
    \label{Willis_coupling_coeff}
    W_0:=c\langle\alpha\,P'\rangle = \langle\beta\,Q'\rangle.
\end{equation} 
We also introduce the effective parameters 
\begin{equation}
    \label{def_E_0_rho0}
    \alpha_0=\langle \alpha(1-cQ'(y))\rangle
\text{ and }
    \beta_0=\langle \beta(1+P'(y))\rangle
\end{equation}
so that the leading-order equation \eqref{eq:identif_0} can finally be written
\begin{equation}
    \label{final_eq_order_0}
    \alpha_0\partial^2_{tt}\mathcal{U}_0-2W_0\partial^2_{tx}\mathcal{U}_0-\beta_0\partial^2_{xx}\mathcal{U}_0=f.
\end{equation}
Upon coming back to dimensionalised variables using \eqref{eq:nondimparam}, setting $f=0$, and looking for wave-like solutions, we obtain the dispersion relation 
\begin{equation}
    \label{DD_order0}
    -\alpha_0\omega^2+2W_0c^\star\omega k +\beta_0 (c^\star)^2k^2 = 0
\end{equation}
where $\omega$ and $k$ are the dimensionalised angular frequency and wavenumber, respectively. Odd orders of space derivatives in the effective equation \eqref{final_eq_order_0}, or in an equivalent manner odd powers of $k$ in the dispersion relation \eqref{DD_order0}, implies that reciprocity is broken since there is no longer symmetry between $k$ and $-k$. However, this occurs only if the Willis coupling coefficient $W_0$ is non-zero.
\begin{remark}\textbf{ What happens at leading order if only one parameter is modulated in time?}\\
    If $\alpha$ does not depend on $y$, we can choose $\rho^\star$ such that $\alpha=1$. Then $W_0=c\langle \alpha P' \rangle =c\langle P' \rangle = 0$, the Willis-coupling term vanishes and the effective equation is the usual one  
    \begin{equation}
    \label{final_eq_order_0_recip}
    \alpha_0\partial^2_{tt}\mathcal{U}_0-\beta_0\partial^2_{xx}\mathcal{U}_0=f.
\end{equation}
In preparation for the following orders, we also notice that $Q=0$ and $\alpha_0=1$ in this case. \\
Similarly, if $\beta=1$, then $W_0=\langle \beta Q'\rangle =\langle Q' \rangle = 0$ and the Willis-coupling term vanishes. We also notice that $P=0$ and $\beta_0=1$.\\
\end{remark}
\vspace*{-0.5cm}
Consequently, the Willis coupling coefficient $W_0$ vanishes if either $\alpha$ or $\beta$ does not depend on $y$, therefore non-reciprocity appears in this leading-order effective equation only if both physical parameters are modulated in space and time. This is consistent with results of the literature of low-frequency homogenization of time-modulated media \cite{Nassar2017,Huidobro2021}. However, we expect non-reciprocity even if one parameter is modulated in time according to Bloch-Floquet analysis and dispersion relations obtained in the literature \cite{Cassedy1963,Cassedy1967}. In the forthcoming sections, we prove that the modulation of both parameters is not a necessary condition for non-reciprocity in the low-frequency setting, but rather a limitation of the leading-order homogenised model.

\subsection{Homogenised equation at order 1}
We start by inserting \eqref{eq:sigma_m} for $j=0,1$ in \eqref{eq:gov_eq_order0}:
\begin{equation}
    \label{eq:gov_eq_order0_u2}
    \partial_y\left[ (\beta-c^2\alpha)\partial_y u_2\right]=\alpha\partial^2_{tt}u_0-\beta\partial^2_{xx}u_0-\alpha c\partial^2_{ty}u_1-c\partial_y(\alpha\partial_t u_1)-\beta\partial^2_{xy}u_1-\partial_y(\beta\partial_x u_1)-f.
\end{equation}
Using \eqref{decompo_u1}, and \eqref{final_eq_order_0} to replace the $\partial^2_{tt}\mathcal{U}_0$ terms, we can write $u_2$ as
\begin{equation}
    \label{form_u2}
    \begin{aligned}
    u_2(x,y,t) &= \mathcal{U}_2(x,t)+P(y)\partial_x\mathcal{U}_1(x,t)+Q(y)\partial_t\mathcal{U}_1(x,t)\\ &+R(y)\partial^2_{xx}\mathcal{U}_0(x,t)+S(y)\partial^2_{xt}\mathcal{U}_0(x,t)+B(y) f(x,t)
    \end{aligned}
\end{equation}
where $R$, $S$ and $B$ are solutions of the following cell problems
\begin{equation}
    \label{cell_pb_R}
    \left\lbrace
    \begin{aligned}
    &\partial_y \left[(\beta-c^2\alpha) R'+\beta P + \frac{\beta_0}{\alpha_0}c\alpha Q \right]=-\beta(1+P')+\frac{\beta_0}{\alpha_0}\alpha(1-cQ') \\
    & \text{1-periodicity of } R \text{, } \langle R \rangle = 0,  \text{ and continuity of $R$ and $(\beta-c^2\alpha)R'+\beta P+\frac{\beta_0}{\alpha_0}\alpha c Q$},
    \end{aligned}
    \right.
\end{equation}
\begin{equation}
    \label{cell_pb_S}
    \left\lbrace
    \begin{aligned}
    &\partial_y \left[(\beta-c^2\alpha) S'+\beta Q+c\alpha P+\frac{2W_0}{\alpha_0}c\alpha Q\right]=-\alpha c P'-\beta Q'+\frac{2W_0}{\alpha_0}\alpha(1-cQ') \\
    & \text{1-periodicity of } S \text{, } \langle S \rangle = 0, \text{ and continuity of $S$ and $(\beta-c^2\alpha)S'+\beta Q+c\alpha P +\frac{2W_0}{\alpha_0}c\alpha Q$},
    \end{aligned}
    \right.
\end{equation}
\begin{equation}
    \label{cell_pb_B}
    \left\lbrace
    \begin{aligned}
    &\partial_y \left[(\beta-c^2\alpha) B'+\frac{c\alpha}{\alpha_0} Q\right]=\frac{\alpha(1-cQ')}{\alpha_0}-1 \\
    & \text{1-periodicity of } B \text{, } \langle B \rangle = 0, \text{ and continuity of $B$ and $(\beta-c^2\alpha) B'+\frac{c\alpha}{\alpha_0} Q$},
    \end{aligned}
    \right.
\end{equation}
with the continuity conditions coming from continuity of $u_2$ and $\sigma_1+c m_1$. Collecting the terms of order $\eta$ in \eqref{eq:non_dim_two_variables} yields: 
\begin{equation}
    \label{eq:order_eta}
    \partial_t m_1 -c\partial_y m_2=\partial_x\sigma_1+\partial_y\sigma_2
\end{equation}
which once averaged gives 
\begin{equation}
    \label{eq:order_eta_averaged}
    \partial_t \langle \alpha(\partial_t u_1-c\partial_y u_2)\rangle = \partial_x\langle\beta(\partial_x u_1+\partial_y u_2)\rangle.
\end{equation}
Inserting \eqref{decompo_u1} and \eqref{form_u2} in \eqref{eq:order_eta_averaged} leads to the final effective equation for the first-order mean field: 
\begin{equation}
    \label{eq:effective_eq_order1}
        \alpha_0\partial^2_{tt}\mathcal{U}_1-2W_0\partial^2_{tx}\mathcal{U}_1-\beta_0\partial^2_{xx}\mathcal{U}_1=\mathcal{F}(\mathcal{U}_0)+\mathcal{A}(f)
\end{equation}
where the first source term reads
\begin{equation}
\begin{aligned}
   \mathcal{F}(g) &= -\langle \alpha Q \rangle \partial^3_{ttt}g+\left[-\langle\alpha P \rangle +c\langle \alpha S'\rangle\right]\partial^3_{ttx}g\\
   &+\left[c\langle\alpha R'\rangle+\langle\beta Q\rangle+\langle\beta S'\rangle\right]\partial^3_{txx}g+\left[ \langle\beta P\rangle +\langle\beta R'\rangle\right]\partial^3_{xxx}g
   \end{aligned}
   \label{eq:def_F}
\end{equation}
for any function $g(X,t)$ and the second one reads
\begin{equation}
   \mathcal{A}(f) = \langle \beta B'\rangle \partial_x f +c\langle \alpha B'\rangle \partial_t f.
   \label{eq:def_M}
\end{equation}

\begin{remark}\textbf{What happens at order 1 if only one parameter is modulated in time?}\\
Let us assume that mass density is constant ($\alpha=1$). Because the cell problem solutions are continuous and have zero mean-values, the source term simplifies to
\begin{equation}
   \mathcal{F}(\mathcal{U}_0) =
   \langle\beta S'\rangle \partial^3_{txx}\mathcal{U}_0+\left[ \langle\beta P\rangle +\langle\beta R'\rangle\right]\partial^3_{xxx}\mathcal{U}_0.
\end{equation}
The reciprocity identities \eqref{eq:IPP3}, \eqref{eq:IPP1} and \eqref{eq:IPPB1}, given in Appendix \ref{Sec:App} allow us to conclude that
\begin{equation}
\label{eq:F_alpha1}
  \mathcal{F}(\mathcal{U}_0) =0
\text{ and }
  \mathcal{A}(f) =0. 
\end{equation}
Similarly, if $\beta=1$, the source terms simplify to 
\begin{equation}
   \mathcal{F}(\mathcal{U}_0)+\mathcal{A}(f) =-\langle\alpha Q\rangle \partial^3_{ttt}\mathcal{U}_0+c\langle\alpha S'\rangle \partial^3_{ttx}\mathcal{U}_0 +c\langle \alpha R'\rangle \partial^3_{txx}\mathcal{U}_0+c\langle \alpha B'\rangle \partial_t f,
\end{equation}
and the reciprocity identities \eqref{eq:IPP4}, \eqref{eq:IPP2}, and  \eqref{eq:IPPB2}, together with \eqref{final_eq_order_0_recip} and $W_0=0$ lead to 
\begin{equation}
    \label{eq:F_beta1}
    \mathcal{F}(\mathcal{U}_0)+\mathcal{A}(f) =0.
\end{equation}
Therefore, as soon as only one parameter is modulated in time, the effective equation reduces to 
\begin{equation}
    \label{order_1_one_modulated}
    \alpha_0\partial^2_{tt}\mathcal{U}_1-\beta_0\partial^2_{xx}\mathcal{U}_1=0 
\end{equation}
and the dispersion relation is the same as at the leading order.
\end{remark}
Consequently, if either $\alpha$ or $\beta$ does not depend on $y$, both the Willis coupling coefficient $W_0$ and the source term $\mathcal{F}(\mathcal{U}_0)$ vanish; then in the absence of source $f$ the equation for the first-order mean field $\mathcal{U}_1$ is the same as for the zeroth-order one. Therefore, up to the first order, non-reciprocity appears in the effective equation only if both physical parameters are modulated in space and time, but we will see in the next section that it is not the case at order 2. 

\subsection{Homogenised equation at order 2}
We start by inserting \eqref{eq:sigma_m} in \eqref{eq:order_eta} to get an equation for $u_3$. We then make use of the expressions obtained for $u_1$ \eqref{decompo_u1} and $u_2$ \eqref{form_u2}.  
With a similar methodology as for the previous order, we replace the terms $\partial^2_{tt}\mathcal{U}_1$, $\partial^3_{ttt}\mathcal{U}_0$, and $\partial^3_{xxx}\mathcal{U}_0$, using \eqref{eq:effective_eq_order1}, \eqref{final_eq_order_0} differentiated with respect to time, and \eqref{final_eq_order_0} differentiated with respect to space, respectively. The field $u_3$ can therefore be written  
\begin{equation}
    \label{eq:form_u3}
    \begin{aligned}
    u_3(x,y,t) &= \mathcal{U}_3(x,t)+P(y)\partial_x\mathcal{U}_2(x,t)+Q(y)\partial_t\mathcal{U}_2(x,t)+R(y)\partial^2_{xx}\mathcal{U}_1(x,t)+S(y)\partial^2_{tx}\mathcal{U}_1(x,t)\\
&+M(y)\partial^3_{ttx}\mathcal{U}_0(x,t)+N(y)\partial^3_{txx}\mathcal{U}_0(x,t)+B_x(y)\partial_x f(x,t)+B_t(y)\partial_t f(x,t)
    \end{aligned}
\end{equation}
with $M$ and $N$ solutions of the following cell problems
\begin{equation}
    \label{cell_pb_T}
    \left\lbrace
    \begin{aligned}
    &\partial_y \left[(\beta-c^2\alpha)M'+c\alpha S + \frac{\alpha_0}{\beta_0}\beta R\right]=\alpha\left(P-cS'+\frac{2W_0}{\alpha_0}Q\right)-\frac{\alpha_0}{\beta_0}(\beta P+\beta R')\\
    & \text{1-periodicity of } M \text{, } \langle M \rangle = 0, \text{ and continuity of $M$ and $(\beta-c^2\alpha) M'+c\alpha S + \frac{\alpha_0}{\beta_0}\beta R$},
    \end{aligned}
    \right.
\end{equation}
\begin{equation}
    \label{cell_pb_V}
    \left\lbrace
    \begin{aligned}
    &\partial_y \left[(\beta-c^2\alpha) N'+c\alpha R+\beta S-\frac{2W_0}{\beta_0}\beta R\right]=-\beta(Q+S')+\alpha\left(\frac{\beta_0}{\alpha_0}Q-cR'\right)+\frac{2W_0}{\beta_0}(\beta P+\beta R') \\
    & \text{1-periodicity of } N \text{, } \langle N \rangle = 0,  \text{ and continuity of $N$ and $(\beta-c^2\alpha) N'+c\alpha R+\beta S-\frac{2W_0}{\beta_0}\beta R$},
    \end{aligned}
    \right.
\end{equation}
\begin{equation}
    \label{cell_pb_Bt}
    \left\lbrace
    \begin{aligned}
    &\partial_y \left[(\beta-c^2\alpha) B_t'+c\alpha B\right]=\frac{\alpha}{\alpha_0}Q-\alpha c B' \\
    & \text{1-periodicity of } B_t \text{, } \langle B_t \rangle = 0, \text{ and continuity of $B_t$ and $(\beta-c^2\alpha) B_t'+c\alpha B$},
    \end{aligned}
    \right.
\end{equation}
\begin{equation}
    \label{cell_pb_Bx}
    \left\lbrace
    \begin{aligned}
    &\partial_y \left[(\beta-c^2\alpha) B_x'-\frac{1}{\beta_0}\beta R+\beta B\right]=\frac{\beta}{\beta_0}(P+R')-\beta B' \\
    & \text{1-periodicity of } B_x \text{, } \langle B_x \rangle = 0, \text{ and continuity of $B_x$ and $(\beta-c^2\alpha) B_x'-\frac{1}{\beta_0}\beta R+\beta B$},
    \end{aligned}
    \right.
\end{equation}
where the continuity conditions come from the continuity of $u_3$ and $\sigma_2+cm_2$. Moreover, when replacing the high-order temporal derivatives, some mean values appear as source terms of the cell problems. They can be shown to cancel out using the reciprocity identitites \eqref{eq:IPP4} and \eqref{eq:IPP1} in \eqref{cell_pb_T}, the identities \eqref{eq:IPP1}, \eqref{eq:IPP2}, \eqref{eq:IPP3} in \eqref{cell_pb_V}, the identities \eqref{eq:IPPB2} in \eqref{cell_pb_Bt}, and finally \eqref{eq:IPP1} and \eqref{eq:IPPB1} in \eqref{cell_pb_Bx}. \\
Collecting the terms of order $\eta^2$ in  \eqref{eq:non_dim_two_variables} and averaging the equation so obtained yields:
\begin{equation}
    \label{eq:order_eta2_avg}
    \partial_t\langle \alpha(\partial_t u_2-c\partial_y u_3)\rangle =\partial_x\langle\beta(\partial_x u_2+\partial_y u_3)\rangle.
\end{equation} 
Using \eqref{form_u2} and \eqref{eq:form_u3} leads to the final effective equation for the mean field at the second order 
\begin{equation}
    \label{final_effective_order2}
      \alpha_0\partial^2_{tt}\mathcal{U}_2-2W_0\partial^2_{tx}\mathcal{U}_2-\beta_0\partial^2_{xx}\mathcal{U}_2=\mathcal{F}(\mathcal{U}_1)+\mathcal{E}(\mathcal{U}_0) + \mathcal{B}(f)
\end{equation}
with $\mathcal{F}$ defined in \eqref{eq:def_F}, the second source term given by
\begin{equation}
    \label{term_source_order2}
    \begin{aligned}
    \mathcal{E}(\mathcal{U}_0)&=-\langle\alpha(S-cM')\rangle \partial^4_{tttx}\mathcal{U}_0-\langle\alpha(R-cN')\rangle\partial^4_{ttxx}\mathcal{U}_0\\&+\langle\beta R\rangle \partial^4_{xxxx}\mathcal{U}_0+\langle\beta(S+N')\rangle\partial^4_{txxx}\mathcal{U}_0+\langle\beta M'\rangle\partial^4_{ttxx}\mathcal{U}_0
    \end{aligned}
\end{equation}
and the third one defined as 
\begin{equation}
    \label{term_source_order2_f}
    \begin{aligned}
\mathcal{B}(f) = \langle \beta(B+B_x')\rangle \partial^2_{xx}f+\langle \beta B_t'+c\alpha B_x'\rangle\partial^2_{tx}f+\langle \alpha (cB_t'-B) \rangle\partial^2_{tt}f.
    \end{aligned}
\end{equation}

We already now that if either $\alpha$ or $\beta$ is constant, $W_0$ vanishes in \eqref{final_effective_order2}. We will show that the same holds for $\mathcal{F}(\mathcal{U}_1)$. The question is therefore to know whether the terms with an odd number of spatial derivatives (associated with non-reciprocial behaviour) in $\mathcal{E}$ cancel out if only one of the parameters is modulated in time: This is the subject of the next two subsections. 
\subsubsection{If $\rho$ is constant }
Let us first consider the case where $\rho$ and therefore $\alpha$ is constant. As a reminder, $\rho^\star$ is chosen such that $\alpha=1$. The reciprocity identities \eqref{eq:IPP3} and \eqref{eq:IPP1} given in Appendix \ref{Sec:App} allow us to conclude that   $\mathcal{F}(\mathcal{U}_1) =0$. Similarly, in this case $B=B_t=0$ so that the source term in \eqref{term_source_order2_f} reduces to 
\begin{equation}
    \label{eq:source_terme_f_alpha1}
    \mathcal{B}(f)=\langle \beta B_x'\rangle \partial^2_{xx}f
\end{equation}
and the effective equation for $\mathcal{U}_2$ is
\begin{equation}
\label{eq:mean_U2_rho1}
         \partial^2_{tt}\mathcal{U}_2-\beta_0\partial^2_{xx}\mathcal{U}_2=\mathcal{E}(\mathcal{U}_0) + \langle \beta B_x'\rangle \partial^2_{xx}f.
         \end{equation}
Since $\langle S \rangle =\langle R \rangle =0$, and $\langle M'\rangle = \langle N' \rangle = 0$ by continuity of $M$ and $N$, the source term \eqref{term_source_order2} then reduces to 
\begin{equation}
    \label{source_term_alpha_1}
     \mathcal{E}(\mathcal{U}_0)=\langle\beta R\rangle \partial^4_{xxxx}\mathcal{U}_0 + \left\langle\beta M'\right \rangle\partial^4_{ttxx}\mathcal{U}_0+\langle\beta(S+N')\rangle\partial^4_{txxx}\mathcal{U}_0.
\end{equation}
\begin{remark}
    If $c=0$, the effective equation for the second-order term reduces to 
     \begin{equation}
     \partial^2_{tt}\mathcal{U}_2-\beta_0\partial^2_{xx}\mathcal{U}_2-\beta_0 \langle P^2\rangle \partial^4_{xxxx}\mathcal{U}_0 =\langle P^2\rangle \partial^2_{xx}f
     \end{equation}
     and we therefore recover the high-order homogenized equation of \cite{Allaire2018}.
\end{remark}
Consequently, non-reciprocity will be achieved if the following coefficient is non zero: 
\begin{equation}
    \label{non_recip_alpha1}
    \mathcal{N}_\beta = \langle \beta (S+N')\rangle.
\end{equation}
Using the reciprocity identity \eqref{eq:IPP8} in the case $\alpha=1$ allows us to get 
\begin{equation}
    \langle\beta N'\rangle =  c \langle R P'\rangle  - c \langle R' P\rangle +\langle \beta S P'\rangle  -\langle \beta S' P\rangle.
    \label{eq:ipp8_alpha1}
\end{equation}
Similarly, \eqref{eq:IPP5} leads to  
\begin{equation}
    \langle \beta S P'\rangle - \langle\beta S'P\rangle=-\langle\beta S\rangle+ c \langle P'R-PR'\rangle.
    \label{ipp5_alpha1}
\end{equation}
Combining \eqref{eq:ipp8_alpha1} and \eqref{ipp5_alpha1} together with the continuity of $P$ and $R$ allows us to simplify $\mathcal{N}_\beta$ to 
\begin{equation}
    \label{N_alpha_PR}
    \mathcal{N}_\beta=-4c\langle PR'\rangle.
\end{equation}
Since $P$ and $R$ appear in \eqref{N_alpha_PR}, let us first rewrite the cell problems \eqref{cell_pb_P} and \eqref{cell_pb_R} when $\alpha=1$:
\begin{equation}
    \label{cell_pb_P_alpha1}
    \left\lbrace
    \begin{aligned}
    &\partial_y \left[(\beta-c^2) P'+\beta\right]=0 \\
    & \text{1-periodicity of } P \text{, } \langle P \rangle = 0, \text{ and continuity for $P$ and $(\beta-c^2)P'+\beta$},
    \end{aligned}
    \right.
\end{equation}
\begin{equation}
    \label{cell_pb_R_alpha1}
    \left\lbrace
    \begin{aligned}
    &\partial_y \left[(\beta-c^2) R'+\beta P\right]=-\beta(1+P')+\beta_0 \\
    & \text{1-periodicity of }R \text{, } \langle R \rangle = 0, \text{ and continuity for $R$ and $(\beta-c^2) R'+\beta P$}.
    \end{aligned}
    \right.
\end{equation}
Integration of \eqref{cell_pb_P_alpha1} leads to 
\begin{equation}
    \label{eq:integration_P}
    (\beta-c^2)P'+\beta = \mathscr{C}_{\partial P},
\end{equation}
where $\mathscr{C}_{\partial P}$ is a constant of integration, which, after averaging this equation on a unit cell, can be shown to be $ \mathscr{C}_{\partial P}=\beta_0$. 
Consequently, we get from \eqref{eq:integration_P} that the first equation of \eqref{cell_pb_R_alpha1} reduces to 
\begin{equation}
    \label{LHS_R_alpha1}
   \partial_y \left[(\beta-c^2) R'+\beta P\right] = -c^2P'.
\end{equation}
It follows that
\begin{equation}
    \label{R'_alpha1}
    R'=-\frac{\beta+c^2}{\beta-c^2}P+\frac{\mathscr{C}_{\partial R}}{\beta-c^2},
\end{equation}
for some integration constant $\mathscr{C}_{\partial R}$, which, using that $\langle R'\rangle = 0$ is given by 
\begin{equation}
    \label{C2}
    \mathscr{C}_{\partial R}=\left\langle \frac{\beta+c^2}{\beta-c^2}P\right\rangle \left\langle \frac{1}{\beta-c^2}\right\rangle^{-1}.
\end{equation}
Therefore $\mathcal{N}_\beta$ can be written in terms of $P$ only: 
\begin{equation}
    \label{N_alpha1_Ponly}
    \mathcal{N}_\beta=-4c\left\lbrace -\left\langle \frac{\beta+c^2}{\beta-c^2}P^2\right\rangle+\left\langle \frac{\beta+c^2}{\beta-c^2} P \right\rangle \left\langle \frac{1}{\beta-c^2}\right\rangle^{-1}\left\langle \frac{1}{\beta-c^2}P\right\rangle \right\rbrace.
\end{equation}
We know that $\beta-c^2$ is of constant sign depending on whether $c_m > \mathrm{max}\, c_h$ or $c_m < \mathrm{min}\, c_h$, and we have now to treat both cases separately. 

\paragraph{Case 1: $\beta-c^2>0$} 

In this subsonic regime, we introduce the following scalar product 
\begin{equation}
    \label{scalar_product_alpha1_case1}
    (f,g)_{\beta,\mathrm{sub}} = \left\langle \frac{1}{\beta-c^2} fg\right\rangle
\end{equation}
where sub denotes the subsonic regime, and we choose $\mu^{\star}$ so that 
\begin{equation}
\label{eq:norm_PS}
    (1,1)_{\beta,\mathrm{sub}} = \left\langle \frac{1}{\beta-c^2}  \right\rangle = 1.
\end{equation}
Then, $\mathcal{N}_\beta$ simplifies to 
\begin{equation}
    \label{N_CS}
    \mathcal{N}_\beta = 4c\left\lbrace \langle P^2\rangle +2c^2 \left[ (P,P)_{\beta,\mathrm{sub}}(1,1)_{\beta,\mathrm{sub}}-(P,1)_{\beta,\mathrm{sub}}^2\right] \right\rbrace.
\end{equation}
We know that $\langle P^2 \rangle > 0$. The term in the bracket, combined with \eqref{eq:norm_PS}, is positive due to Cauchy-Schwarz inequality. Consequently $\mathcal{N}_\beta>0$ and we get non-reciprocity.
\paragraph{Case 2: $\beta-c^2<0$}
In this supersonic case, we introduce the following scalar product
\begin{equation}
    \label{eq:scal_prod_alpha1_case2}
   (f,g)_{\beta,\mathrm{sup}}=\left\langle \frac{\beta}{c^2-\beta}fg\right\rangle 
\end{equation}
and we choose $\mu^{\star}$ so that 
$(1,1)_{\beta,\mathrm{sup}}=\left\langle \frac{\beta}{c^2-\beta}\right\rangle=1.$
Using $\langle P\rangle=0$, the terms in $\mathcal{N}_\beta$ can be written as:
\begin{equation}
    \label{terms_N_alpha1_case2}
    \left\lbrace
    \begin{aligned}
    & \left\langle \frac{1}{\beta-c^2}\right \rangle = \frac{1}{c^2} \left[-1+\left\langle\frac{\beta}{\beta-c^2}\right\rangle \right]=-\frac{2}{c^2} \, \text{, } \, - \left\langle \frac{\beta+c^2}{\beta-c^2}P^2\right\rangle = \langle P^2\rangle+2\left\langle\frac{\beta}{c^2-\beta}P^2\right\rangle \\
        & -\left\langle \frac{\beta+c^2}{\beta-c^2}P\right\rangle = 2\left\langle\frac{\beta}{c^2-\beta}P\right\rangle \text{, and }\, \left\langle \frac{1}{\beta-c^2 }P\right\rangle =- \frac{1}{c^2}\left\langle \frac{\beta}{c^2-\beta}P\right\rangle
    \end{aligned}
    \right.
\end{equation}
so that $\mathcal{N}_\beta$ simplifies to 
\begin{equation}
    \label{N_alpha1_case2_CS}
    \mathcal{N}_\beta=-4c\left\lbrace \langle P^2\rangle +2 \left[(P,P)_{\beta,\mathrm{sup}}(1,1)_{\beta,\mathrm{sup}}-\frac{1}{2}(P,1)_{\beta,\mathrm{sup}}^2\right] \right\rbrace.
\end{equation}
Consequently 
\begin{equation}
    \label{N_less_alpha1}
    \mathcal{N}_\beta \leq -4c\left\lbrace \langle P^2\rangle +2 \left[(P,P)_{\beta,\mathrm{sup}}(1,1)_{\beta,\mathrm{sup}}-(P,1)_{\beta,\mathrm{sup}}^2\right] \right\rbrace <0
\end{equation}
where the last inequality comes from the fact that $\langle P^2\rangle >0$ and $(P,P)_{\beta,\mathrm{sup}}(1,1)_{\beta,\mathrm{sup}}-(P,1)_{\beta,\mathrm{sup}}^2 \geq 0$ with Cauchy-Schwarz inequality.
\subsubsection{If $\mu$ is constant}
Let us now consider the second case where $\mu$, and consequently $\beta$, is constant. As a reminder, $\mu^{\star}$ is chosen such that $\beta=1$. In this case, the reciprocity identities \eqref{eq:IPP4}, \eqref{eq:IPP2} together with \eqref{order_1_one_modulated} leads to $\mathcal{F}(\mathcal{U}_1)=0$, while the source term \eqref{term_source_order2_f} reduces to 
\begin{equation}
    \label{terme_source_order_2f_beta1}
    \mathcal{B}(f) = c\langle \alpha B'_x\rangle \partial^2_{tx}f -\langle\alpha(B-cW'_t)\rangle \partial^2_{tt}f
\end{equation}
so that the mean equation for $\mathcal{U}_2$ reads 
\begin{equation}
\label{eq:mean_U2_alpha1}
         \alpha_0\partial^2_{tt}\mathcal{U}_2-\partial^2_{xx}\mathcal{U}_2=\mathcal{E}(\mathcal{U}_0) + c\langle \alpha B'_x\rangle \partial^2_{tx}f -\langle\alpha(B-cW'_t)\rangle \partial^2_{tt}f.
         \end{equation}
Eventually, the source term \eqref{term_source_order2} reduces to 
\begin{equation}
    \label{eq:source_term_alpha1}
    \mathcal{E}(\mathcal{U}_0)=-\langle \alpha(S-cM')\rangle \partial^4_{tttx}\mathcal{U}_0-\langle \alpha (R-cN')\rangle \partial^4_{ttxx}\mathcal{U}_0
\end{equation}
so that what matters is to prove that $ \mathcal{N}_\alpha = -\langle \alpha (S-cM')\rangle $ is non zero. Using \eqref{eq:IPP9} and \eqref{eq:IPP5} together with continuity of $Q$ and $R$, we get $ \mathcal{N}_\alpha = -4\alpha_0 \langle QR'\rangle $
with $Q$ \eqref{cell_pb_Q} and $R$ \eqref{cell_pb_R} solutions of the following simpler cell problems when $\beta=1$
\begin{equation}
    \label{cell_pb_Q_beta1}
    \left\lbrace
    \begin{aligned}
    &\partial_y \left[(1-c^2\alpha )Q'+c\alpha\right]=0 \\
    & \text{1-periodicity of } Q \text{, } \langle Q \rangle = 0, \text{ and continuity for $Q$ and $(1-c^2\alpha)Q'+c\alpha$},
    \end{aligned}
    \right.
\end{equation}
\begin{equation}
    \label{cell_pb_R_beta1}
    \left\lbrace
    \begin{aligned}
    &\partial_y \left[(1-c^2 \alpha) R'+\frac{c\alpha}{\alpha_0}Q\right]=-1+\frac{\alpha}{\alpha_0}(1-cQ') \\
    & \text{1-periodicity of }R \text{, } \langle R \rangle = 0, \text{ and continuity for $R$ and $(1-c^2\alpha)R'+\frac{\alpha c}{\alpha_0} Q$}.
    \end{aligned}
    \right.
\end{equation}
Similarly to the case $\alpha=1$, we get 
    $(1-c^2\alpha)Q'+c\alpha=\mathscr{C}_{\partial Q}$
with $\mathscr{C}_{\partial Q}=c\alpha_0$ 
and 
\begin{equation}
    R'=-\frac{1+\alpha c^2}{c\alpha_0(1-\alpha c^2)}Q+\frac{\mathscr{C}_{\partial R}}{1-\alpha c^2}
   \quad \text{ with } \quad
    \mathscr{C}_{\partial R}=\frac{1}{c\alpha_0}\langle \frac{1+\alpha c^2}{1-\alpha c^2}Q\rangle \langle \frac{1}{1-\alpha c^2}\rangle^{-1}.
\end{equation}
Consequently, 
\begin{equation}
    \label{N_q_beta1}
    \mathcal{N}_\alpha=\frac{4}{c}\left[ \left\langle \frac{1+\alpha c^2}{1-\alpha c^2}Q^2\right\rangle-\langle \frac{1}{1-\alpha c^2}Q\rangle\langle \frac{1}{1-\alpha c^2}\rangle^{-1}\langle \frac{1+\alpha c^2}{1-\alpha c^2}Q\rangle \right].
\end{equation}
We know that $1-\alpha c^2$ has constant sign and we need to consider two cases again. 
\paragraph{Case 1: $\alpha c^2>1$}
In this case, we introduce the scalar product 
\begin{equation}
    \label{scalar_product_beta1_case1}
 (f,g)_{\alpha,\mathrm{sup}}=\langle \frac{1}{\alpha c^2-1} fg \rangle 
\end{equation}
and we choose $\rho^\star$ such that $
    (1,1)_{\alpha,\mathrm{sup}} = 1.$
Consequently $\mathcal{N}_\alpha$ writes 
\begin{equation}
    \label{N_CS_beta1_case1}
    \mathcal{N}_\alpha=\frac{4}{c}\left\lbrace 2\left[-(Q,Q)_{\alpha,\mathrm{sup}}(1,1)_{\alpha,\mathrm{sup}}+(Q,1)_{\alpha,\mathrm{sup}}^2 \right]-\langle Q^2\rangle \right\rbrace.
\end{equation}
Cauchy-Schwarz inequality for the first bracket and $\langle Q^2\rangle >0$ yields 
$\mathcal{N}_\alpha<0.$
\paragraph{Case 2: $\alpha c^2<1$}
In this last case, we introduce the scalar product 
\begin{equation}
(f,g)_{\alpha,\mathrm{sub}}=\langle \frac{\alpha c^2}{1-\alpha c^2} fg \rangle 
\end{equation} 
and we choose $\rho^\star$ so that 
$(1,1)_{\alpha,\mathrm{sub}} = 1.$
Consequently, we get the following simplifications, using $\langle Q \rangle = 0$:
\begin{equation}
    \left\lbrace
    \begin{aligned}
        & \langle \frac{1}{1-\alpha c^2} \rangle = 2 \,\text{, }\quad  \langle \frac{1+\alpha c^2}{1-\alpha c^2 }Q^2\rangle = \langle Q^2\rangle +2\langle \frac{\alpha c^2}{1-\alpha c^2}Q^2\rangle, \\
        & \langle \frac{1}{1-\alpha c^2} Q \rangle = \langle \frac{\alpha c^2}{1-\alpha c^2}Q \rangle \,\text{ and }\, \langle \frac{1+\alpha c^2}{1-\alpha c^2} Q \rangle = 2 \langle \frac{\alpha c^2}{1-\alpha c^2} Q \rangle 
    \end{aligned}
    \right.
\end{equation}
so that $\mathcal{N}_\alpha$ writes 
\begin{equation}
    \mathcal{N}_\alpha = \frac{4}{c}\left \lbrace  \langle Q^2\rangle +2\left[ (Q,Q)_{\alpha,\mathrm{sub}}(1,1)_{\alpha,\mathrm{sub}}-\frac{1}{2}(Q,1)_{\alpha,\mathrm{sub}}^2\right]\right \rbrace.
\end{equation}
Therefore,
\begin{equation}
    \mathcal{N}_\alpha \geq \frac{4}{c}\left \lbrace  \langle Q^2\rangle +2\left[ (Q,Q)_{\alpha,\mathrm{sub}}(1,1)_{\alpha,\mathrm{sub}}-(Q,1)_{\alpha,\mathrm{sub}}^2\right]\right \rbrace >0
\end{equation}
where we used for the last inequality that $\langle Q^2\rangle >0$ and Cauchy-Schwarz inequality. \\
Consequently, in each case, we recover the fact that non-reciprocity is present through the source term that appears in the second-order homogenised model even if only one of the physical parameters is modulated in time. We will illustrate this property in the next section through the expression of the dispersion relation. 
\subsection{Total mean field and dispersion relation}
The second-order mean field $\bar{u}_2$ is defined by
\begin{equation}
    \bar{u}_2(x,t)=\mathcal{U}_0(x,t)+\eta\mathcal{U}_1(x,t)+\eta^2\mathcal{U}_2(x,t).
\end{equation}
Collecting \eqref{final_eq_order_0}, \eqref{eq:effective_eq_order1} and \eqref{final_effective_order2}, and neglecting the higher-order terms, this second-order mean field satisfies
\begin{equation}
    \label{total_mean_eq}
     \alpha_0\partial^2_{tt}\bar{u}_2-2W_0\partial^2_{tx}\bar{u}_2-\beta_0\partial^2_{xx}\bar{u}_2=\eta \mathcal{F}(\bar{u}_2)+\eta^2\mathcal{E}(\bar{u}_2)+f+\eta \mathcal{A}(f)+\eta^2\mathcal{B}(f).
\end{equation}
Since our main interest is to investigate non-reciprocity for only one varying parameter, we will distinguish this case in the next two subsections. Our interest being mainly on the non-reciprocal behaviour, we will consider $f=0$ and look at the associated dispersion relations. 
\subsubsection{Dispersion relation in the case of $\rho$ being constant}
Using \eqref{final_eq_order_0_recip}, \eqref{order_1_one_modulated}, \eqref{eq:mean_U2_rho1} and \eqref{source_term_alpha_1} for $f=0$, \eqref{total_mean_eq} reduces to 
\begin{equation}
    \label{total_mean_eq_alpha1}
     \partial^2_{tt}\bar{u}_2-\beta_0\partial^2_{xx}\bar{u}_2=\eta^2\left[ \left\langle\beta\left(\frac{1}{\beta_0} R+ M'\right)\right\rangle\partial^4_{ttxx}\bar{u}_2+\mathcal{N}_\beta\partial^4_{txxx}\bar{u}_2 \right]
\end{equation}
which, in dimensionalised coordinates, gives for the dimensionalized mean field $\bar{U}_2=\bar{u}_2/k^\star $
\begin{equation}
     \partial^2_{TT}\bar{U}_2-\beta_0(c^\star)^2\partial^2_{XX}\bar{U}_2=h^2\left[ \left\langle\beta\left(\frac{1}{\beta_0} R+ M'\right)\right\rangle\partial^4_{TTXX}\bar{U}_2+\mathcal{N}_\beta c^\star\partial^4_{TXXX}\bar{U}_2 \right].
\end{equation}
Considering a harmonic solution $\bar{U}_2=\mathrm{exp}(\mathrm{i}(\omega T - k X))$ yields the dispersion relation
\begin{equation}
    \label{dispersion_relation_alpha1}
    -\omega^2+\beta_0 (c^\star)^2k^2=\left\langle\beta\left(\frac{1}{\beta_0} R+ M'\right)\right\rangle h^2k^2\omega^2 -\mathcal{N}_\beta c^\star h^2 \omega k^3.
\end{equation}
Since $\mathcal{N}_\beta \neq 0$, the term in $k^3$ is non zero, which ensures non-reciprocity. For numerical examples later on, we write explicitly the different coefficients in the dispersion relation. \\
Firstly, we know from \eqref{eq:integration_P} that $\beta_0=\mathscr{C}_{\partial P}$. Moreover, we deduce from $\langle P' \rangle =0$ that 
\begin{equation}
    \label{E0_alpha1}
    \beta_0 = \left\langle \frac{\beta}{\beta-c^2} \right\rangle\left \langle \frac{1}{\beta-c^2}\right\rangle ^{-1}.
\end{equation}
Secondly, integration of $P'$ and $\langle P \rangle = 0$ leads to the following expression of $P$
\begin{equation}
    \label{eq:P_explicit_alpha1}
    P(y) =\int_0^y\frac{\beta_0-\beta(z)}{\beta(z)-c^2}\mathrm{d}z-\int_0^1 \left(\int_0^s \frac{\beta_0-\beta(z)}{\beta(z)-c^2}\mathrm{d}z\right)\mathrm{d}s,
\end{equation}
and we want to write the right-hand-side terms of \eqref{dispersion_relation_alpha1} in terms of $P$ only. It has already been done for $\mathcal{N}_\beta$ in \eqref{N_alpha1_Ponly}, but we still have to compute $\langle \beta ( \frac{1}{\beta_0}R+M')\rangle$. Integration of \eqref{R'_alpha1} together with $\langle R \rangle=0$ gives
\begin{equation}
\label{R_anal_alpha1}
    R(y) = \int_0^y\frac{-(\beta(z)+c^2)P(z)+\mathscr{C}_{\partial R}}{\beta(z)-c^2}\mathrm{d}z+\int_0^1\left(\int_0^s\left(\frac{(\beta(z)+c^2)P(z)-\mathscr{C}_{\partial R}}{\beta(z)-c^2}\right)\mathrm{d}z\right)\mathrm{d}s,
\end{equation}
with $\mathscr{C}_{\partial R}$ given by \eqref{C2}. 
Concerning $\langle \beta M'\rangle$, the reciprocity identity \eqref{eq:IPP10} together with \eqref{R'_alpha1} yields 
\begin{equation}
    \label{beta_T'_alpha1}
    \langle \beta M'\rangle = \left\langle \left(1-\frac{\beta}{\beta_0}\right)P^2-\frac{\beta}{\beta_0}\left(-\frac{\beta+c^2}{\beta-c^2}P+\frac{\mathscr{C}_{\partial R}}{\beta-c^2}\right)P\right\rangle -2c\langle S'P\rangle +\frac{1}{\beta_0}\langle \beta RP'\rangle,
\end{equation}
where $S$ is solution of 
\begin{equation}
    \label{cell_pb_S_alpha1}
    \left\lbrace
    \begin{aligned}
    &\partial_y \left[(\beta-c^2 ) S'+cP\right]=-cP' \\
    & \text{1-periodicity of }S \text{, } \langle S \rangle = 0, \text{ and continuity of $S$ and $(\beta-c^2) S'+cP$}.
    \end{aligned}
    \right.
\end{equation}
We deduce from this system together with $\langle S'\rangle = 0$ that 
\begin{equation}
    \label{I_S'}
    S'=-\frac{2c}{\beta-c^2}P+\frac{\mathscr{C}_{\partial S}}{\beta-c^2} \quad \text{ with } \quad
    \mathscr{C}_{\partial S}=2c\left\langle \frac{1}{\beta-c^2}P\right\rangle \left\langle \frac{1}{\beta-c^2}\right\rangle^{-1}.
\end{equation}
Consequently the dispersion relation \eqref{dispersion_relation_alpha1} is given by 
   \begin{equation} \label{dispersion_relation_alpha1_bis}
    -\omega^2+\beta_0 (c^\star)^2k^2=\mathcal{L}(P)h^2 k^2\omega^2 -\mathcal{N}_\beta c^\star h^2 \omega k^3,
    \end{equation}
with $\beta_0$ given by \eqref{E0_alpha1}, $\mathcal{N}_\beta$ by \eqref{N_alpha1_Ponly} and 
\begin{equation}
    \label{L_P}
    \mathcal{L}(P)=\left\langle\left(1+\frac{2c^2}{\beta-c^2}\left(\frac{\beta}{\beta_0}+2\right)\right)P^2\right\rangle-\left\langle \left( \frac{\beta}{\beta_0}\mathscr{C}_{\partial R}+2c \mathscr{C}_{\partial S}\right)\frac{1}{\beta-c^2}P\right\rangle+\frac{1}{\beta_0}\langle\beta R(P'+1)\rangle,  
\end{equation}
with $\mathscr{C}_{\partial R}$ given by \eqref{C2}, $\mathscr{C}_{\partial S}$ by \eqref{I_S'}, $R$ by \eqref{R_anal_alpha1}, $P'$ by \eqref{eq:integration_P} and $P$ by \eqref{eq:P_explicit_alpha1}. 

\subsubsection{Dispersion relation in the case of $\mu$ being constant}
The same approach is followed in this case. It results in the following equation in dimensionalised variables 
\begin{equation}
     \alpha_0\partial^2_{TT}\bar{U}_2-(c^\star)^2\partial^2_{XX}\bar{U}_2=h^2\left[- \left\langle\alpha\left(R-cN'\right)\right\rangle\partial^4_{TTXX}\bar{U}_2+\frac{\mathcal{N}_\alpha}{c^\star}\partial^4_{TTTX}\bar{U}_2 \right],
\end{equation}
with associated dispersion relation 
   \begin{equation} \label{dispersion_relation_beta1_bis}
    -\alpha_0\omega^2+ (c^\star)^2k^2=\mathcal{M}(Q) h^2k^2\omega^2 -\frac{\mathcal{N}_\alpha}{ c^\star}h^2 \omega^3 k, 
    \end{equation}
where 
\begin{equation}
    \label{details_DD}
    \left\lbrace
    \begin{aligned}
        &\mathcal{M}(Q) = \left\langle \left( -1+\tfrac{2\alpha}{\alpha_0}\tfrac{1}{1-c^2\alpha}+\tfrac{4}{1-c^2\alpha}\right)Q^2\right\rangle       +\left\langle\alpha R\left( \tfrac{-1+c^2\alpha_0}{1-c^2\alpha}\right)\right\rangle -\left\langle (c\mathscr{C}_{\partial R}\alpha+2\mathscr{C}_{\partial S}) \tfrac{1}{1-c^2\alpha}Q\right\rangle \\
          &Q(y) =c\int_0^y\tfrac{\alpha_0-\alpha(z)}{1-c^2\alpha(z)}\mathrm{d}z+\mathscr{C}_{Q} \text{ and }
          R(y) = -\int_0^y\tfrac{1+c^2\alpha(z)}{c\alpha_0(1-c^2\alpha(z))}Q(z)\mathrm{d}z+\mathscr{C}_{\partial R}\int_0^y\tfrac{1}{1-c^2\alpha(z)}\mathrm{d}z+\mathscr{C}_{R} \\
            &\mathscr{C}_{\partial R} =\frac{1}{c\alpha_0}\left\langle\frac{1+c^2\alpha}{1-c^2\alpha}Q \right\rangle \left\langle\frac{1}{1-c^2\alpha}\right\rangle^{-1} \text{ and }\,
         \mathscr{C}_{\partial S} = \left\langle\frac{2}{1-c^2\alpha}Q\right\rangle\left\langle\frac{1}{1-c^2\alpha}\right\rangle^{-1} \\
          &\alpha_0 = \left\langle \frac{\alpha}{1-c^2\alpha}\right\rangle\left\langle\frac{1}{1-c^2\alpha}\right\rangle^{-1} \text{ and }\,
         \mathscr{C}_{Q} =c\int_0^1\left(\int_0^s\frac{\alpha(z)-\alpha_0}{1-c^2\alpha(z)}\mathrm{d}z\right)\mathrm{d}s \\
        &\mathscr{C}_{R} = \int_0^1\left(\int_0^s\frac{1+c^2\alpha(z)}{c\alpha_0(1-c^2\alpha(z))}Q(z)\mathrm{d}z\right)\mathrm{d}s-\mathscr{C}_{\partial R}\int_0^1\left(\int_0^s\frac{1}{1-c^2\alpha(z)}\mathrm{d}z\right)\mathrm{d}s.
       \end{aligned}
    \right.
\end{equation}

\section{Numerical experiments}\label{SecNum}
In this section, we illustrate on a bilayered example the results obtained through asymptotic developments in the previous section. We first focus on the dispersion diagrams associated with both the space-time modulated laminate and the homogenised models. We then illustrate the results expected through time-domain simulations in the microstructured medium. 
\subsection{Dispersion diagrams}
To start with, we consider a bilayered medium of periodicity $h=20$ with piecewise constant parameters given by 
\begin{equation}
(\rho,\,\mu)(y)=
\left\{
\begin{array}{ll}
(\rho_A,\,\mu_A) & \mathrm{ if } \, y\in (0,0.25)\\ [6pt]
(\rho_B,\,\mu_B) &   \mathrm{ if } \,y\in (0.25,1),
\end{array}
\right.
\label{Bilayer}
\end{equation}  
and a modulation velocity $c_m$. We also choose the reference parameters such that 
\begin{equation}
    \label{eq:param_typ_num}
    \rho^\star=\langle \rho \rangle \text{ and } \mu^\star=\langle 1/\mu \rangle^{-1} \text{, hence } {c}^\star=1/\sqrt{\langle \rho \rangle\langle 1/\mu \rangle}.
\end{equation}
\vspace*{-0.5cm}
\begin{table}[ht]
\caption{Physical parameters for the bilayered medium \eqref{Bilayer} with both parameters modulated. Mass densities are given in $\mathrm{kg}\cdot\mathrm{m}^{-3}$, shear moduli in $\mathrm{Pa}$, and velocities in  $\mathrm{m}\cdot\mathrm{s}^{-1}$.  }
\begin{center}
\begin{tabular}{ |c|c|c|c|c|c|c|c|c|} 
 \hline
Fig. number &$\rho_A$ & $\rho_B$ & $\mu_A$ & $\mu_B$ &$c_A$ &$c_B$&$c_B/c_A$ & $c_m$ \\ \hline 
Fig. \ref{fig:both_config1} &$10^3$ & $1.5\times 10^3$ & $10^9$ & $6\times 10^9$ &$10^3$ &$2\times 10^3$&$2$ & $5\times 10^2$\\  
\hline 
Fig. \ref{fig:both_config2} &$10^3$ & $1.5\times 10^3$ & $10^9$ & $6\times 10^9$ &$10^3$ &$2\times 10^3$&$2$ & $3\times 10^3$\\ 
\hline 
Fig. \ref{fig:both_config3} &$10^3$ & $10^2$ & $10^9$ & $6\times 10^9$ &$10^3$ &$7.74\times 10^3$&$7.74$ & $5\times 10^2$\\ 
 \hline
\end{tabular}
\end{center}
\label{table:both}
\end{table}

\vspace*{-1cm}
\subsubsection{Leading-order model}

\begin{figure}[h!]
\subfloat[]{\includegraphics[trim = 5mm 0mm 5mm 0mm, clip, width = 0.32\linewidth]{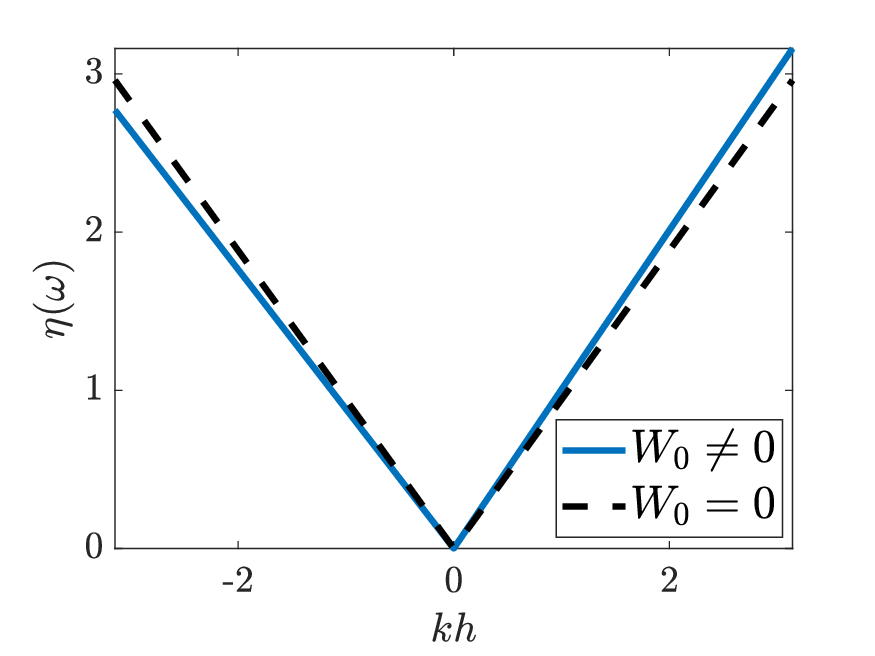}\label{fig:both_config1}}
\hspace{0.1cm}
\subfloat[]{\includegraphics[trim = 5mm 0mm 5mm 0mm, clip, width = 0.32\linewidth]{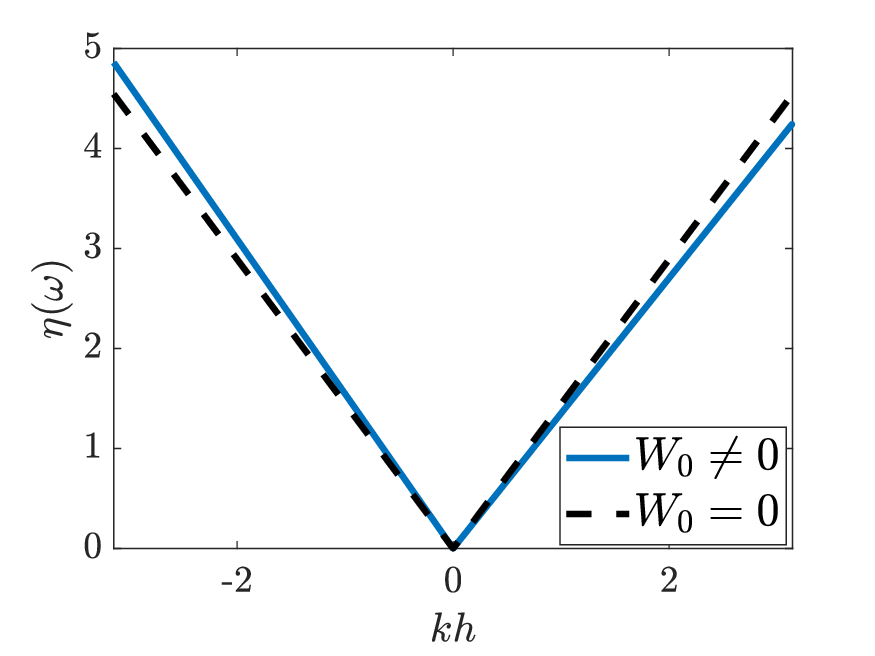}\label{fig:both_config2}} 
\hspace{0.1cm}
\subfloat[]{\includegraphics[trim = 5mm 0mm 5mm 0mm, clip, width = 0.32\linewidth]{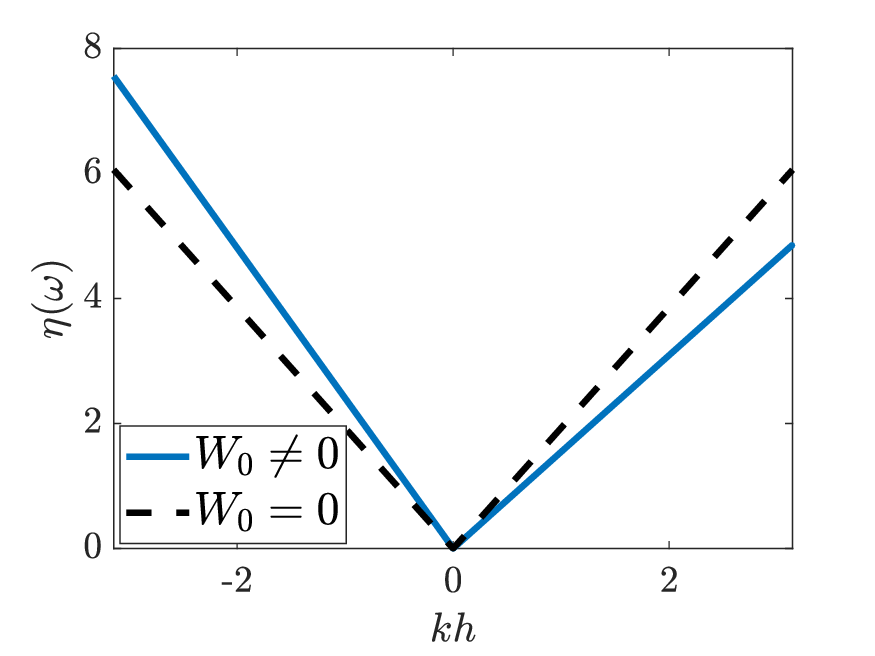}\label{fig:both_config3}} 
\caption{Dispersion diagrams for the leading-order model. The plain line denotes a configuration where both parameters are modulated described by Table \ref{table:both}, while the dotted line denotes the case where $W_0=0$. The $y$-axis is given in terms of the small parameter $\eta=k^\star h = \omega h/ c^\star$.}
\label{fig:DD_both}
\end{figure}

Fig. \ref{fig:DD_both} displays the dispersion diagrams associated with this configuration for the leading-order model \eqref{DD_order0} previously obtained in the literature \cite{Nassar2017}. Different values of the parameters are chosen and summarized in Table \ref{table:both}. In each subfigure, the plain line denotes the dispersion diagram for the case described by Table \ref{table:both} where both parameters are modulated, and shows asymmetry with respect to $k=0$ underlying therefore non-reciprocity. Through the different examples, we can see that non-reciprocity happens in both supersonic and subsonic regimes, and increases as the contrast of the parameters increases. The dotted line denotes the limit case $W_0=0$ which happens as soon as one of the parameters is constant leading therefore to reciprocity. 

\vspace*{-0.5cm}
\subsubsection{Second-order model}
Now we consider the case where only $\mu$ or $\rho$ are modulated and compare the exact dispersion diagrams, and the homogenised one at both leading and second order. \\
Firstly, the exact dispersion relation for a bilaminate alternating two phases of parameters $(\rho_i,\mu_i)$ and length $h_i$ ($i=A,B$) and modulated in time in a wave-like fashion, such as in Fig. \ref{fig:laminate}, is given by the zeros of the following dispersion function \cite{lurie2007introduction,Nassar2017}:
\begin{equation}
    \label{Disp_Func_bilaminate}
    \begin{aligned}
   &\mathrm{Disp}(\omega,k) =\cos\left[kh-c_m(\omega-c_m k)\left( \tfrac{h_A}{c_A^2-c_m^2}+\tfrac{h_B}{c_B^2-c_m^2}\right)\right]\\&-\cos\left( \tfrac{\omega-c_m k}{c_A^2-c_m^2}c_A h_A\right)\cos\left( \tfrac{\omega-c_m k}{c_B^2-c_m^2}c_B h_B\right)  
   + \tfrac{1}{2}\left( \tfrac{Z_A}{Z_B}+\tfrac{Z_B}{Z_A}\right)\sin\left( \tfrac{\omega-c_m k}{c_A^2-c_m^2}c_A h_A\right)\sin\left( \tfrac{\omega-c_m k}{c_B^2-c_m^2}c_B h_B\right),
   \end{aligned}
\end{equation}
with $c_i=\sqrt{\mu_i/\rho_i}$ and $Z_i=\sqrt{\rho_i\mu_i}$ for $i=A,B$. Secondly, the homogenised dispersion relation at the leading (or first) order is computed solving for \eqref{DD_order0} with $W_0=0$. Thirdly, the homogenised dispersion relation associated to the second-order model developped in this paper is plotted solving either \eqref{dispersion_relation_alpha1_bis} or \eqref{dispersion_relation_beta1_bis} for the cases $\alpha$ constant and $\beta$ constant, respectively.

\begin{figure}[h!]
\subfloat[]{\includegraphics[trim = 0mm 0mm 0mm 0mm, clip, width = 0.49\linewidth]{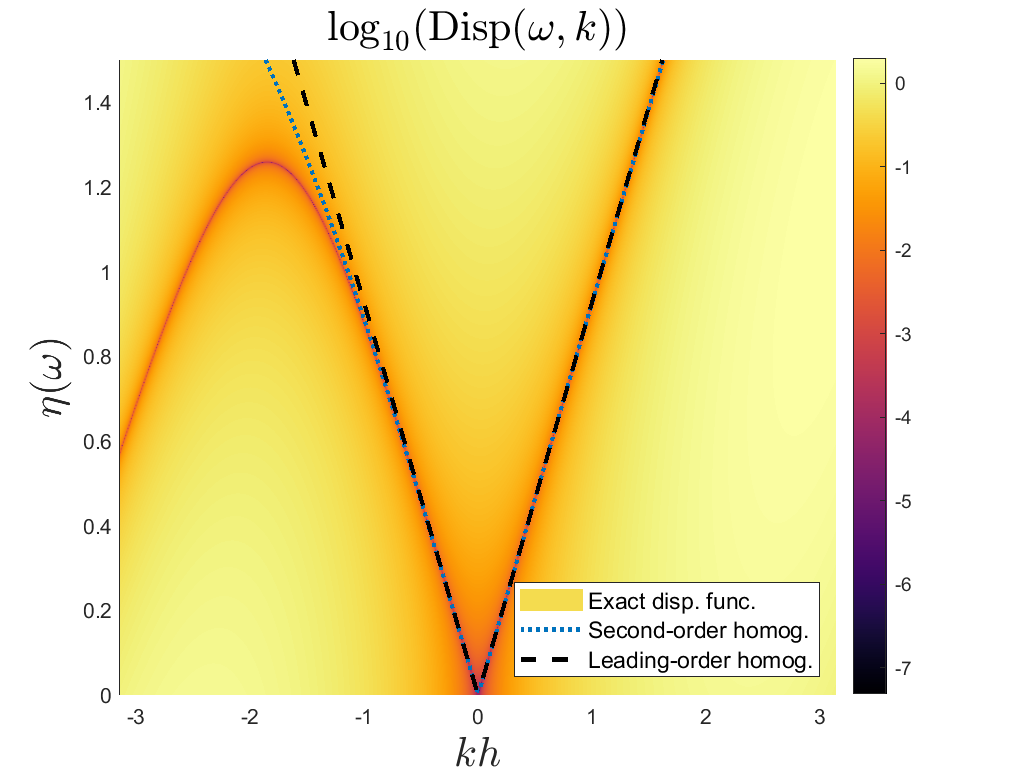}\label{fig:alpha1_a}}
\hspace{0.1cm}
\subfloat[]{\includegraphics[trim = 0mm 0mm 0mm 0mm, clip, width = 0.49\linewidth]{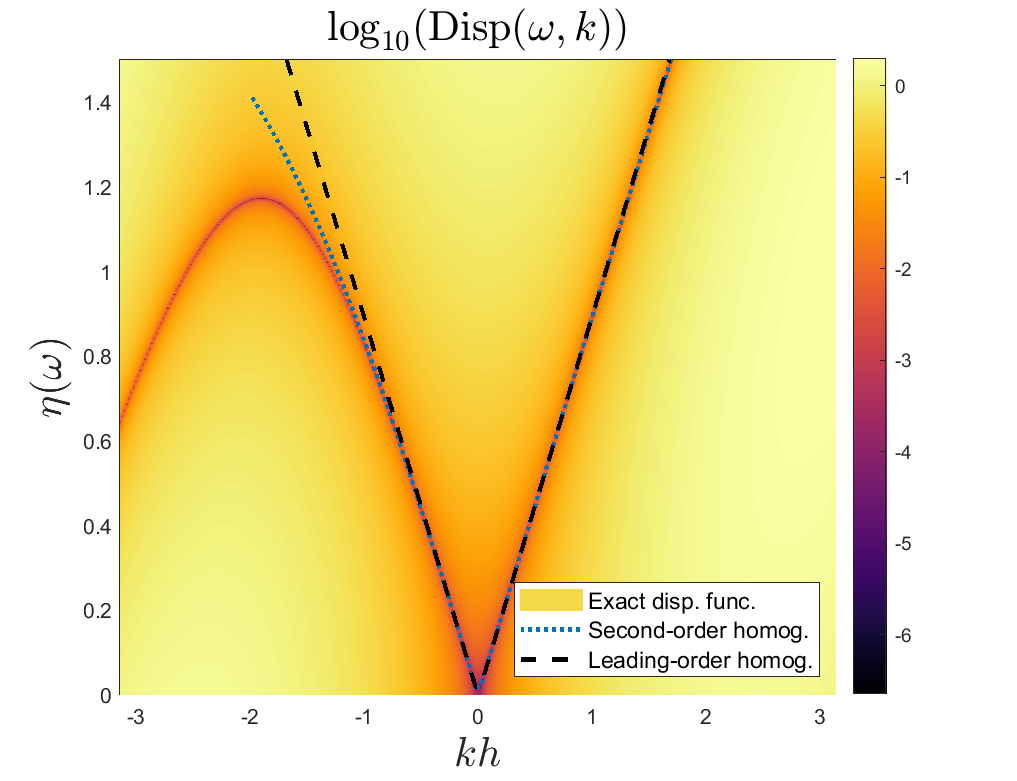}\label{fig:alpha1_b}} \\
\subfloat[]{\includegraphics[trim = 0mm 0mm 0mm 0mm, clip, width = 0.49\linewidth]{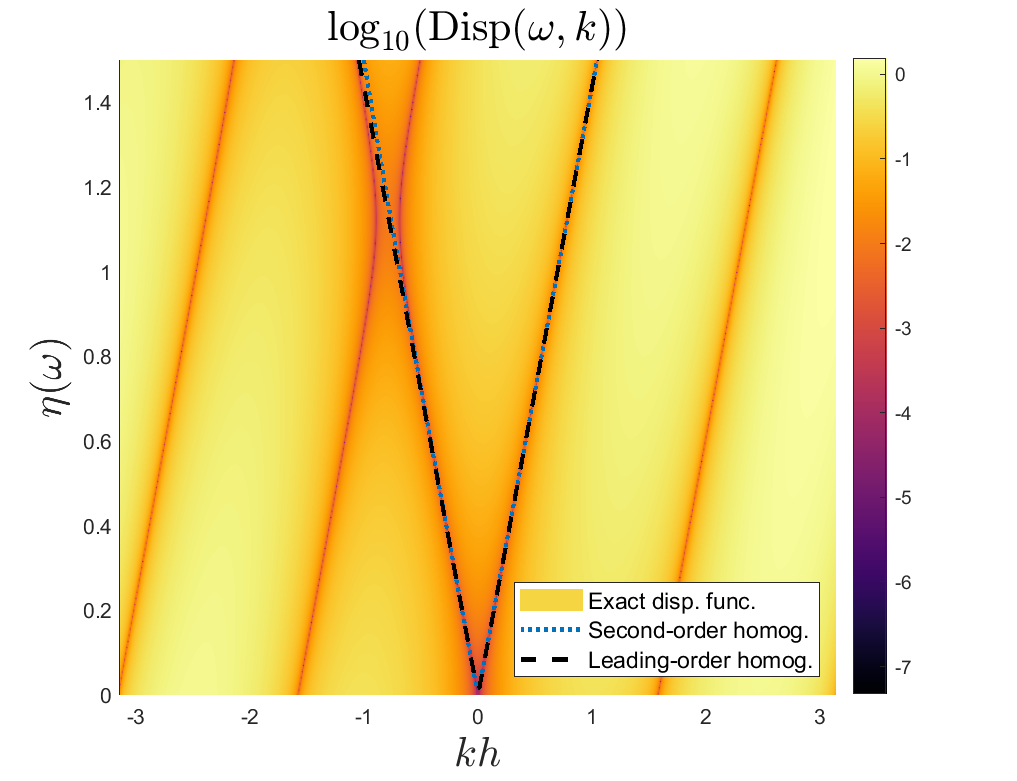}\label{fig:alpha1_c}} 
\subfloat[]{\includegraphics[trim = 0mm 0mm 0mm 0mm, clip, width = 0.49\linewidth]{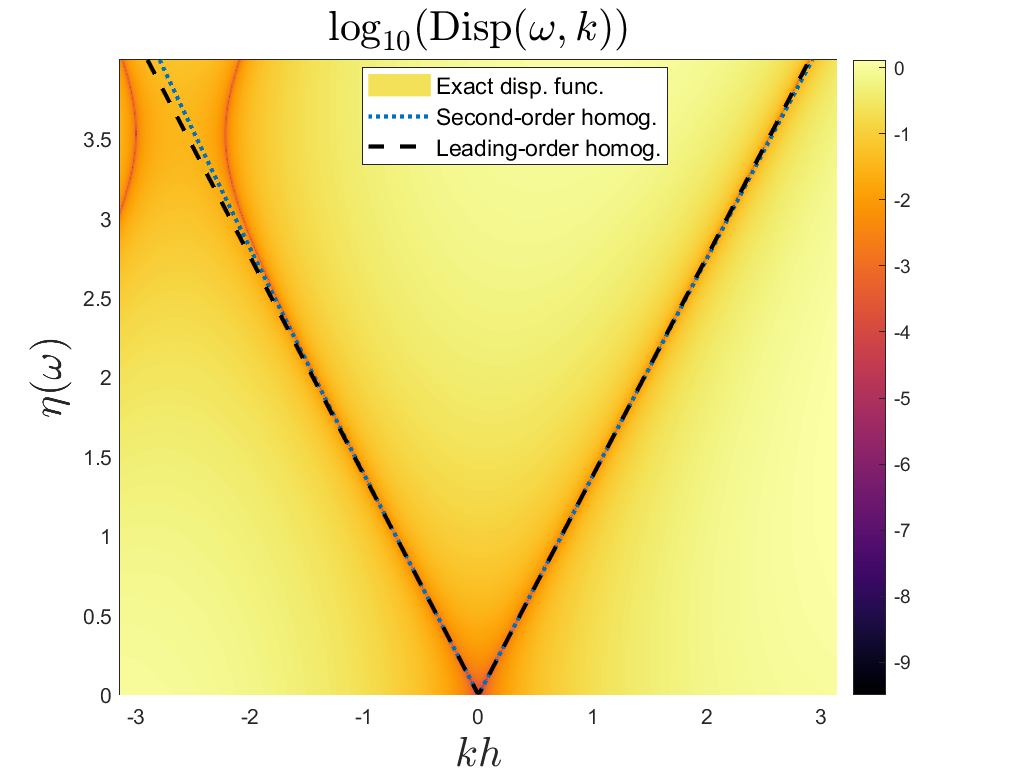}\label{fig:alpha1_d}} 
\caption{Dispersion diagrams when modulating only $\mu$. The colormap denotes the logarithm of the dispersion function for the microstructured configuration, the dark lines therefore represent the exact dispersion diagram. The dotted blue line denotes the dispersion diagram obtained at the second order in \eqref{dispersion_relation_alpha1_bis}. The dashed black line denotes the dispersion diagram obtained at the leading \cite{Nassar2017} or first order. Parameters are given in Table \ref{table:alpha1}}
\label{fig:DD_alpha1}
\end{figure}

\begin{figure}[h!]
\subfloat[]{\includegraphics[trim = 0mm 0mm 0mm 0mm, clip, width = 0.49\linewidth]{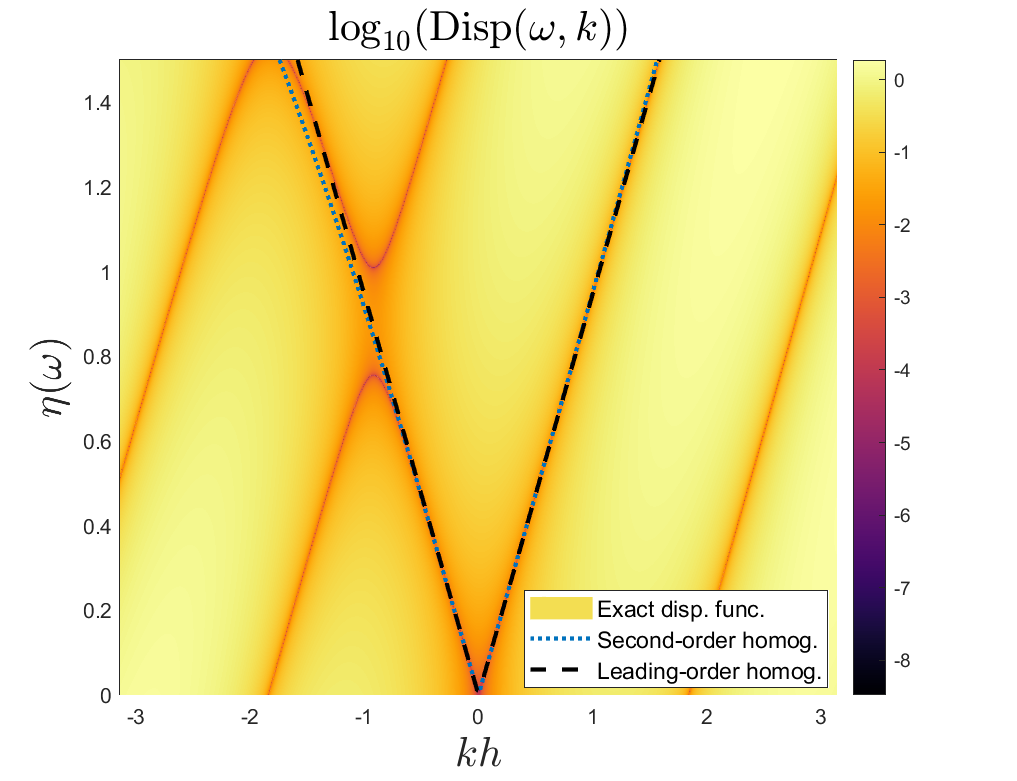}\label{fig:beta1_a}}
\hspace{0.1cm}
\subfloat[]{\includegraphics[trim = 0mm 0mm 0mm 0mm, clip, width = 0.49\linewidth]{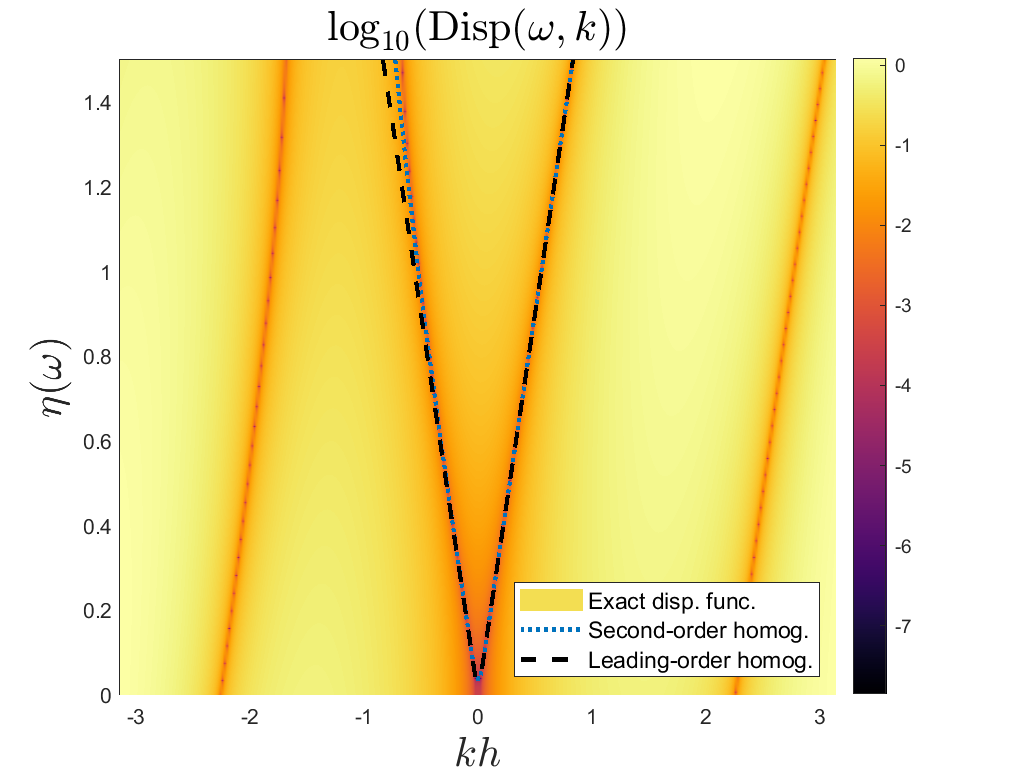}\label{fig:beta1_b}} 

\caption{Dispersion diagrams when modulating only $\rho$. The colormap denotes the logarithm of the dispersion function for the microstructured configuration, the dark lines therefore represent the exact dispersion diagram. The dotted blue line denotes the dispersion diagram obtained at the second order in \eqref{dispersion_relation_beta1_bis}. The dashed black line denotes the dispersion diagram obtained at the leading \cite{Nassar2017} or first order. Parameters are given in Table \ref{table:beta1}}
\label{fig:DD_beta1}
\end{figure}

The case $\alpha=1$ is described by Table \ref{table:alpha1} and Fig. \ref{fig:DD_alpha1}, while the case $\beta=1$ is described by Table \ref{table:beta1} and Fig. \ref{fig:DD_beta1} for different choices of parameter contrasts or modulation velocities. In the figures, the colormap denotes the logarithm of the exact dispersion function \eqref{Disp_Func_bilaminate}. Consequently, the dark lines represent its zeros which are solutions of the exact dispersion relation for the time-modulated bilaminate. We recover the existence of asymetric band-gaps in the subsonic regime, and of stop-bands in wavenumber instead of frequency (the so called $k$-gaps) in the supersonic regime \cite{Cassedy1963,Cassedy1967,Nassar2017}. The dispersion relation associated to the leading-order (or first-order) model and the second-order model are also plotted with a dashed black line, and a dotted blue line, respectively. The $y-$axis is given in terms of the small parameter $\eta(\omega)=\omega h/c^\star$ to investigate the quality of the approximation as we move away from the assumption $\eta \ll 1$ of the asymptotic process.

\begin{figure}[h!]
\subfloat[]{\includegraphics[trim = 0mm 0mm 0mm 8.5mm, clip, width = 0.49\linewidth]{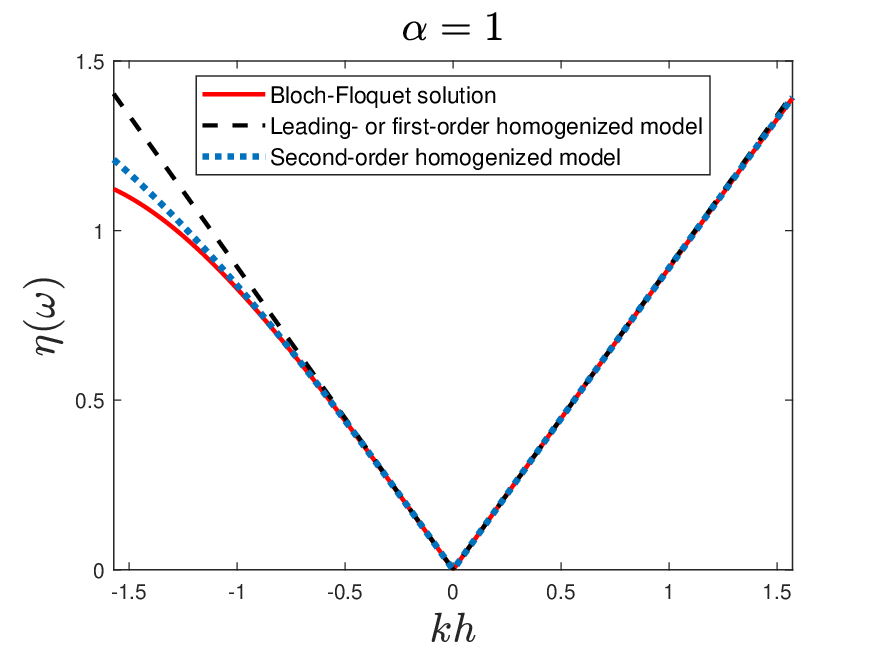}\label{fig:alpha1_a_roots}}
\hspace{0.1cm}
\subfloat[]{\includegraphics[trim = 0mm 0mm 0mm 8.5mm, clip, width = 0.49\linewidth]{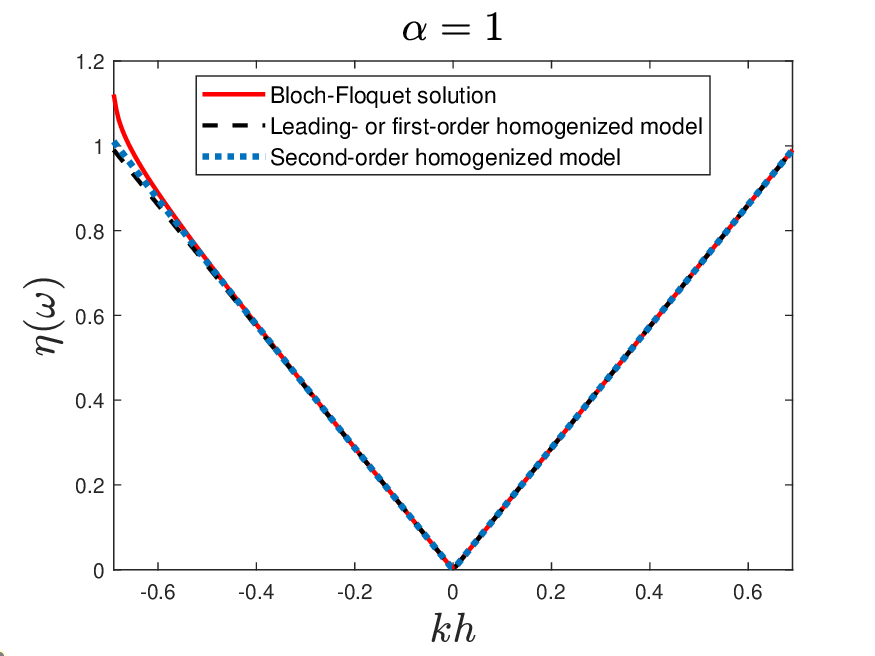}\label{fig:alpha1_b_roots}} 
\caption{Comparison of dispersion diagrams when modulating only $\mu$. Parameters are given in Table \ref{table:alpha1}}
\label{fig:DD_alpha1_roots}
\end{figure}

\begin{figure}[h!]
\subfloat[Negative wavenumbers]{\includegraphics[trim = 0mm 0mm 0mm 0mm, clip, width = 0.49\linewidth]{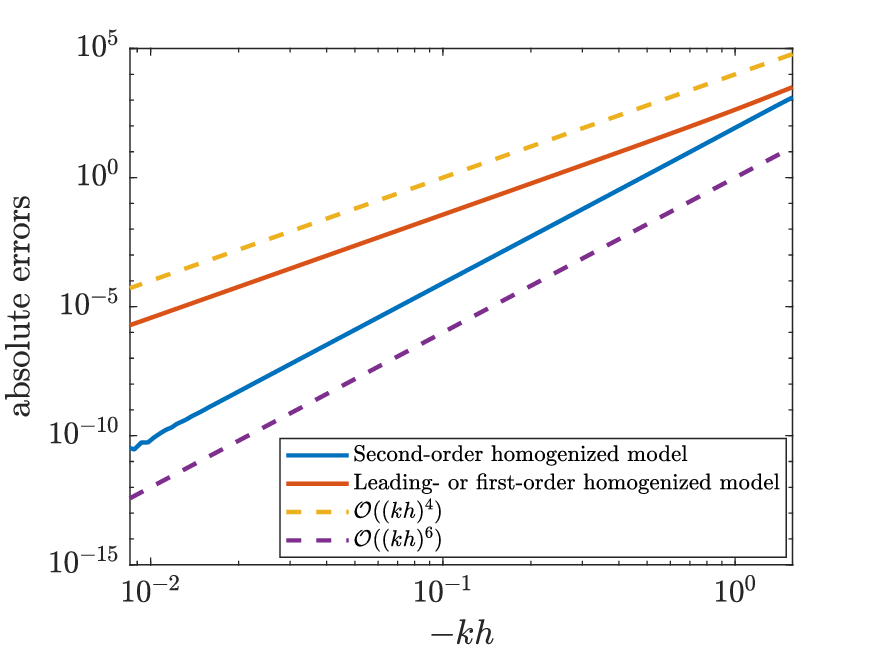}\label{fig:alpha1_error_a}}
\hspace{0.1cm}
\subfloat[Positive wavenumbers]{\includegraphics[trim = 0mm 0mm 0mm 0mm, clip, width = 0.49\linewidth]{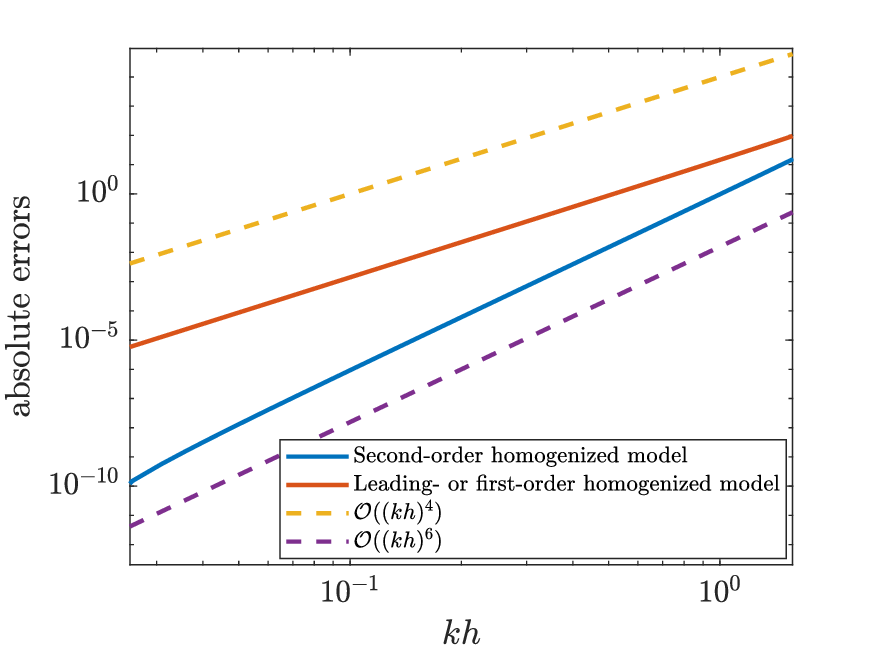}\label{fig:alpha1_error_b}} 
\caption{Errors (in a log-log scale) made on the dispersion curves  for the physical parameters of Fig. \ref{fig:alpha1_a}. Plain lines denote the error $ |\omega_\mathrm{micro}^2-\omega_{\mathrm{homog},i}^2|$ between the frequency obtained by Bloch-Floquet analysis and the one obtained through homogenization at order $i$. Dashed lines denote reference orders of convergence.}
\label{fig:alpha1_error}
\end{figure}

\begin{figure}[h!]
\subfloat[]{\includegraphics[trim = 0mm 0mm 0mm 8.5mm, clip, width = 0.49\linewidth]{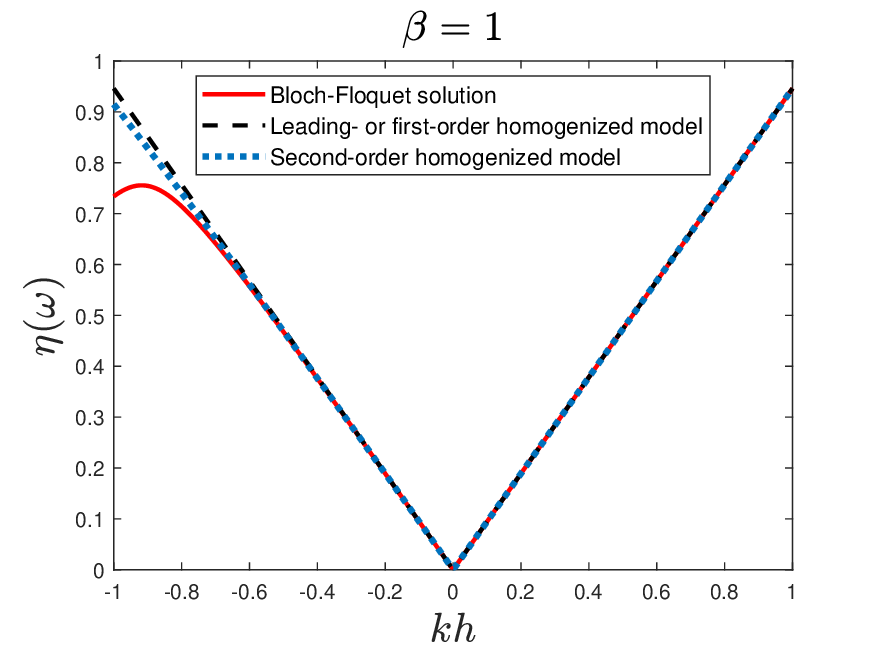}\label{fig:beta1_roots_a}} 
\subfloat[]{\includegraphics[trim = 0mm 0mm 0mm 8.5mm, clip, width = 0.49\linewidth]{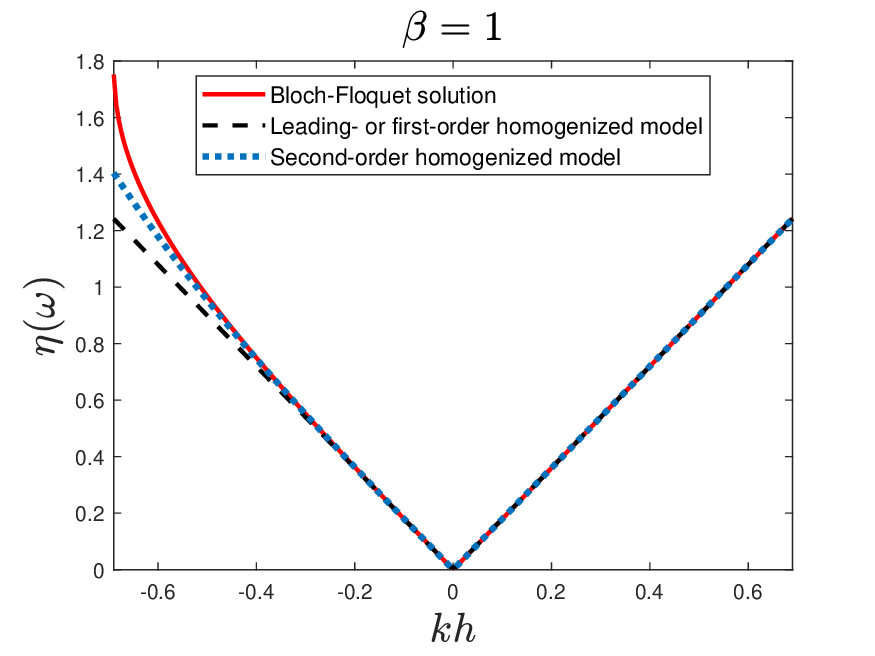}\label{fig:beta1_roots_b}} 
\caption{Comparison of dispersion diagrams when modulating only $\rho$. Parameters are given in Table \ref{table:beta1}}
\label{fig:DD_beta1_roots}
\end{figure}

To investigate more closely, the agreement between the dispersion diagrams of the microstructured configuration and of the homogenised models, dispersion diagrams close to the origins are obtained by root finding and plotted in Fig. \ref{fig:DD_alpha1_roots} and \ref{fig:DD_beta1_roots}. These figures underline that the second-order homogenised model investigated in this paper allows to get a better approximation of the exact dispersion diagram compared to the leading-order or first-order model, especially as we go further away from the origin. In practice, one notices that, as long as we do not enter a bang-gap or a $k$-gap, the agreement is even quite good as $\eta$ approaches 1 while the assumption of the asymptotic process is $\eta \ll 1$. For more quantitative measures, we introduce the following relative errors between the microstructured dispersion relation $\omega_\mathrm{micro}$ and the homogenised ones $\omega_{\mathrm{homog},i}$ ($i=0,2$ for the leading/first-order and the second-order one, respectively):
\begin{equation}
    \label{def_error_measure}
 \epsilon_i=\sqrt{\frac{\int_\mathcal{I} (\omega_\mathrm{micro}(k)-\omega_{\mathrm{homog},i}(k))^2\mathrm{d}k}{\int_\mathcal{I}\omega_\mathrm{micro}(k)^2\mathrm{d}k}}\in (0,1).
\end{equation}
Integrations are performed on an interval $\mathcal{I}$ defined by 
$\mathcal{I}=\left \lbrace k \text{ such that } k\in [-k_\mathrm{lim};k_\mathrm{lim}] \right \rbrace$
where $k_\mathrm{lim}$ is chosen so that both $kh \ll \pi/2$ and the interval stops about the entrance of any band-gap or $k$-gap, for which both homogenised models are no longer accurate, as noticed in Fig. \ref{fig:DD_alpha1} and \ref{fig:DD_beta1}. The values of the chosen $k_\mathrm{lim}$ and of these relative errors are given in the first columns of Tables \ref{table:alpha1_measures} and \ref{table:beta1_measures}, and confirm the better agreeement of the second-order homogenised model. To validate the developments of the high-order homogenization, we plot the absolute error $[\omega_\mathrm{micro}(k)^2-\omega_{\mathrm{homog},i}(k)^2|$  as a function of both positive and negative wavenumbers. Fig. \ref{fig:alpha1_error} represents the results for the very first example of Table \ref{table:alpha1}. We can see that we recover an error of order 4 for the leading-order model, and of order 6 for the second-order model. This validates the developments of the higher-order terms conducted in this paper.    \\
This second-order model captures dispersive effects that are missed at the previous orders, and more importantly non-reciprocal behaviour even if only one of the parameters is space-time dependent. To measure this non-reciprocal behaviour more quantitatively we introduce the following ratio: 
\begin{equation}
    \label{def_NR_measure}
    \mathcal{R}=\sqrt{\frac{\int_\mathcal{I} \omega_\mathrm{odd}(k)^2\mathrm{d}k}{\int_\mathcal{I}\omega_{\mathrm{homog},2}(k)^2\mathrm{d}k}}\in (0,1)
\end{equation}
where $\omega_\mathrm{odd}(k)=(\omega_{\mathrm{homog},2}(k)-\omega_{\mathrm{homog},2}(-k))/{2}$ is the odd part of the frequency as a function of the wavenumber. In the reciprocal case, $\omega(k)$ is an even function and $\mathcal{R}$ is consequently zero. On the contrary, the closer to 1 $\mathcal{R}$ is, the more asymmetric the dispersion curve is. These measures of non-reciprocity are given in the last columns of Tables \ref{table:alpha1_measures} and \ref{table:beta1_measures}. It confirms that the second-order homogenised model captures non-reciprocal effects: as soon as one parameter is modulated, $\mathcal{R}$ is non-zero, while at the leading order it vanishes unless both parameters are modulated. These measurements also give an insight on how reciprocity evolves together with the parameters: it increases as the material contrast increases, and as the modulation velocity approaches the velocity range of the material.

\begin{table}[ht]
\caption{Physical parameters for the bilayered medium \eqref{Bilayer} with only shear modulus modulated (mass density is constant $\rho_A=\rho_B=1.37\times 10^3\,\mathrm{kg}\cdot\mathrm{m}^{-3}$). }
\begin{center}
\begin{tabular}{ |c|c|c|c|c|c|c|c|c|c|} 
 \hline
 Fig. number & $\mu_A$ (Pa) & $\mu_B$ (Pa) &$c_A$  ($\mathrm{m}\cdot\mathrm{s}^{-1}$) &$c_B$ ($\mathrm{m}\cdot\mathrm{s}^{-1}$)&$c_B/c_A$ & $c_m$ ($\mathrm{m}\cdot\mathrm{s}^{-1}$)\\
  
\hline 
Fig. \ref{fig:alpha1_a}  & $10^9$ & $6\times 10^9$ &$8.53\times 10^2$ &$2.09\times 10^3$&$2.45$ & $5\times 10^2$ \\

\hline 
Fig. \ref{fig:alpha1_b} and \ref{fig:alpha1_a_roots} & $10^9$ & $1.5\times 10^{10}$ &$8.53\times 10^2$ &$3.31\times 10^3$&$3.87$ & $5\times 10^2$\\ \hline 

Fig. \ref{fig:alpha1_c}  and \ref{fig:alpha1_b_roots}  & $10^9$ & $6\times 10^{9}$ &$8.53\times 10^2$ &$2.09\times 10^3$&$2.45$ & $2.5\times 10^3$\\ \hline 

Fig. \ref{fig:alpha1_d}  & $10^9$ & $6\times 10^{9}$ &$8.53\times 10^2$ &$2.09\times 10^3$&$2.45$ & $3.5\times 10^3$\\ \hline 
\end{tabular}
\end{center}
\label{table:alpha1}
\end{table}

\begin{table}[ht]
\caption{Case of a bilayered medium \eqref{Bilayer} with only shear modulus modulated (physical parameters are given in Table \ref{table:alpha1}). Relative errors \eqref{def_error_measure} and measure of non-reciprocity \eqref{def_NR_measure}. }
\begin{center}
\begin{tabular}{|c|c|c|c|c|} 
 \hline
 Fig. number &$k_\mathrm{lim}$&$\epsilon_0$&$\epsilon_2$ & $\mathcal{R}$\\
  
\hline 
Fig. \ref{fig:alpha1_a}  & $\pi/2$ &  7.4761\% &2.7001\%& 3.2335\% \\

\hline 
Fig. \ref{fig:alpha1_b} and \ref{fig:alpha1_a_roots}& $\pi/2$  &  9.4515\% &  2.4897\%&  4.6307\%\\ \hline 

Fig. \ref{fig:alpha1_c}  and \ref{fig:alpha1_b_roots}  & $0.69$ & 2.7655\% & 1.9600\%& 0.6545\%\\ \hline 
Fig. \ref{fig:alpha1_d}  & $\pi/2$ & 0.7259\% &   0.1905\%&  0.3804\%\\ \hline 
\end{tabular}
\end{center}
\label{table:alpha1_measures}
\end{table}

\begin{table}[ht]
\caption{Physical parameters for the bilayered medium \eqref{Bilayer} with only mass density modulated (shear modulus is constant $\mu_A=\mu_B=2.66\times 10^9\,\mathrm{Pa}$). Mass densities are given in $\mathrm{kg}\cdot\mathrm{m}^{-3}$.}
\begin{center}
\begin{tabular}{ |c|c|c|c|c|c|c|} 
 \hline
 Fig. number & $\rho_A$ & $\rho_B$ &$c_A$ ($\mathrm{m}\cdot\mathrm{s}^{-1}$)&$c_B$ ($\mathrm{m}\cdot\mathrm{s}^{-1}$)&$c_A/c_B$ & $c_m$ ($\mathrm{m}\cdot\mathrm{s}^{-1}$) \\

 \hline 

Fig. \ref{fig:beta1_a} \& \ref{fig:beta1_roots_a} & $10^3$ & $6\times 10^3$ &$1.63\times 10^3$ &$6.66\times 10^2$&$2.45$ & $5\times 10^2$ \\
\hline 
Fig. \ref{fig:beta1_b} \& \ref{fig:beta1_roots_b} & $10^3$ & $6\times 10^3$ &$1.63\times 10^3$ &$6.66\times 10^2$&$2.45$& $1.8\times 10^3$\\ \hline 
\end{tabular}
\end{center}
\label{table:beta1}
\end{table}


\begin{table}[ht]
\caption{Case of a bilayered medium \eqref{Bilayer} with only mass density modulated (physical parameters are given in Table \ref{table:beta1}). Relative errors \eqref{def_error_measure} and measure of non-reciprocity \eqref{def_NR_measure}.}
\begin{center}
\begin{tabular}{|c|c|c|c|c|} 
 \hline
 Fig. number & $k_\mathrm{lim}$&$\epsilon_0$&$\epsilon_2$ & $\mathcal{R}$\\
 \hline 

Fig. \ref{fig:beta1_a} and \ref{fig:beta1_roots_a}& $1$ & 7.2641\% &  5.7946\%& 1.1042\% \\ 
\hline 
Fig. \ref{fig:beta1_b} and \ref{fig:beta1_roots_b} & $0.69$ & 9.8916\%  &  4.8875\%& 3.8788\%\\ \hline 
\end{tabular}
\end{center}
\label{table:beta1_measures}
\end{table}

\subsection{Numerical setup for time-domain simulations}\label{SecNumSetup}
We now focus on time-domain simulations in the microstructured medium. The system \eqref{dTPsi} is solved in the moving frame. For this purpose, a ADER (Arbitrary high order using Derivatives) scheme is used on a uniform grid with mesh $\Delta x$ and time step $\Delta t$. This two-time step finite-difference scheme is fourth-order accurate in space and time. It is stable under the Courant–Friedrichs–Lewy condition $\gamma=\max c \Delta t/\Delta\leq 1$; in practice, one uses $\gamma=0.95$. Outgoing conditions ensure that no waves are reflected by the edges of the computational domain. An immersed interface method \cite{LombardSIAM03} is implemented to discretize the jump conditions \eqref{JCmove} on a Cartesian grid. This method ensures a subcell resolution of interfaces inside the meshing and maintains the accuracy of the scheme despite the non-smoothness of the solution across interfaces.

The computational domain $[-2500, 1500]$ is discretized on 4000 grid nodes, hence $\Delta x=1$. A finite heterogeneous slab with 101 interfaces is considered on $[-1000, 0]$. It is a bilayered medium of periodicity $h=20$ with piecewise constant parameters \eqref{Bilayer}. The parameters in the two media surrounding the slab are the leading-order effective parameters $\rho_0=\rho^\star\alpha_0$ and $\mu_0=\mu^\star\beta_0$, with $\alpha_0$ and $\beta_0$ defined in \eqref{def_E_0_rho0}, respectively. Doing so minimizes the reflection at the edges of the slab and puts the focus on the interaction with the bulk. 

\begin{figure}[htbp]
\begin{center}
\begin{tabular}{cc}
\hspace{1cm}(a) Source on the left ($t=0.15$ s) & \hspace{1cm}(b) Source on the right ($t=0.20$ s)\\
\includegraphics[scale=0.3]{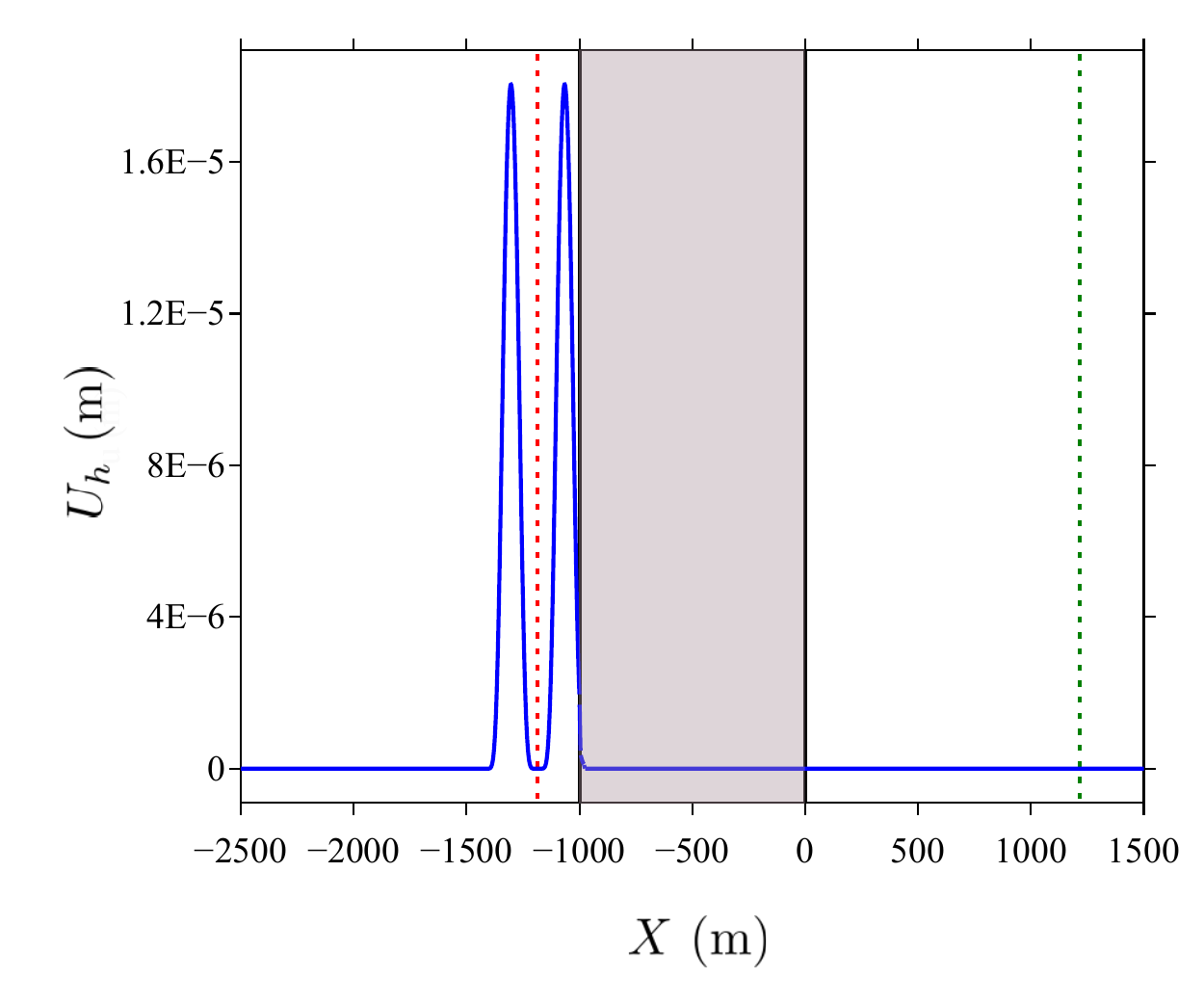} &
\includegraphics[scale=0.3]{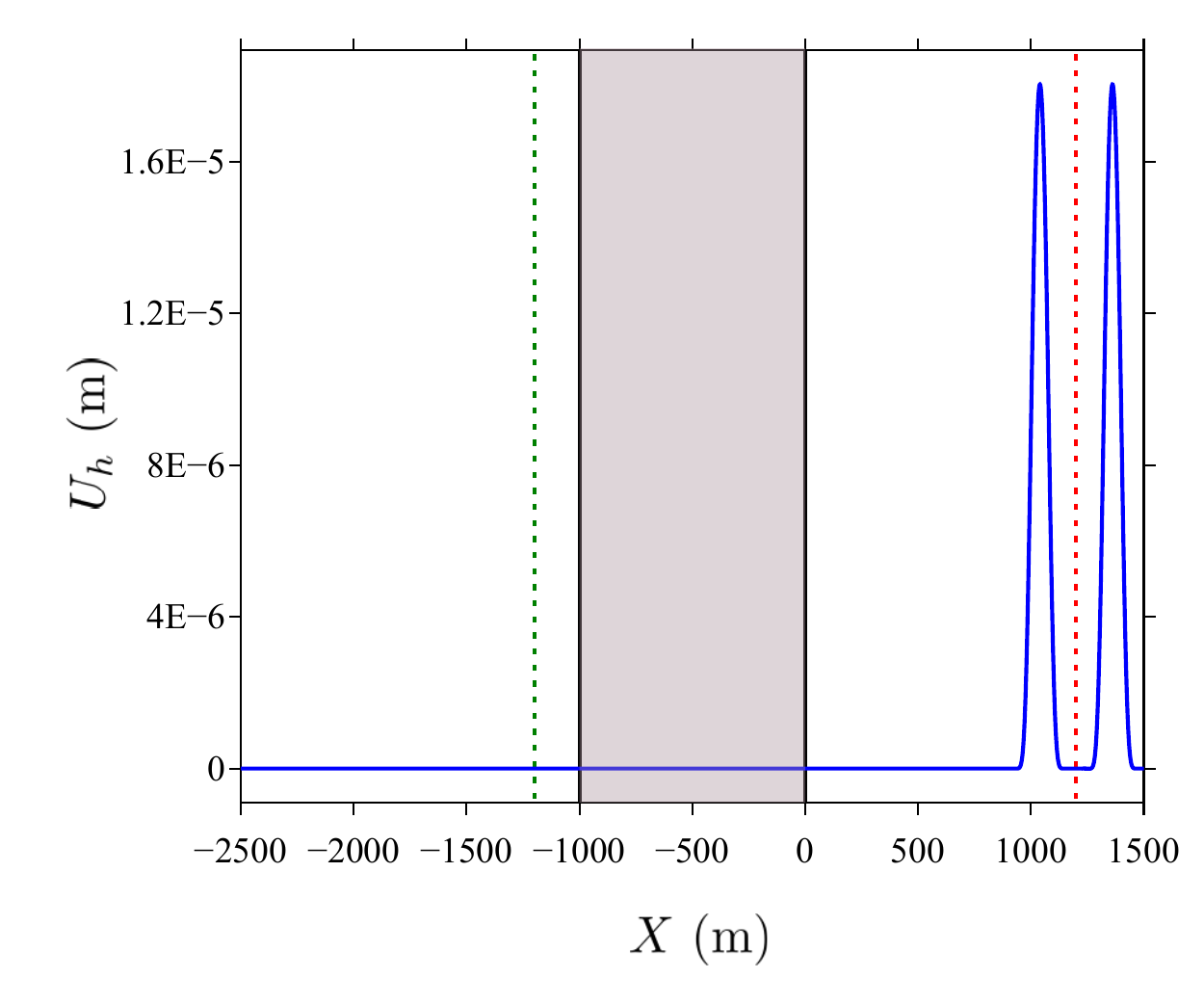} \\
\hspace{1cm}(c) Source on the left ($t=0.76$ s) & \hspace{1cm}(d) Source on the right ($t=1.00$ s)\\
\includegraphics[scale=0.3]{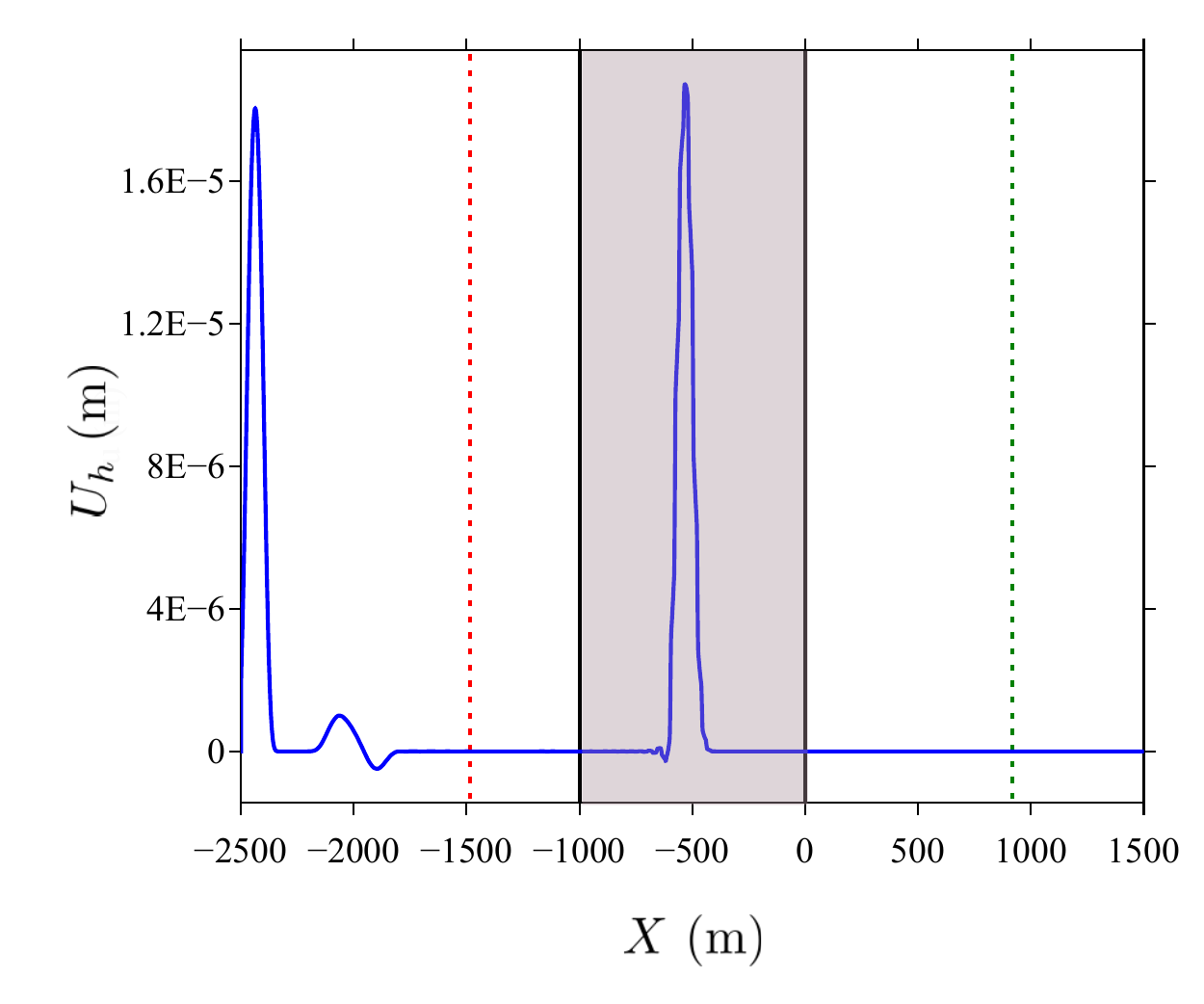} &
\includegraphics[scale=0.3]{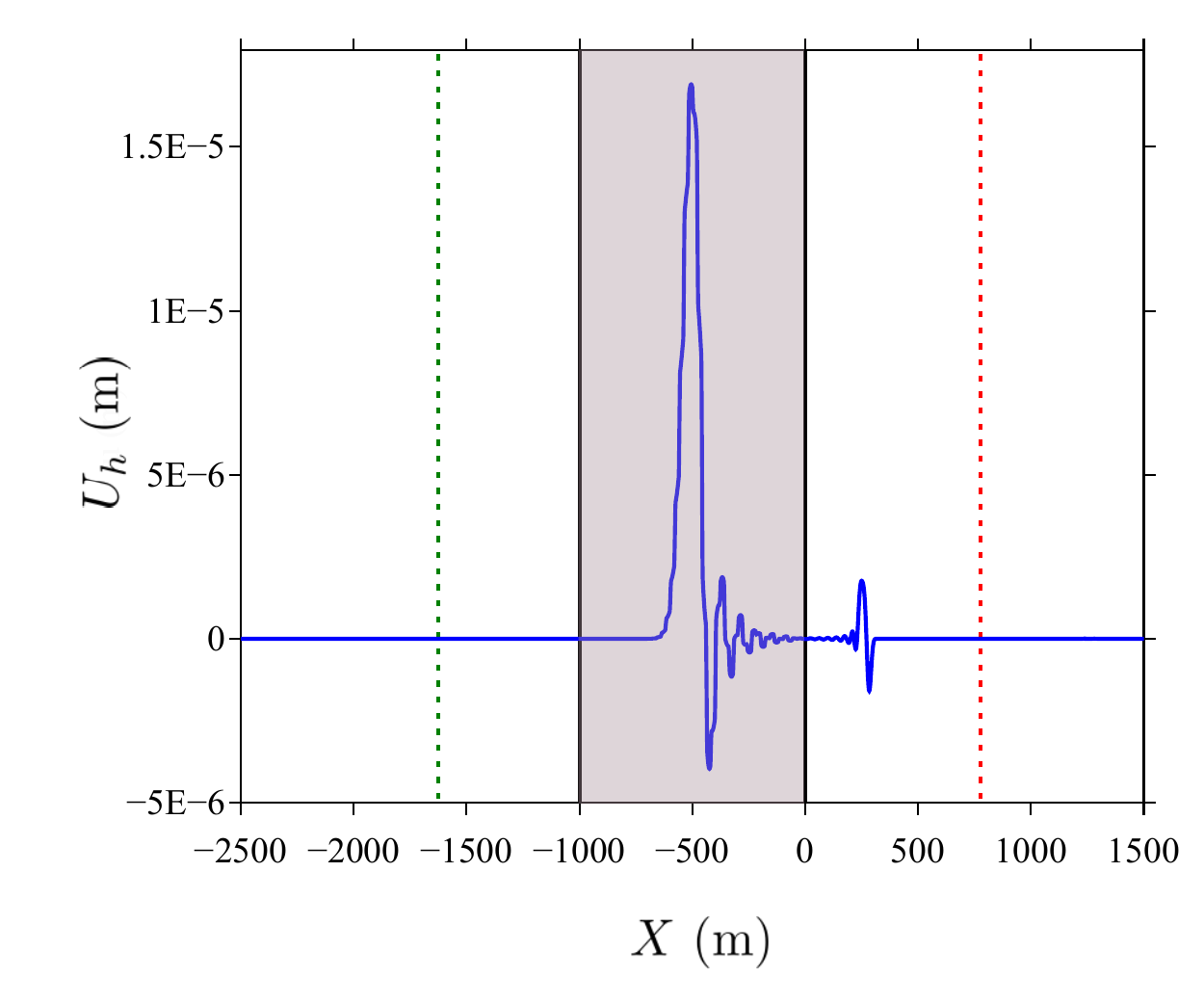} \\
\hspace{1cm}(e) Source on the left ($t=2.50$ s) & \hspace{1cm}(f) Source on the right ($t=2.05$ s)\\
\includegraphics[scale=0.3]{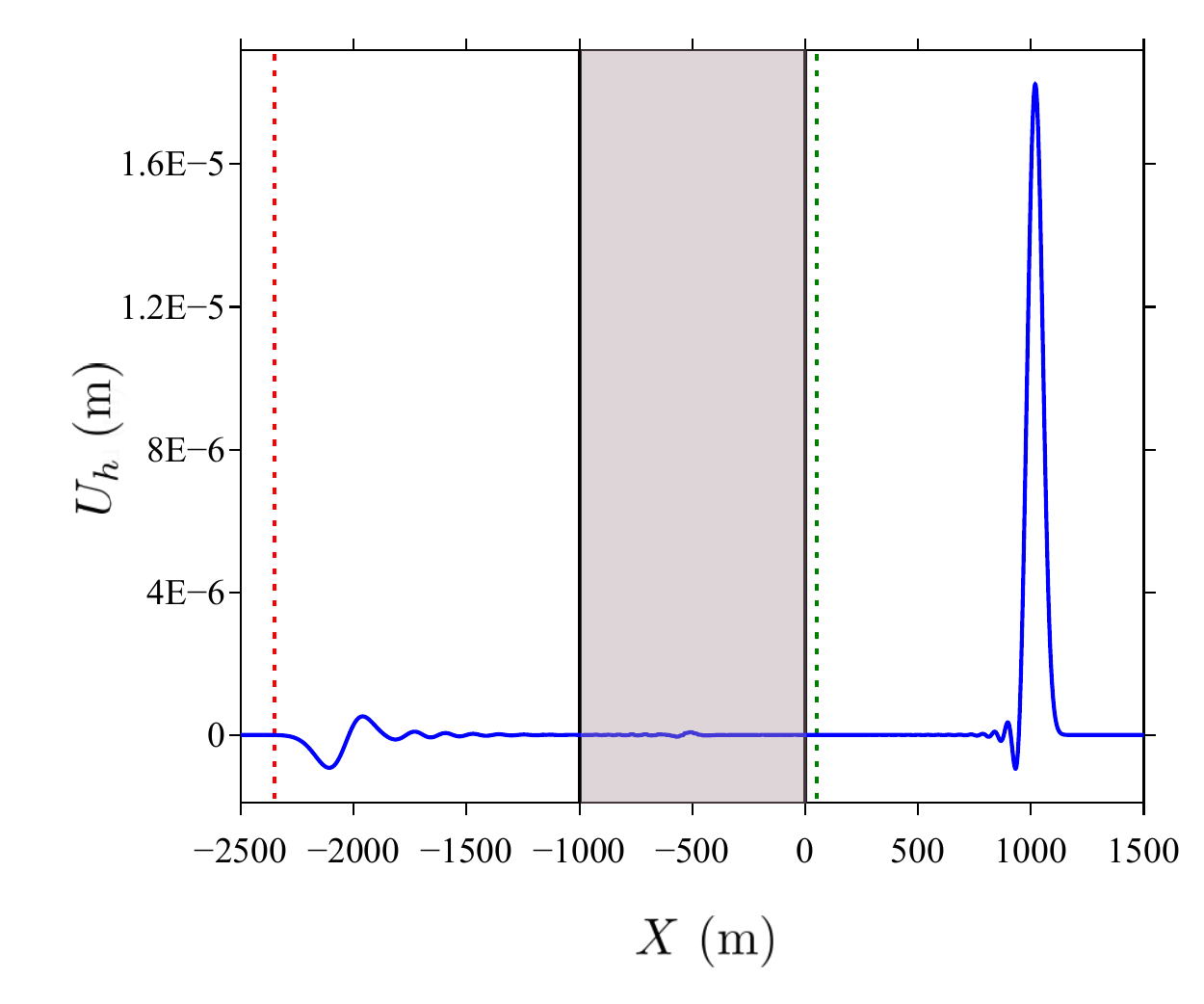} &
\includegraphics[scale=0.3]{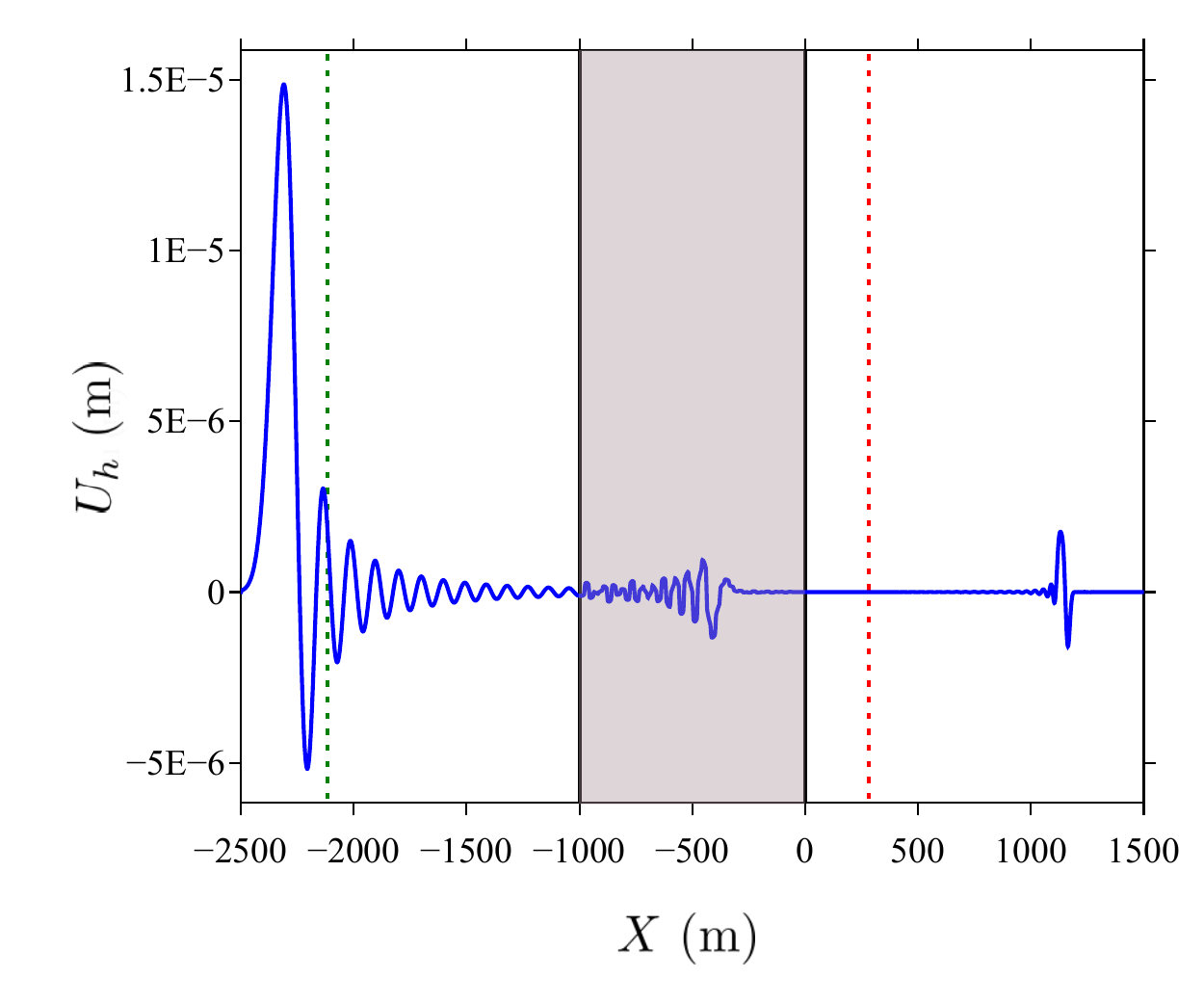} \\
\end{tabular}
\end{center}
\vspace{-0.6cm}
\caption{\label{FigSnapshotsU}Snapshots of $U_h$ at different times, with source such as $\eta_c=0.54$ \eqref{eta_numerique} on the left (a-c-e) or on the right (b-d-f) of the modulated slab (grey zone). Simulations are performed in the moving frame, with a left-going source (red dotted line) and a left-going receiver (green dotted line).} 
\end{figure}

A pulse is emitted by a Dirac source point at $X=X_s$ and with a central frequency $f_c$. The time evolution of the source is a smooth combination of sinusoids with bounded
support:
\begin{equation}
s(t)=
\ds \sum_{m=1}^4 a_m\,\sin(b_m \omega_\mathrm{c} t)\, \text{ if }\, 0<t<\frac{1}{f_\mathrm{c}},
 \, \quad 0 \text{ otherwise},
\label{Source}
\end{equation}
where $b_m=2^{m-1}$, the coefficients $a_m$ are $a_1=1$, $a_2=-21/32$, $a_3=63/768$, $a_4=-1/512$, ensuring $C^6$ smoothness (differentiable 6 times and the 6-th derivative is continuous). The typical small parameter \eqref{eq:nondimparam} for this type of signal is chosen to be the one associated with the central frequency i.e. 
\begin{equation}
    \label{eta_numerique}
    \eta_c=2\pi h f_c/c^\star,
\end{equation}
with $c^\star$ given by \eqref{eq:param_typ_num}.

In the moving frame, this source moves at speed $-c_m$.  Similarly, a receiver is considered at $X=X_r$; it moves at speed $-c_m$ and records the fields $V_h$ and $\Sigma_h$ at each time step. Numerical integration provides the displacement $U_h$. Lastly, the simulations are stopped before the time where the source or receiver exits from the computational domain or enters into the slab.


\subsection{Numerical results}\label{SecNumResults}
\begin{figure}[htbp]
\begin{center}
\begin{tabular}{cc}
\hspace{1cm}$\eta_c=0.27$ ($f_c$=3 Hz) & \hspace{1cm}$\eta_c=0.54$ ($f_c$=6 Hz)\\
\includegraphics[scale=0.33]{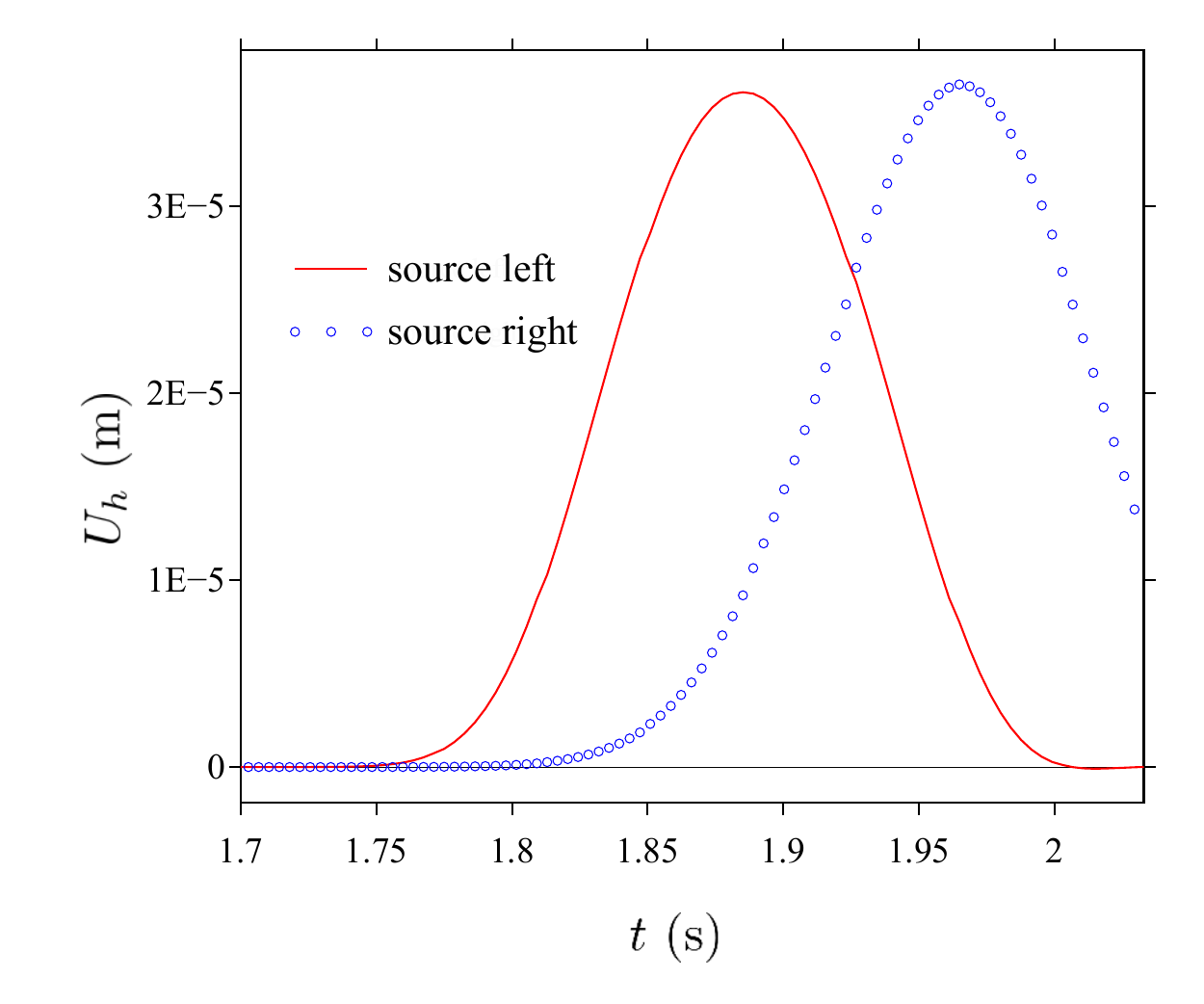} &
\includegraphics[scale=0.33]{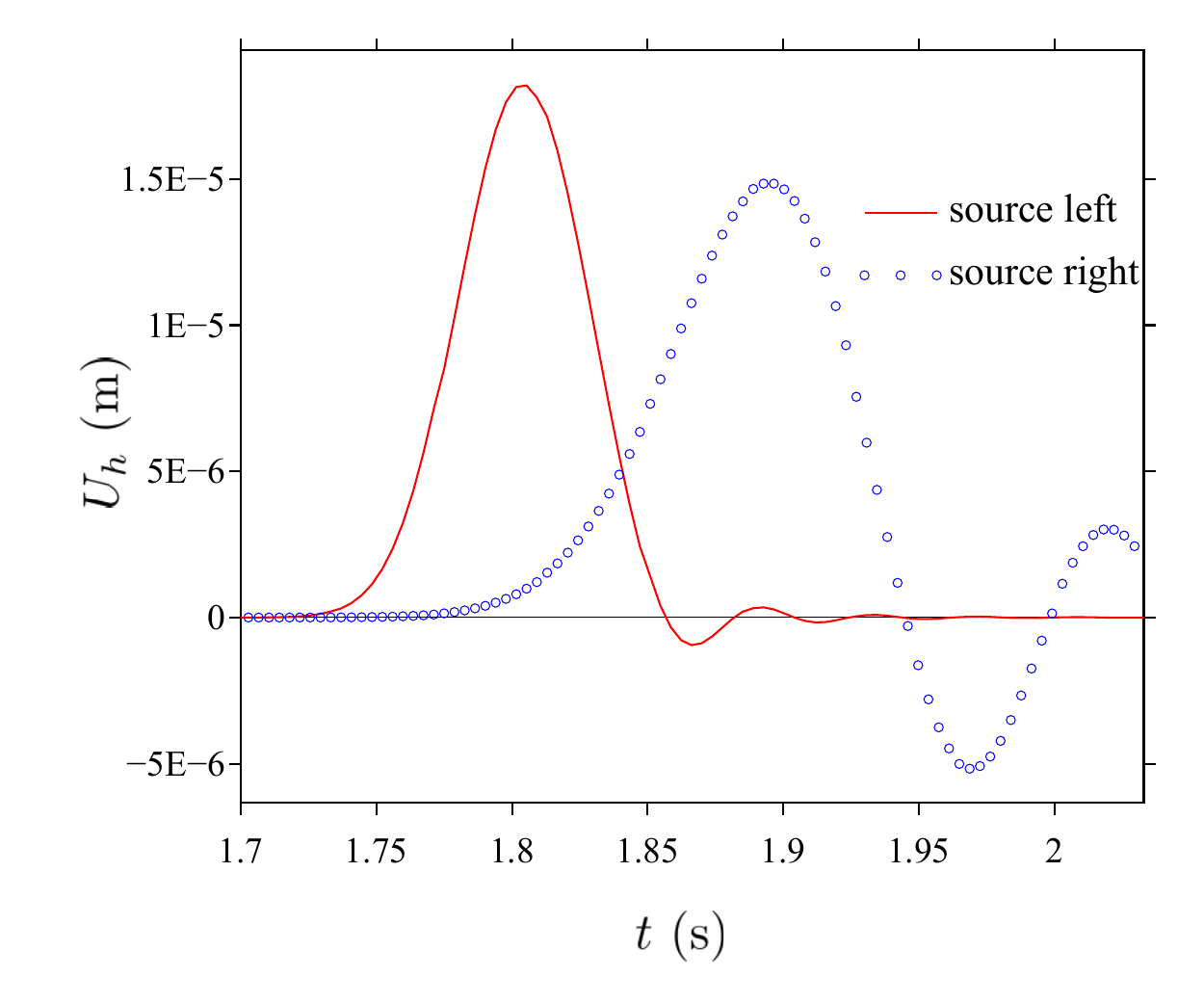} \\
\hspace{1cm}$\eta_c=0.81$ ($f_c$=9 Hz) & \hspace{1cm}$\eta_c=1.08$ ($f_c$=12 Hz)\\
\includegraphics[scale=0.33]{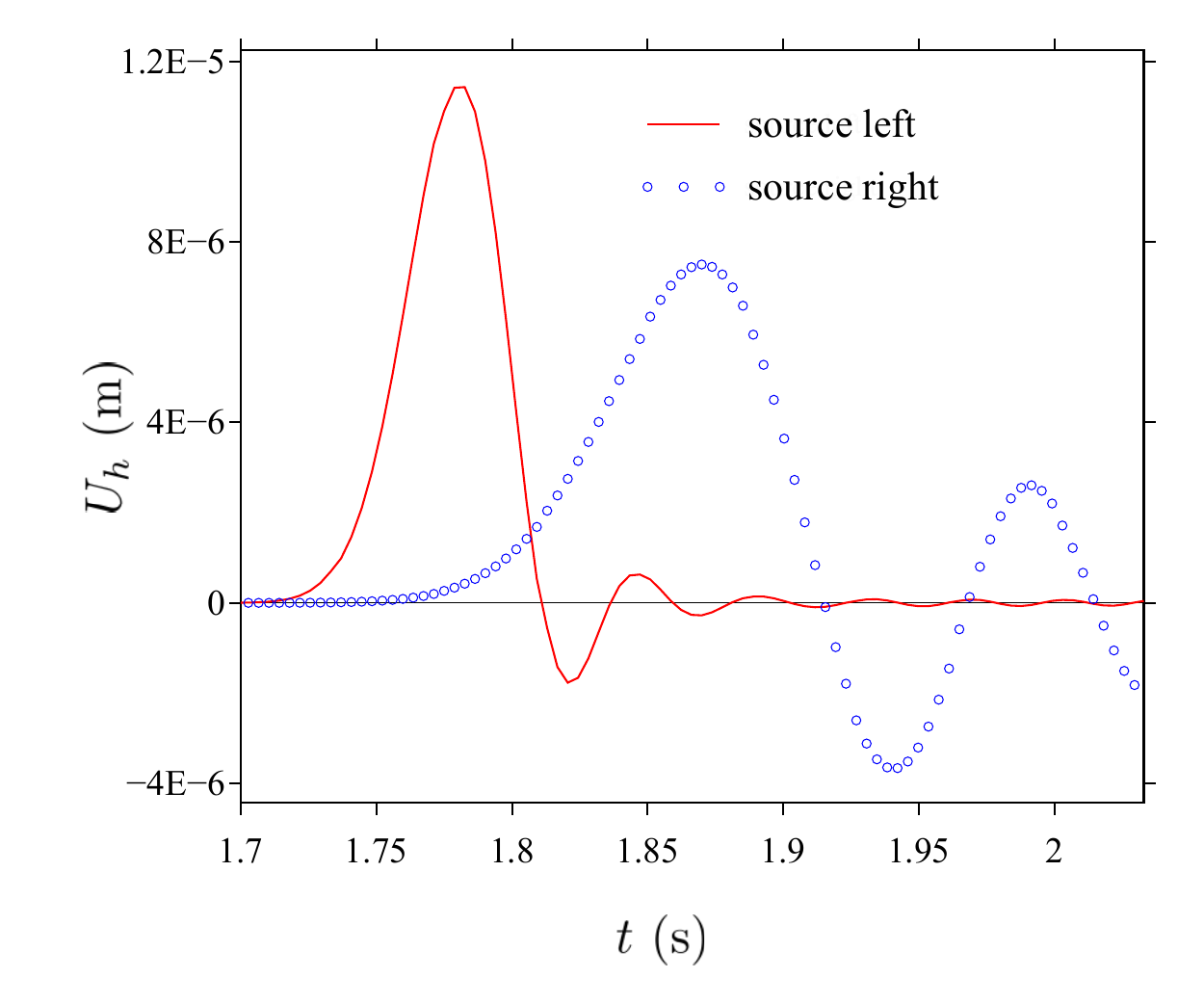}&
\includegraphics[scale=0.33]{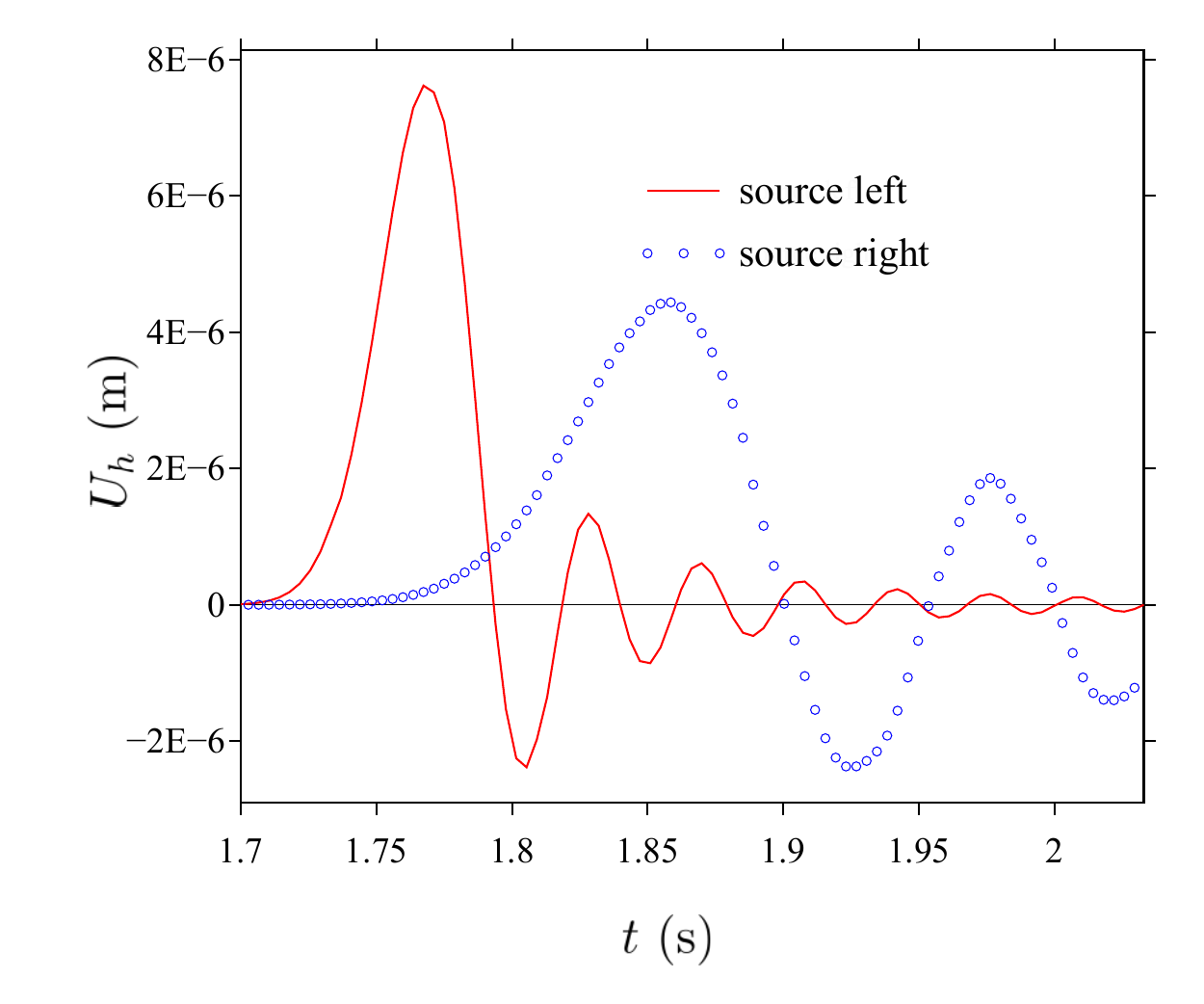} \\
\end{tabular}
\end{center}
\vspace{-0.6cm}
\caption{\label{FigTM-RE} Modulation of both parameters. Time history of $U_h$ measured at the receiver, for different small parameters $\eta_c$ \eqref{eta_numerique}. Plain and dotted lines denote the case of a source on the left and on the right of the slab, respectively.} 
\end{figure}

First we consider the case where both parameters are modulated given by the following values of the parameters in \eqref{Bilayer}:
\begin{equation}
    \label{param_Time_both}
  (\rho_A,\mu_A)=(10^3 \mathrm{kg}\cdot\mathrm{m}^{-3}, 10^9\mathrm{Pa}) \quad \text{ and } \quad (\rho_B,\mu_B)=(1.5\times 10^3 \mathrm{kg}\cdot\mathrm{m}^{-3},6\times 10^9\mathrm{Pa}),
\end{equation}
with the modulation velocity being $c_m=5\times 10^2\mathrm{m}\cdot\mathrm{s}^{-1}<\min(c_A,c_B)$, which corresponds to a subsonic case.  
Fig. \ref{FigSnapshotsU} displays snapshots of $U_h$ for $f_c=6$Hz, i.e. $\eta_c=0.54$  \eqref{eta_numerique}, at three different times. In the left column (a-c-e), the source is on the left of the slab, and the receiver is on the right. In the right column (b-d-f), the situation is reversed. Qualitatively, one observes the difference in scattered fields in both cases. Fig. \ref{FigTM-RE} displays the time evolution of the recorded $U_h$, for various central frequencies $f_c$. The differences by exchanging source and receiver are clearly observed.

\begin{figure}[htbp]
\begin{center}
\begin{tabular}{cc}
\hspace{1cm}$\eta_c=0.25$ ($f_c$=3 Hz) & \hspace{1cm}$\eta_c=0.50$ ($f_c$=6 Hz)\\
\includegraphics[scale=0.33]{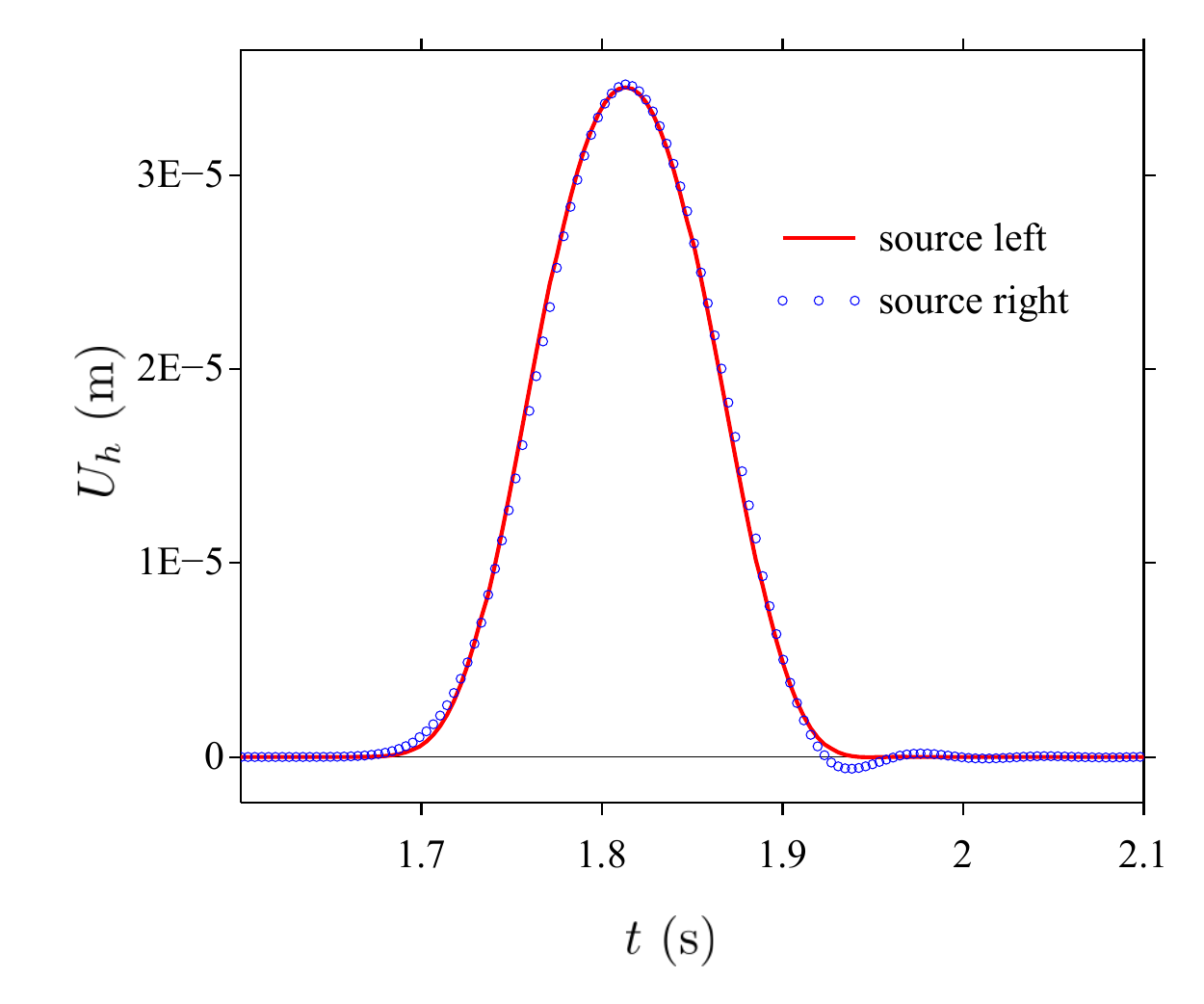} &
\includegraphics[scale=0.33]{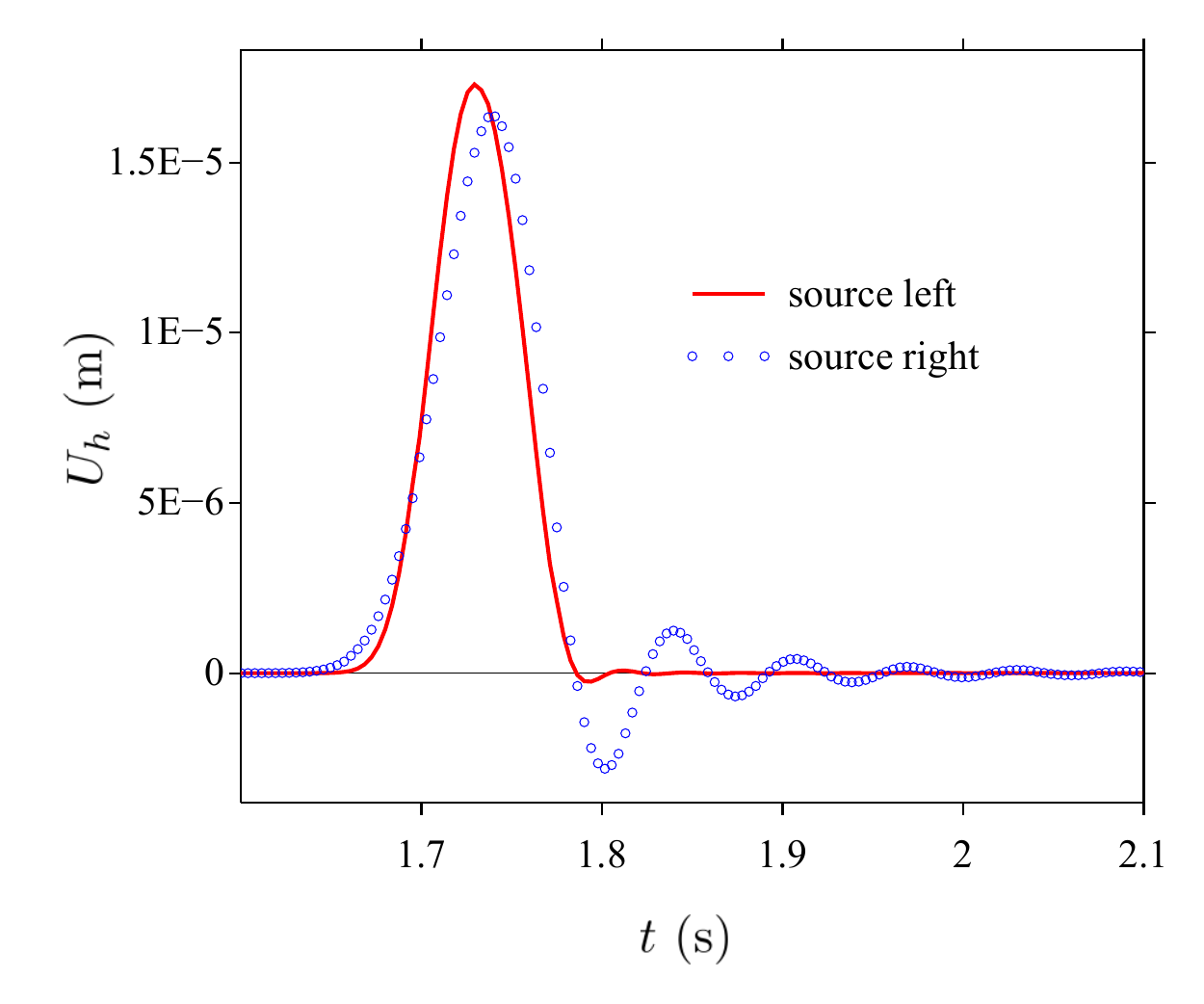} \\
\hspace{1cm}$\eta_c=0.75$ ($f_c$=9 Hz) &\hspace{1cm} $\eta_c=1.00$ ($f_c$=12 Hz)\\
\includegraphics[scale=0.33]{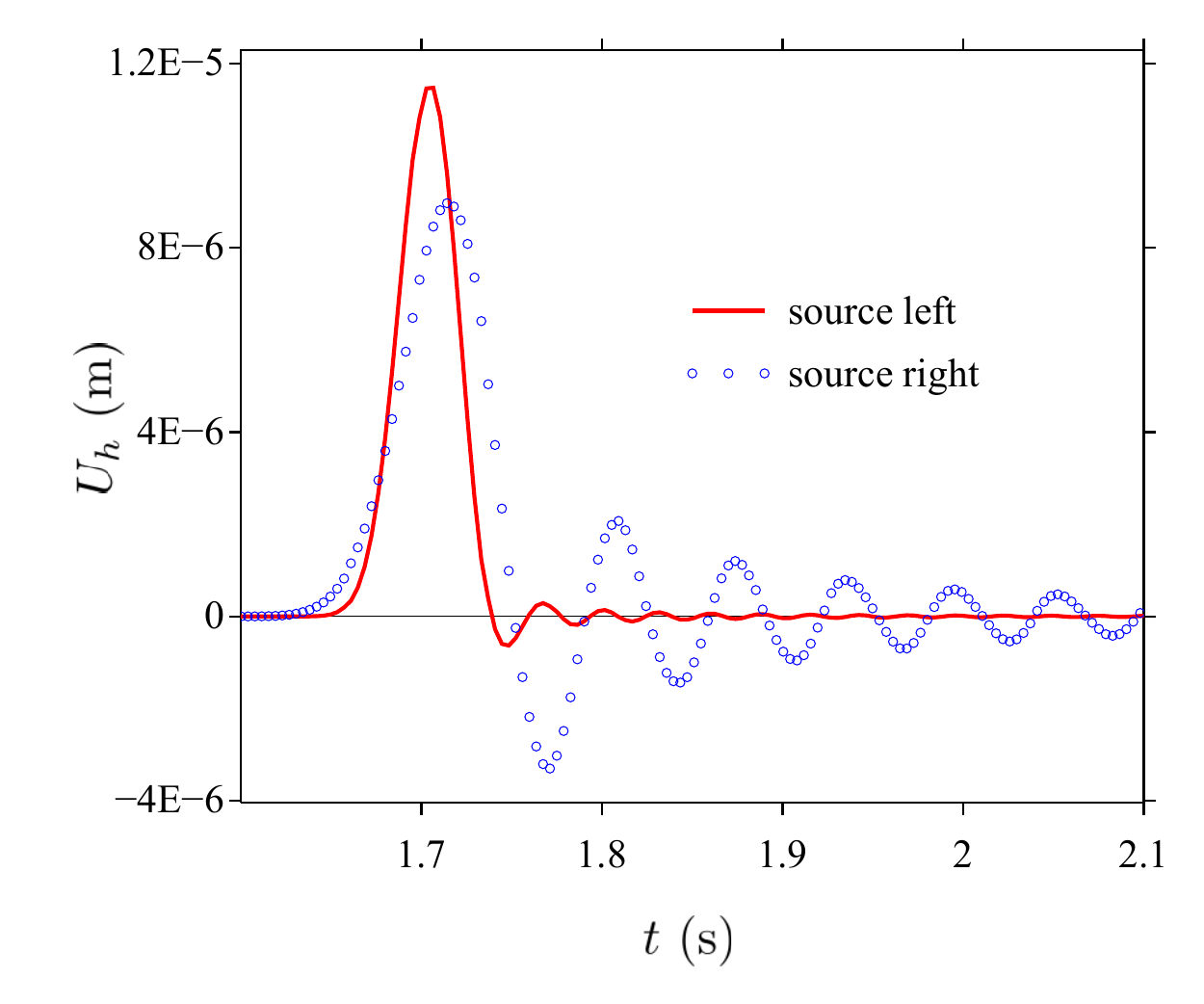} &
\includegraphics[scale=0.33]{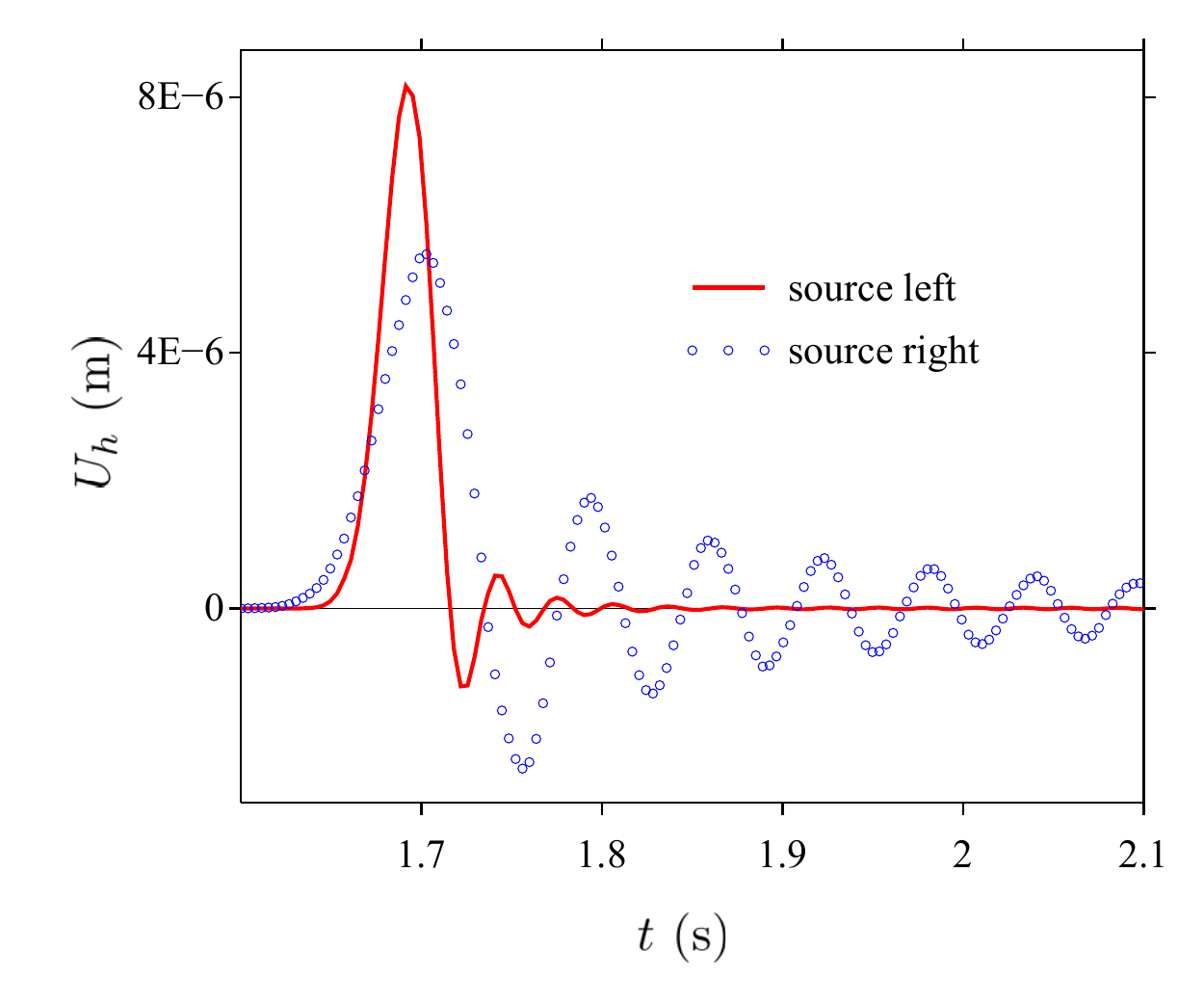} \\
\end{tabular}
\end{center}
\vspace{-0.6cm}
\caption{\label{FigTM-E}Modulation of $\mu_h$ with constant $\rho_h$. Time history of $U_h$ measured at the receiver, for different small parameters $\eta_c$ \eqref{eta_numerique}. Plain and dotted lines denote the case of a source on the left and on the right of the slab, respectively. } 
\end{figure}
Fig. \ref{FigTM-E} illustrates the case where only $\mu_h$ is modulated with a velocity $c_m=5\times 10^2\mathrm{m}\cdot\mathrm{s}^{-1}$. The parameters of the bilaminate for this case are given by:
\begin{equation}
    \label{param_Time_alpha1}
(\rho_A,\mu_A)=(1.375 \times 10^3 \mathrm{kg}\cdot\mathrm{m}^{-3},1.375\times 10^9\mathrm{Pa}) \text{, }  (\rho_B,\mu_B)=(1.375 \times 10^3\mathrm{kg}\cdot\mathrm{m}^{-3},5.5\times 10^9\mathrm{Pa}).
\end{equation}

\begin{figure}[htbp]
\begin{center}
\begin{tabular}{cc}
\hspace{1cm}$\eta_c=0.24$ ($f_c$=3 Hz) & \hspace{1cm}$\eta_c=0.49$ ($f_c$=6 Hz)\\
\includegraphics[scale=0.33]{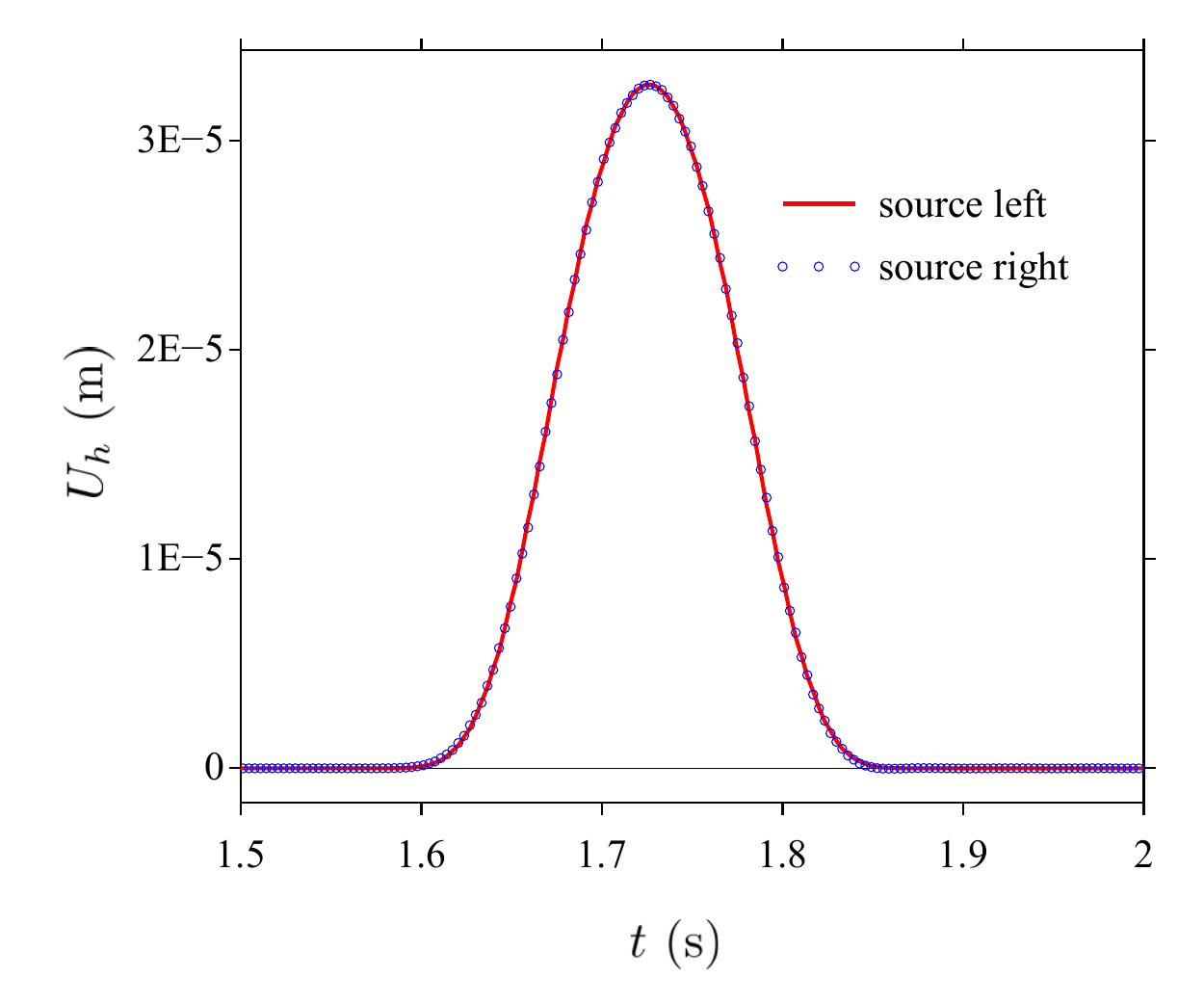} &
\includegraphics[scale=0.33]{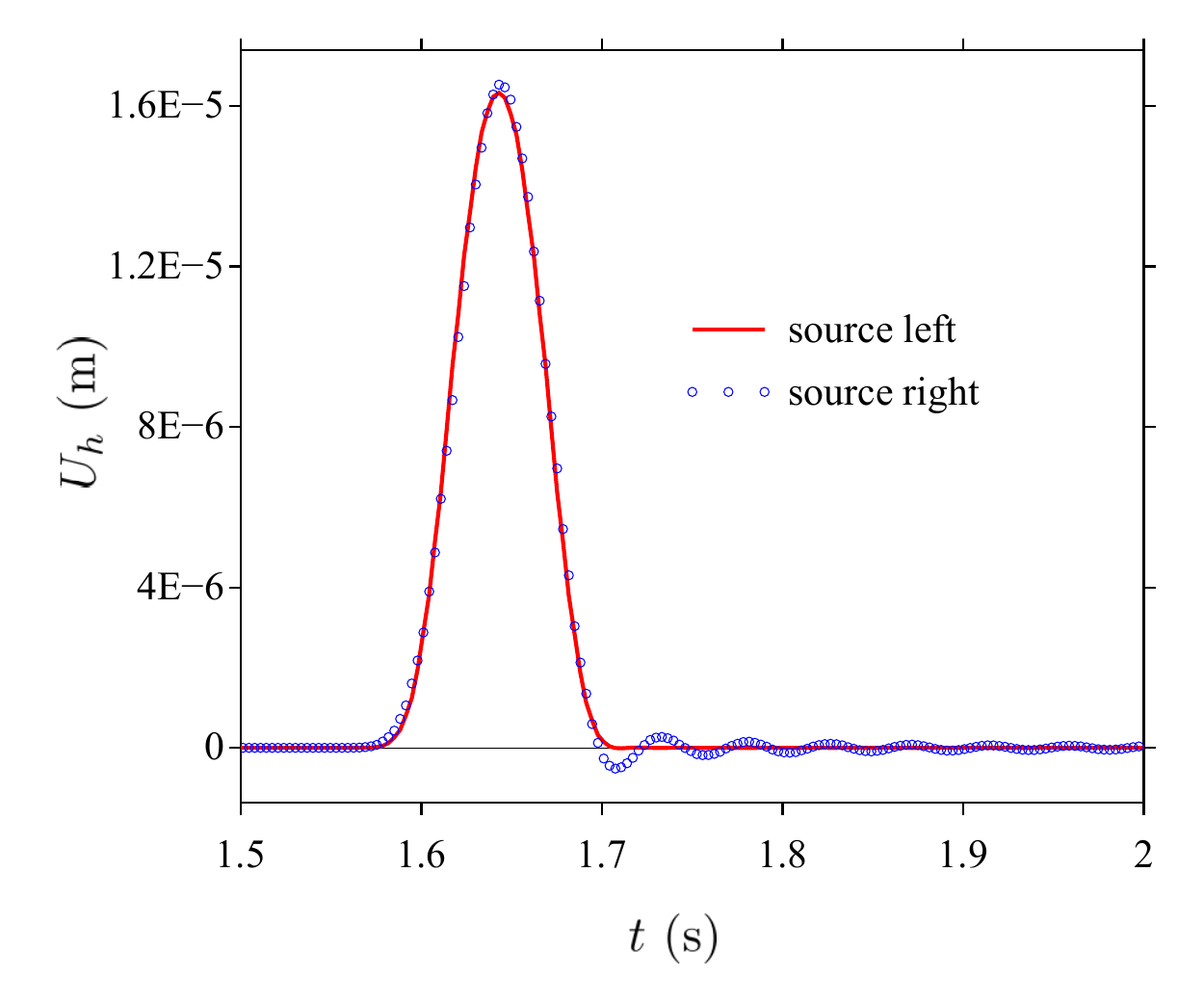} \\
\hspace{1cm}$\eta_c=0.73$ ($f_c$=9 Hz) & \hspace{1cm}$\eta_c=0.97$ ($f_c$=12 Hz)\\
\includegraphics[scale=0.33]{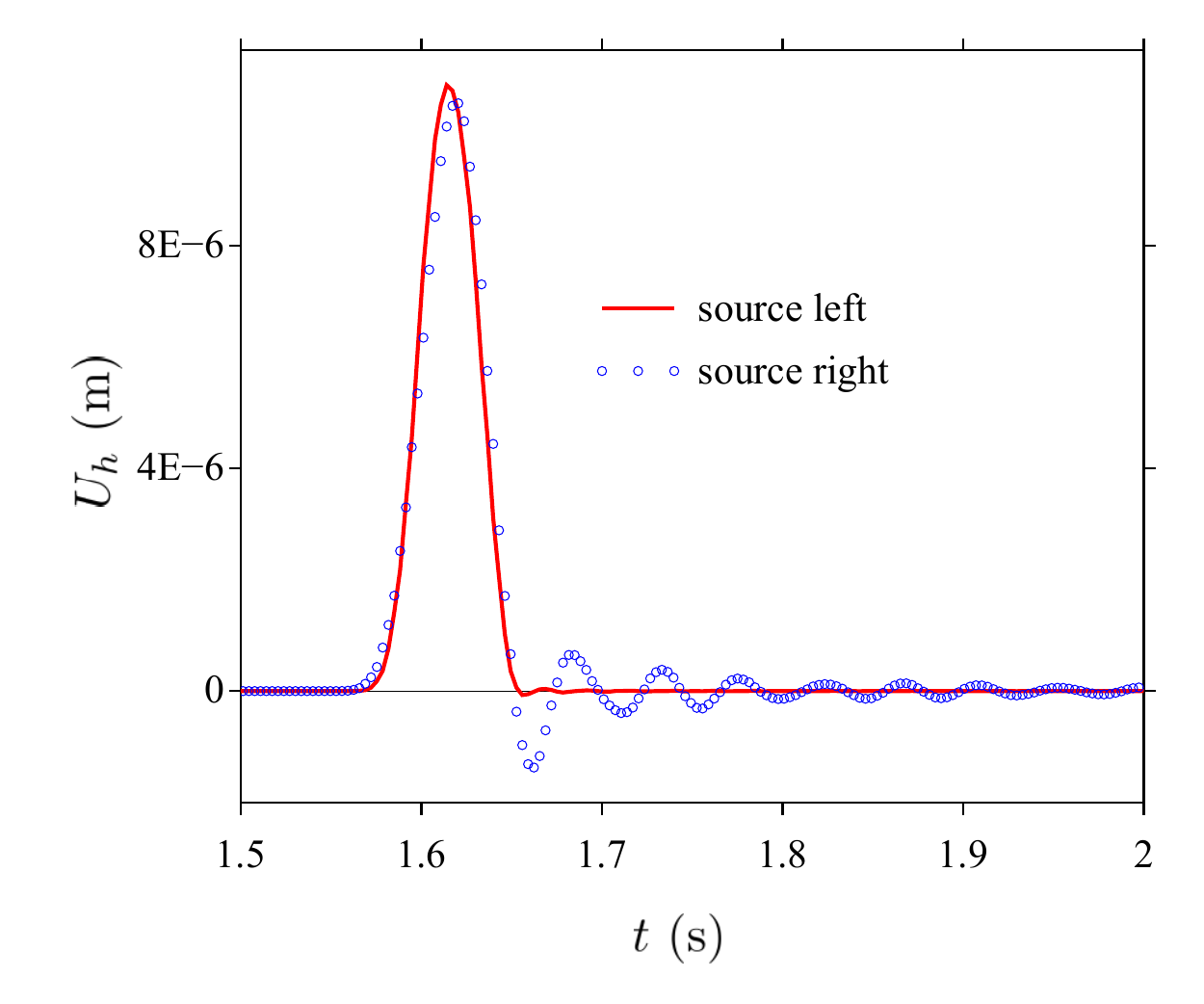} &
\includegraphics[scale=0.33]{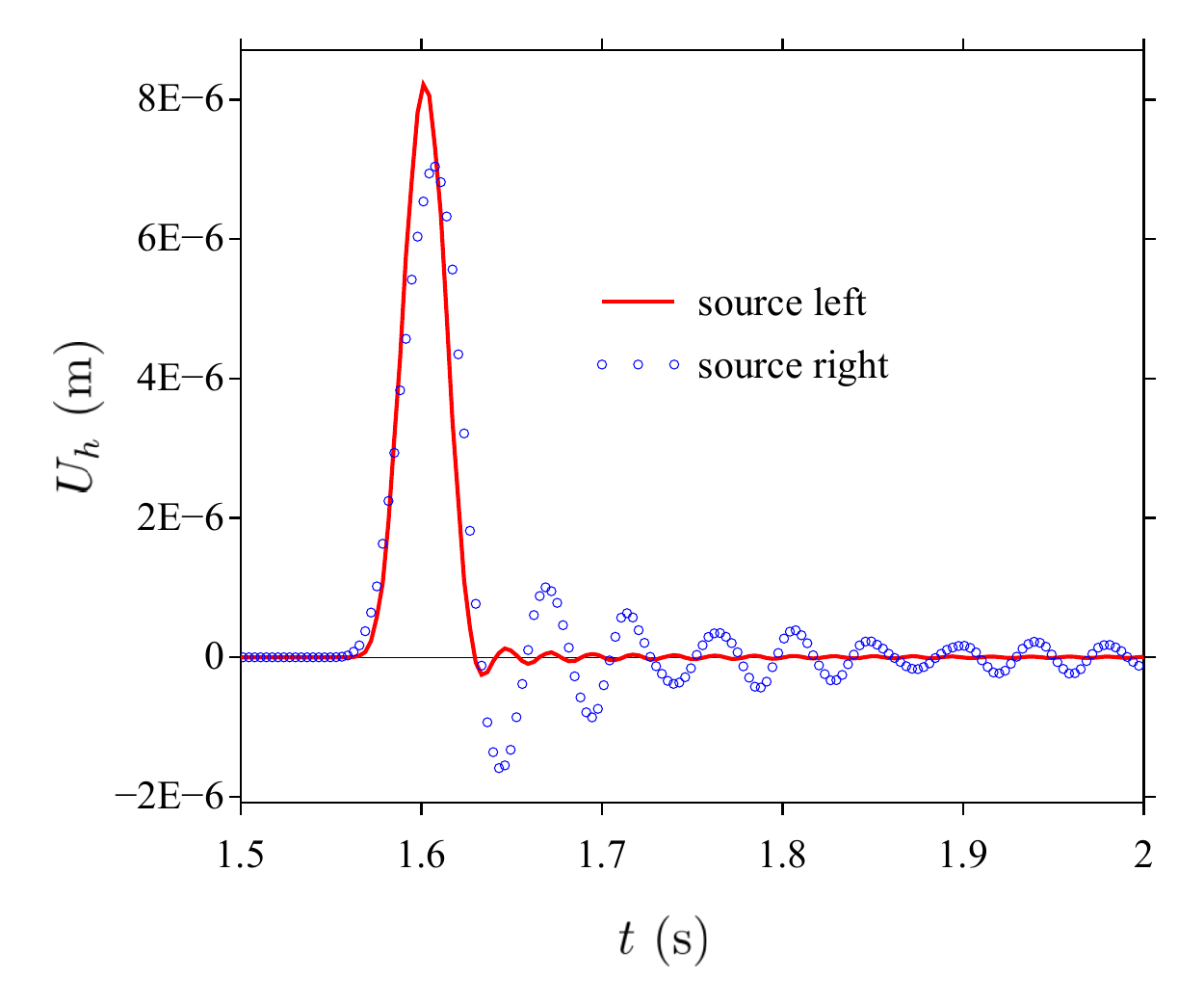} \\
\end{tabular}
\end{center}
\vspace{-0.6cm}
\caption{\label{FigTM-R}Modulation of $\rho_h$ with constant $\mu_h$. Time history of $U_h$ measured at the receiver, for different small parameters $\eta_c$ \eqref{eta_numerique}. Plain and dotted lines denote the case of a source on the left and on the right of the slab, respectively.} 
\end{figure}

In this case, the differences obtained by exchanging source and receiver are still clearly observed, which is a signature of non-reciprocity. Similar observations have been done by modulating only $\rho_h$ following
\begin{equation}
    \label{param_Time_alpha1}
    (\rho_A,\mu_A)=(10^3 \mathrm{kg}\cdot\mathrm{m}^{-3},6\times 10^9\mathrm{Pa})\quad \text{ and } \quad (\rho_B,\mu_B)=(3\times 10^3 \mathrm{kg}\cdot\mathrm{m}^{-3},6\times 10^9\mathrm{Pa}),
\end{equation}
and the results of the corresponding time-domain simulations are given by Fig. \ref{FigTM-R}. In both figures, we observe a greater dispersive behaviour for the left-going waves (the ones emited by the source on the right), which is consistent with the dispersion diagrams of Fig. \ref{fig:DD_alpha1} and \ref{fig:DD_beta1}.

\section{Conclusion}
In the present paper, we performed high-order homogenisation of the 1D wave equation when the physical parameters  are modulated in space and time in a wave-like fashion.
Approximating fields up to the second order allows to take into account dispersive effects, and most importantly, non-reciprocity that appears at this order does not vanish as we vary only one of the parameters. We therefore proved that it is enough to modulate a single parameter in order to achieve low-frequency non-reciprocity for the acoustic wave equation. 
Similar properties then hold for the corresponding equations in the case of transverse electric (\ref{maxwell4}) and transverse magnetic (\ref{maxwell6}) waves. The natural choice for SH waves is to modulate in space and time the shear modulus $\mu$ \cite{Danas2012} and keep constant the mass density $\rho$. For transverse electromagnetic waves, it is to time-modulate the relative permittivity $\varepsilon$ and assume the relative permeability $\mu=1$ (i.e. no magnetism) in (\ref{maxwell4}) and (\ref{maxwell6}). \\
Thus, non-reciprocity does not require modulation of both parameters as was assumed in elasticity and electromagnetism \cite{Nassar2017,Huidobro2021}. We hope this will foster efforts towards an experimental demonstration, e.g. of the Fresnel drag effect in time-modulated dielectric layers \cite{Huidobro2019}.\\
Some theoretical aspects have been disregarded in this paper, such as the boundary effects in the case of a finite medium \cite{BeneteauPhd,Vinoles,Maurel2018b,Cornaggia2023}, or the behaviour for long times \cite{LAMACZ2011,Allaire2022}; they could be investigated in future works. Field pattern space-time composites with PT-symmetry such as chessboards as introduced by Milton and Mattei \cite{milton2017field} also represent a further challenge. We could further explore optimal design of time-modulated media \cite{cherkaev2023optimal}.
As another perspective, our asymptotic analysis can be applied to time-modulated conductivity, hydrodynamic and elasticity equations (we note in passing the fascinating work \cite{movchan2022frontal} on temporal elastic laminates with imperfect chiral interfaces), as well as other governing equations in physics described by linear partial differential equations, where we expect that similar non-reciprocal effects can be unveiled in the homogenisation regime. 
Another perspective concerns the case of locally-resonant media \cite{Felbacq2005,Auriault2012}, to combine both low-frequency non-reciprocity and low-frequency resonances for more important macroscopic effects. Finally, high-frequency homogenisation \cite{Craster2010} could be applied to get effective properties around the opening of stop-bands either in frequency or in wavenumber. \vskip6pt

\enlargethispage{20pt}

\appendix 
\section{Reciprocity identities}\label{Sec:App}
The cell problems introduced to express $u_1$, $u_2$ and $u_3$ lead to a family of reciprocity identities that are listed here and extensively used in the homogenisation process. \\
Integration by parts of $\langle\eqref{cell_pb_Q}\times P - \eqref{cell_pb_P}\times Q\rangle$ leads to
\begin{equation}
    \label{eq:IPP6}
    \langle\beta Q'\rangle = c\langle\alpha P'\rangle
\end{equation}
Integration by parts of $\langle\eqref{cell_pb_R}\times P - \eqref{cell_pb_P}\times R\rangle$ leads to
\begin{equation}
    \label{eq:IPP1}
    \langle \beta R'\rangle = -\langle \beta P\rangle+\frac{\beta_0}{\alpha_0}\langle \alpha(1-cQ')P\rangle +\frac{\beta_0}{\alpha_0}c\langle\alpha QP'\rangle
\end{equation}
Integration by parts of $\langle\eqref{cell_pb_R}\times Q - \eqref{cell_pb_Q}\times R\rangle$ leads to
\begin{equation}
    \label{eq:IPP2}
    c\langle\alpha R'\rangle = -\langle\beta Q\rangle +\langle\beta P Q'\rangle +\langle\beta Q P'\rangle +\frac{\beta_0}{\alpha_0}\langle \alpha Q\rangle
\end{equation}
Integration by parts of $\langle\eqref{cell_pb_S}\times P - \eqref{cell_pb_P}\times S\rangle$ leads to
\begin{equation}
    \label{eq:IPP3}
    \langle \beta S' \rangle =-\langle\beta Q'P\rangle +\langle \beta Q P'\rangle +\frac{2W_0}{\alpha_0}\langle\alpha(1-cQ')P+c\alpha Q P'\rangle 
\end{equation}
Integration by parts of $\langle\eqref{cell_pb_Q}\times S - \eqref{cell_pb_S}\times Q\rangle$ leads to
\begin{equation}
    \label{eq:IPP4}
    c\langle\alpha S'\rangle = -c\langle\alpha P'Q\rangle+c\langle\alpha PQ'\rangle +\frac{2W_0}{\alpha_0}\langle \alpha Q\rangle 
\end{equation}
Integration by parts of $\langle\eqref{cell_pb_P}\times B - \eqref{cell_pb_B}\times P\rangle$ leads to
\begin{equation}
    \label{eq:IPPB1}
    \langle \beta B' \rangle =\frac{1}{\alpha_0}\left[ \langle\alpha P\rangle +c\langle\alpha(QP'-PQ') \rangle\right]
\end{equation}
Integration by parts of $\langle\eqref{cell_pb_Q}\times B - \eqref{cell_pb_B}\times Q\rangle$ leads to
\begin{equation}
    \label{eq:IPPB2}
    c\langle\alpha B'\rangle = \frac{1}{\alpha_0}\langle \alpha Q \rangle
\end{equation}
Integration by parts of $\langle\eqref{cell_pb_R}\times S - \eqref{cell_pb_S}\times R\rangle$ leads to
\begin{equation}
    \label{eq:IPP5}
    \begin{aligned}
    &-\langle\beta(1+P')S\rangle+\langle\beta PS'\rangle +\frac{\beta_0}{\alpha_0}\langle\alpha(1-cQ')S\rangle+\frac{\beta_0}{\alpha_0}c\langle\alpha QS'\rangle \\&=-\langle (\alpha cP' +\beta Q')R\rangle
   + \langle (c\alpha P+\beta Q)R'\rangle+\frac{2W_0}{\alpha_0}\langle\alpha(1-cQ')R\rangle+\frac{2W_0}{\alpha_0}c\langle\alpha QR'\rangle
    \end{aligned}
\end{equation}
Integration by parts of $\langle\eqref{cell_pb_V}\times Q - \eqref{cell_pb_Q}\times N\rangle$ leads to
\begin{equation}
    \label{eq:IPP7}
    c\langle\alpha N'\rangle = \langle (\frac{\beta_0}{\alpha_0}\alpha-\beta) Q^2\rangle + c \langle \alpha ( R Q' - R'Q) \rangle  +\langle\beta (SQ'-S'Q) \rangle +\frac{2W_0}{\beta_0}\langle  \beta P Q + \beta(R'Q-RQ')\rangle 
\end{equation}
Integration by parts of $\langle\eqref{cell_pb_V}\times P - \eqref{cell_pb_P}\times N\rangle$ leads to
\begin{equation}
    \label{eq:IPP8}
\langle \beta N'\rangle = \langle (\frac{\beta_0}{\alpha_0}\alpha-\beta )PQ\rangle +c\langle \alpha (RP'-R'P)\rangle +\langle \beta(SP'-S'P)\rangle+\frac{2W_0}{\beta_0}\langle \beta P^2+\beta(R'P-RP')\rangle
\end{equation}
Integration by parts of $\langle\eqref{cell_pb_Q}\times M - \eqref{cell_pb_T}\times Q\rangle$ leads to
\begin{equation}
    \label{eq:IPP9}
    c\langle\alpha M'\rangle = \langle \alpha PQ\rangle +c\langle \alpha (SQ'-S'Q) \rangle+\frac{2W_0}{\alpha_0}\langle\alpha Q^2\rangle-\frac{\alpha_0}{\beta_0}\langle\beta PQ +\beta(R'Q-RQ') \rangle
\end{equation}
Integration by parts of $\langle\eqref{cell_pb_P}\times M - \eqref{cell_pb_T}\times P\rangle$ leads to 
\begin{equation}
    \label{eq:IPP10}
    \langle \beta M'\rangle = \langle (\alpha-\frac{\alpha_0}{\beta_0}\beta)P^2\rangle +c\langle\alpha(SP'-S'P)\rangle +\frac{2W_0}{\alpha_0}\langle\alpha PQ\rangle +\frac{\alpha_0}{\beta_0}\langle \beta(RP'-R'P)\rangle
\end{equation}

\section{Transverse electromagnetic waves}\label{Sec:EM}
We now consider the two-scale homogenisation of the vector Maxwell system of a time-modulated layered medium. Let us consider a fixed space-time Cartesian coordinate system $({\bf x},t)=(x_1,x_2,x_3,t)$ and a time modulated layered periodic medium. The propagation is along the direction $x_1$ of stacking of layers.
In what follows, the subscript denotes dependence of the field upon the periodicity $\eta$ in the space-time variable $x_1-c_1 t$, where $c_1$ is the modulation speed along $x_1$.

We first consider the Maxwell-Faraday equation
that relates the magnetic inductance field ${\bf B}_\eta$ and the electric field ${\bf E}_\eta$ via
\begin{equation}
\rot{\bf E}_\eta +\frac{\partial}{\partial t}{\bf B}_\eta={\bf 0}.
\label{maxwell1}  
\end{equation}
This equation implies that $\nabla\cdot{\bf B}_\eta({\bf x},t)=0$, from which it follows
that there is a vector magnetic potential field
such that
\begin{equation}
{\bf B}_\eta({\bf x},t)=\rot {\bf A}_\eta({\bf x},t).
\label{vecpot1}  
\end{equation}
We use this vector potential as an unknown rather than the electric and magnetic fields.
Combining (\ref{maxwell1}) and (\ref{vecpot1}), it follows that ${\bf E}_\eta+\frac{\partial}{\partial t}{\bf A}_\eta$ is curl free and thus there exists a scalar electric potential $\phi_\eta$ such that
\begin{equation}
{\bf E}_\eta({\bf x},t)=-\frac{\partial}{\partial t}{\bf A}_\eta({\bf x},t) - \nabla\phi_\eta({\bf x},t).
\label{vecpot2}  
\end{equation}

Besides, the magnetic field ${\bf H}_\eta$ and the electric displacement field ${\bf D}_\eta$ satisfy Maxwell-Ampère's equation  
\begin{equation}
\rot{\bf H}_\eta -\frac{\partial}{\partial t}{\bf D}_\eta={\bf 0} \; .
\label{maxwell2} 
\end{equation}
Furthermore, the electric displacement ${\bf D}_\eta$ (resp. magnetic inductance ${\bf B}_\eta$) is related to the electric field ${\bf E}_\eta$ (resp. magnetic field ${\bf H}_\eta$) via
\begin{equation}
{\bf D}_\eta=\varepsilon_0\varepsilon\left(\frac{x_1-c_1 t}{\eta}\right){\bf E}_\eta \; ,\quad \; {\bf B}_\eta=\mu_0\mu\left(\frac{x_1-c_1 t}{\eta}\right){\bf H}_\eta \; ,
\label{maxwellconstitute}
\end{equation}
where $\varepsilon_0\mu_0=1/c_0^2$, $c_0$ being the speed of light in vacuum, and 
where $\varepsilon$ and $\mu$ respectively denote the relative permittivity and permeability. These are $1$ periodic functions of $x_1$ and $1/c_1$ periodic functions of $t$.

From (\ref{vecpot1})-(\ref{maxwellconstitute}), we deduce that the magnetic potential ${\bf A}_\eta$ and the electric potential $\phi_\eta$ satisfy
\begin{equation}
\rot\left[\mu^{-1}\left(\frac{x_1-c_1 t}{\eta}\right)\rot {\bf A}_\eta \right] 
=-\frac{1}{c_0^2}\frac{\partial}{\partial t}\left[\varepsilon\left(\frac{x_1-c_1 t}{\eta}\right)\frac{\partial}{\partial t}{\bf  A}_\eta\right]
-\frac{1}{c_0^2}\frac{\partial}{\partial t}\left[\varepsilon\left(\frac{x_1-c_1 t}{\eta}\right)\nabla{\phi}_\eta\right]
\label{maxwell3} 
\end{equation}

Let us further assume that the electric field is orthogonal to the plane $(x_1x_2)$ of the layering (transverse electric, TE, case). Eq. (\ref{vecpot2}) implies that $
(0,0,E_\eta(x_1,t))^T
=-(\frac{\partial}{\partial x_1}\phi_\eta(x_1,t),\frac{\partial}{\partial x_2}\phi_\eta(x_1,t),\frac{\partial}{\partial t}A_\eta(x_1,t)+\frac{\partial}{\partial x_3}\phi_\eta(x_1,t))^T=-(0,0,\frac{\partial}{\partial t}A_\eta(x_1,t))^T$, where $A_\eta$ is the only non-zero component of ${\bf A}_\eta$, and so $\nabla\phi_\eta$ must be a null vector field in (\ref{maxwell3}).

We deduce that in TE polarisation and for the considered layering, (\ref{maxwell3}) reduces to
\begin{equation}
\frac{\partial}{\partial x_1}\left[\mu^{-1}\left(\frac{x_1-c_1 t}{\eta}\right)\frac{\partial}{\partial x_1} {A}_\eta \right] 
= \frac{1}{c_0^2}\frac{\partial}{\partial t}\left[\varepsilon\left(\frac{x_1-c_1 t}{\eta}\right)\frac{\partial}{\partial t}{A}_\eta\right].
\label{maxwell4} 
\end{equation}
We note that (\ref{non_dim_equation}) and (\ref{maxwell4}) are identical upon the replacement of $\alpha$ by
$\mu^{-1}$ and of $\beta$ by $c_0^{-2}\varepsilon$.

Maxwell's equations are fully symmetric, so we can now interchange the roles of the vector magnetic potential ${\bf A}_\eta$ and the scalar electric potential $\phi_\eta$ with a vector electric potential ${\bf V}_\eta$ and a scalar electric potential $\psi$. More precisely, (\ref{maxwell2}) implies that $\nabla\cdot{\bf D}_\eta({\bf x},t)=0$, from which it follows
that there is a vector magnetic potential field
such that
\begin{equation}
{\bf D}_\eta({\bf x},t)=\rot {\bf V}_\eta({\bf x},t).
\label{vecpot1b}  
\end{equation}

Combining (\ref{maxwell2}) and (\ref{vecpot1b}),
we know there exists a scalar magnetic potential $\psi_\eta$ such that
\begin{equation}
{\bf H}_\eta({\bf x},t)=-\frac{\partial}{\partial t}{\bf V}_\eta({\bf x},t) - \nabla\psi_\eta({\bf x},t).
\label{vecpot2b}  
\end{equation}

From (\ref{maxwell1}), (\ref{maxwellconstitute}), (\ref{vecpot1b}) and (\ref{vecpot2b}), we deduce that
${\bf V}_\eta$
and
$\psi_\eta$ satisfy
\begin{equation}
\rot\left[\varepsilon^{-1}\left(\frac{x_1-c_1 t}{\eta}\right)\rot {\bf V}_\eta \right] 
=-\frac{1}{c_0^2}\frac{\partial}{\partial t}\left[\mu\left(\frac{x_1-c_1 t}{\eta}\right)\frac{\partial}{\partial t}{\bf  V}_\eta\right]
-\frac{1}{c_0^2}\frac{\partial}{\partial t}\left[\mu\left(\frac{x_1-c_1 t}{\eta}\right)\nabla{\psi}_\eta\right]
\label{maxwell5} 
\end{equation}

Let us now assume that the magnetic field is orthogonal to the plane $(x_1x_2)$ of the layering (transverse magnetic, TM, case). From Eq. (\ref{vecpot1b})
we deduce that $\nabla\psi_\eta$ is a null vector field in (\ref{maxwell5}).

Therefore, in TM polarisation and for the considered layering, (\ref{maxwell5}) reduces to
\begin{equation}
\frac{\partial}{\partial x_1}\left[\varepsilon^{-1}\left(\frac{x_1-c_1 t}{\eta}\right)\frac{\partial}{\partial x_1} {V}_\eta \right] 
= \frac{1}{c_0^2}\frac{\partial}{\partial t}\left[\mu\left(\frac{x_1-c_1 t}{\eta}\right)\frac{\partial}{\partial t}{V}_\eta\right]
\label{maxwell6} 
\end{equation}

We note that (\ref{non_dim_equation}) and (\ref{maxwell6}) are identical upon the replacement of $\alpha$ by $\varepsilon^{-1}$ and of $\beta$ by $c_0^{-2}\mu$. Therefore, it is enough to modulate the permittivity $\varepsilon$ in Eq. (\ref{maxwell4}) and Eq. (\ref{maxwell6}) to achieve some non-reciprocity in the homogenisation frequency regime in TE and TM polarizations. Our work thus complements leading-order homogenisation results in \cite{Huidobro2019, Huidobro2021} wherein space-time modulation of both permittivity and permeability was a prerequisite for non-reciprocity.

\section*{Acknowledgments} 
SG and RVC were funded by UK Research and Innovation (UKRI) under the UK government’s Horizon Europe funding guarantee [grant number 10033143]. MT and RA would like to thank the Isaac Newton Institute for Mathematical Sciences, Cambridge, for support and hospitality during the programme Mathematical theory and applications of multiple wave scattering where work on this paper was undertaken.

\bibliographystyle{unsrt}  
\bibliography{sample}  
\end{document}